\documentclass[preprint,12pt]{emulateapj}
\usepackage{graphicx}
\usepackage{natbib}
\usepackage{amssymb}
\usepackage{amsbsy}
\usepackage{natbib}
\usepackage[mathcal]{euscript}
\usepackage{lscape}
\bibpunct{(}{)}{;}{a}{}{,}

\shorttitle{Photometric monitoring of $\sigma$ Orionis}

\begin{document}
 
\def\etal{et al.\ }

\title{Precision photometric monitoring of very low mass $\sigma$ Orionis cluster members: variability and 
rotation at a few Myr }
\author{Ann Marie Cody\altaffilmark{1} and Lynne Hillenbrand}
\affil{California Institute of Technology, Department of Astrophysics, MC 249-17, Pasadena, CA 91125}

\altaffiltext{1}{amc@astro.caltech.edu; visiting astronomer, Cerro Tololo Inter-American Observatory. CTIO 
is operated by the Association of Universities for Research in Astronomy, Inc., under contract to the 
National Science Foundation.}
%\altaffiltext{2}{lah@astro.caltech.edu}

\begin{abstract}

We present high-precision photometry on 107 variable low-mass stars and brown dwarfs in the $\sim$3~Myr 
$\sigma$~Orionis open cluster. We have carried out $I$-band photometric monitoring within two fields, 
encompassing 153 confirmed or candidate members of the low-mass cluster population, from 0.02 to 
0.5~$M_\odot$. We are sensitive to brightness changes on time scales from 10 minutes to two weeks with 
amplitudes as low as 0.004 magnitudes, and find variability on these time scales in nearly 70\% of 
cluster members. We identify both periodic and aperiodic modes of variability, as well as semi-periodic 
rapid fading events that are not accounted for by the standard explanations of rotational modulation of 
surface features or accretion. We have incorporated both optical and infrared color data to uncover 
trends in variability with mass and circumstellar disks. While the data confirm that the lowest-mass 
objects ($M<0.2M_\odot$) rotate more rapidly than the 0.2--0.5$M_\odot$ members, they do not support a 
direct connection between rotation rate and the presence of a disk. Finally, we speculate on the origin 
of irregular variability in cluster members with no evidence for disks or accretion.

\end{abstract}


\section{Introduction}

Stars and brown dwarfs in the $\sim$1-15~Myr age range occupy a pivotal position in the stellar evolution 
sequence, characterized by emergence from molecular cloud birthplaces, ongoing dissipation of primordial 
circumstellar disks, and assembly of planet systems. The evolutionary stage also involves dramatic 
changes in internal structure as well as radius and angular momentum. Some circumstellar and stellar 
changes during this epoch are interconnected, through deposition of accreting material on the central 
object, as well as possible transfer of angular momentum to the surrounding disk. Although the physics 
governing these processes remains difficult to probe directly, accompanying photometric variability 
offers a valuable tracer of the prevalence of various underlying phenomena at work.

It has long been known that pre-main-sequence T~Tauri stars with masses near solar exhibit variability on 
levels of $\sim$1-50\% \citep{1949ApJ...110..424J}. At visible and near-infrared wavelengths, prominent 
phenomena causing photometric variability include modulations of the stellar brightness by rotation of cool 
magnetic surface spots, sporadic flux variations due to accretion, extinction fluctuations due to clumpy 
circumstellar material, and eclipses by companions. Data derived from temporal variability studies complement 
single-epoch surveys of stellar populations spanning a range of spectral types and ages in nearby young 
clusters by contributing information on changes occurring much faster than the evolutionary time scale. 
Photometric monitoring campaigns have thus become an integral part of our toolbox in the investigation of 
young cluster members.

Among the most appreciated stellar parameters accessed through time series monitoring is the rotational 
angular momentum. For objects with periodic brightness changes that can be attributed to the passage of cool 
surface spots, photometric variability analyses yield rotation rates. Recent work has established the overall 
angular momentum trends from the pre-main-sequence (PMS) through ages of 500 Myr, as reviewed by 
\citet{2007prpl.conf..297H}, \citet{2007IAUS..243..231B}, and \citet{2009AIPC.1094...61S}. Of particular 
interest is the 1-10~Myr regime, which is the first opportunity to measure the cumulative effect of the 
formation process on rotation rates after the embedded phases of protostellar development. During these early 
stages, a large portion of the initial angular momentum is carried off by outflows and jets, and additional 
amounts subsequently may be deposited into surrounding disks via magnetic interaction with the central star. 
The growing census of young stars and brown dwarfs has allowed recent studies to probe rotation rates in a 
number of 1-10~Myr old clusters, including Chamaeleon I \citep{2003ApJ...594..971J}, IC 348 
\citep{2004AJ....127.1602C,2005MNRAS.358..341L, 2006ApJ...649..862C}, Taurus \citep{2009ApJ...695.1648N}, the 
Orion Nebula Cluster \citep{1999AJ....117.2941S,2002A&A...396..513H}, $\sigma$ Orionis 
\citep{2004A&A...419..249S}, $\epsilon$ Orionis \citep{2005A&A...429.1007S}, NGC~2363 
\citep{2008MNRAS.384..675I}, and NGC 2264 \citep{2005A&A...430.1005L}.

Observations to date find that the majority of rotation rates at ages of a few Myr correspond to periods 
between 1 and 10 days, with a smaller population of slower rotators extending to periods of $\sim$25 days. In 
addition, the distribution appears to be highly mass-dependent: earlier than spectral type M2.5 (or 
$\sim$0.3--0.4~$M_{\odot}$, depending on the theoretical model used), typical rotation periods lie between 
$\sim$2 and 10 days, and in some cases display a bimodal distribution 
\citep{2002A&A...396..513H,2005A&A...430.1005L} However, where data is available at lower mass, the 
distribution peaks near 1--3 days and steadily declines toward longer periods \citep[e.g., 
][]{2007ApJ...671..605C}. At first glance the slow rotation rates are somewhat surprising, given that these 
stars are recently accreting material and still undergoing pre-main-sequence contraction. Stellar evolution 
theory alone predicts approximately an order of magnitude increase in angular velocity during the PMS phase, 
whereas rotation rate distributions in clusters of different age remain roughly constant out to $\sim$30~Myr 
\citep{2009IAUS..258..363I}. Current evidence suggests that at least among the higher mass objects, rotation 
rates are strongly linked to the presence or lack of a disk, as indicated by long-wavelength infrared excesses 
\citep{2006ApJ...646..297R,2007ApJ...671..605C}.

Despite the wealth of data, many open questions remain, which we will address in this work. The mechanism for 
removal of angular momentum during the protostar stages is not well understood, and the role of circumstellar 
disks in rotation rate regulation remains controversial among the low-mass stars at spectral type M2.5 and 
later \citep{1999AJ....117.2941S,2001AJ....121.1676R}. Furthermore, the lower limit to rotation periods in 
young clusters is not well established. Photometric derivations of rotation rate or pulsation period are 
complicated by the variety of variable phenomena operating in young stars.  Notably, aperiodic variability due 
to stochastic accretion can appear as a semi-periodic phenomenon when sampling is sparse or when hot spots 
produced by columns of accreting material produce transient signals at the period of rotation 
\citep{1989A&A...211...99B,1996A&A...310..143F,2007prpl.conf..297H}. A number of authors claim evidence for a 
pattern of faster rotation as masses decrease into the brown dwarf regime 
\citep{2001A&A...367..218B,2003A&A...408..663Z,2009A&A...502..883R}. In some cases, periods as short as a few 
hours are inferred for brown dwarfs (BDs) and very low mass stars (VLMSs), implying that they may be spinning 
at close to break-up velocity. \citet{2005A&A...432L..57P} suggested that variability in these particular 
short-period objects may represent a completely different effect-- pulsation powered by deuterium burning. 
Detection of this phenomenon is one motivation for our work.

We recently initiated a campaign to probe low-amplitude photometric variability on short ($\sim$1-hour) time 
scales, obtaining rotation periods and searching for pulsation among young BDs and VLMSs ($<0.1 M_{\odot}$). 
In this paper, we present results of photometric monitoring on members of the $\sim$3~Myr 
\citep{2008AJ....135.1616S} cluster around $\sigma$ Orionis.  At a distance of 440~pc 
\citep{2008AJ....135.1616S}, spatial extent of $\sim$1 square degree, [Fe/H] of -0.02 
\citep{2008A&A...490.1135G}, and low extinction \citep[$E$($B$-$V$)=0.05;][]{1968ApJ...152..913L}, the cluster 
is a convenient target for photometric and spectroscopic studies. Indeed, prior surveys have revealed a rich 
population of 338 confirmed members \citep[and references therein]{2008A&A...478..667C}, along with some 
$\sim$300 additional candidates from photometry, proper motions, and x-ray detections 
\citep[e.g.,][]{2007ApJ...662.1067H,2009A&A...505.1115L,2004AJ....128.2316S,2006A&A...446..501F}. Of 
particular interest to our pulsation search is that $\sigma$~Orionis is one of few young clusters with very 
low mass members claimed to exhibit periodic variability on time scales of 2-5 hours, as reported by 
\citet{2001A&A...367..218B,2003A&A...408..663Z}, and \citet{2004A&A...419..249S}. However, apart from the 
latter study which presented 23 periodic objects in the northern reaches of the cluster, no comprehensive 
variability studies have been carried out in the main portion of the cluster. A campaign by 
\citet{2004A&A...424..857C} resulted in the measurement of three rotation periods from a sample of 28 
candidate brown dwarfs, while the studies by \citet{2001A&A...367..218B} and \citet{2003A&A...408..663Z} 
contributed another two. Other work by \citet{2007ApJ...662.1067H} and \citet{2009A&A...505.1115L} present 
evidence for generic variability based on sparsely sampled photometry over year time scales.

We have taken advantage of the numerous prior single-pointing surveys to select a sample of $\sim$150 likely 
young BDs and VLMSs distributed throughout $\sigma$~Orionis. We collected photometry on these objects with the 
Cerro Tololo Interamerican Observatory (CTIO) 1.0-meter telescope and Y4KCam detector, operated by the SMARTS 
consortium. We obtained data on two observing runs of nearly two weeks each, benefiting from uninterrupted 
clear skies and probing to magnitudes of $I$=21, well beyond the substellar boundary ({\em I}$\sim$17). The 
excellent precision of our dataset (a few percent or better for the majority of targets) and continuous 
monitoring offers an unprecedented window into low-amplitude variability on 15-minute to two-week time scales 
in VLMSs and BDs, encompassing multiple rotation periods for many of these objects. Based on these 
observations, we present 65 new rotation rates-- more than tripling the number for confirmed and likely 
$\sigma$ Ori members-- as well as provide a new assessment of the period distribution among late-type objects. 
In addition, we show evidence for other types of variability, including possible rapid circumstellar 
extinction events associated with very low mass stars. We identify several new candidate members of the 
cluster based on their variability and colors.

The outline for this paper is as follows: In $\S$2 and $\S3$, we respectively describe the selection of 
photometric targets in $\sigma$ Orionis and basic data acquisition and reduction procedures. In $\S$4, we 
detail several different photometry techniques tested to minimize night-to-night photometric systematics and 
achieve the lowest possible noise scatter in our time series. In $\S$5 and $\S$6, we discuss our methods for 
identifying both periodic and aperiodic variability in the light curves, as well as the corresponding 
detection limits as a function of magnitude and frequency (in the case of periodic variability). In $\S$7, we 
present an overview of the types of variability found in our sample, as well as analyze the connections to 
parameters such as color, mass, time scale, and circumstellar disk indicators. Finally, in $\S$8, we present 
our main findings concerning young cluster variability in the context of prior studies. The Appendix includes 
a detailed list of all previously identified $\sigma$~Orionis variables that fall in our fields of view, 
along with redetections where applicable.

\section{Target Fields}

The $\sigma$ Orionis cluster was first identified by \citet{1996PhDT........63W} and 
\citet{1997MmSAI..68.1081W} via clustered sources of x-ray emission in ROSAT observations. Possibly associated 
with the Orion OB1b subgroup, the cluster of low-mass stars surrounds the O9.5V binary star $\sigma$ Ori~AB. 
\citet{1999ApJ...521..671B} and \citet{2000Sci...290..103Z} presented an initial sample of candidate low-mass 
cluster members, for most of which spectral types were later determined by \citet{2003A&A...404..171B}. 
Subsequent surveys \citep[e.g.,][]{2004AJ....128.2316S,2005MNRAS.356.1583B,2005MNRAS.356...89K} have augmented 
the list of low-mass candidate members via photometric selection in the near-IR, spectroscopic analysis of 
H$\alpha$, Na I, and Li lines, as well as characterization of mid-IR excesses indicative of disks 
\citep[e.g.,][]{2007ApJ...662.1067H}. While most of these methods do not rule out the presence of foreground 
and background sources, the contamination rate from photometry alone is expected to be relatively low 
($\sim$15\% based on the color-magnitude distribution of a non-cluster field; \citet{2009A&A...505.1115L}).

\begin{figure}
\begin{center}
\includegraphics[scale=0.45]{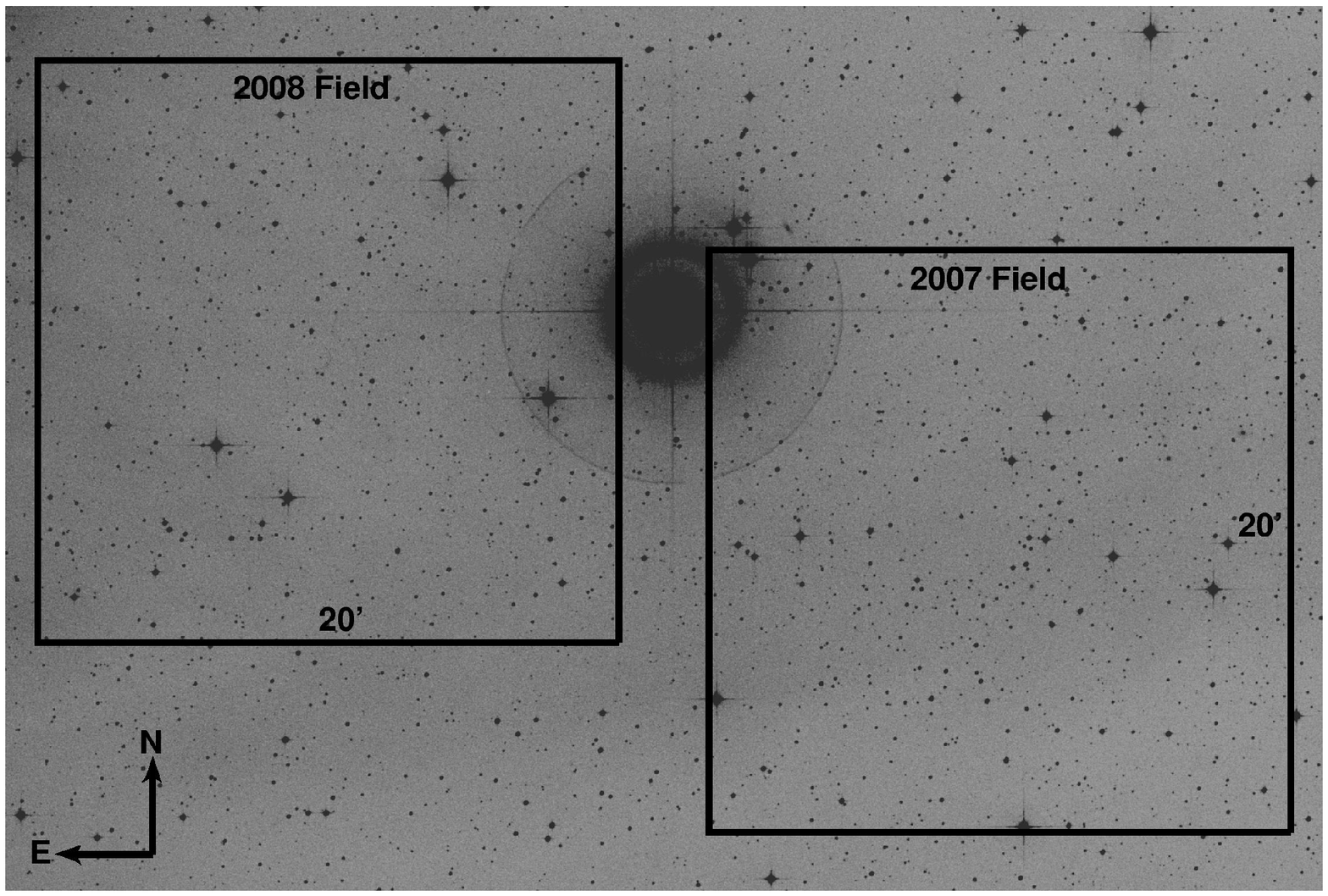}
\includegraphics[scale=0.45]{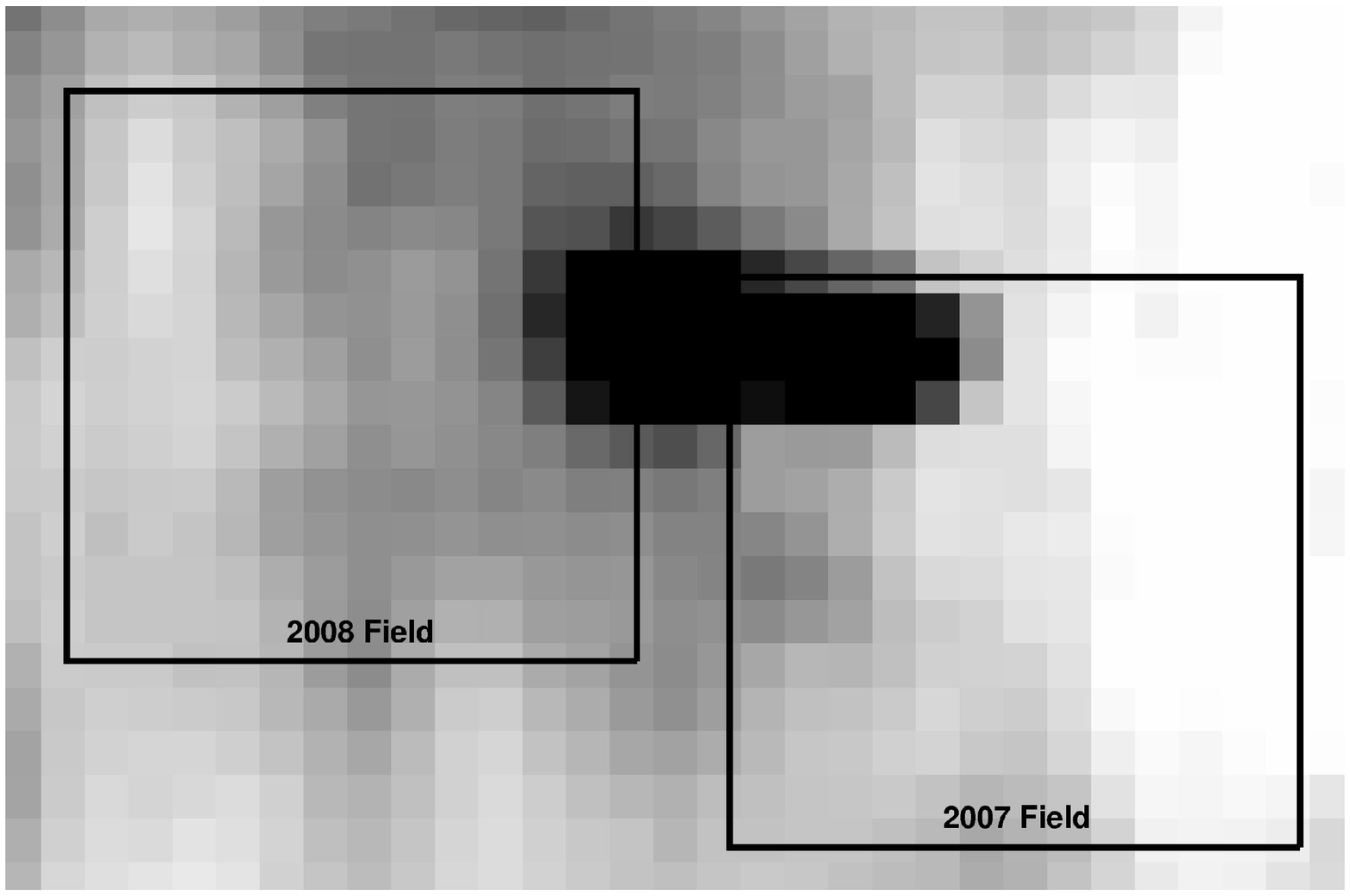}
%\plotone{f2.eps}
\end{center}
\caption[]{Observed fields are superimposed on a Palomar Observatory Sky Survey 2 (POSS2) red image
(top) obtained from the Digitized Sky Survey (DSS)\footnotemark[3] and an Infrared Astronomical
Satellite (IRAS) 100 $\mu$m image\footnotemark[4] (bottom). The 2007 field is centered at
RA=05$^{\rm h}$38$^{\rm m}$00.6$^{\rm s}$ and Dec=-02$\arcdeg$43$\arcmin$46.3$\arcsec$, while the
coordinates of the 2008 field are RA=05$^{\rm h}$39$^{\rm m}$31.2$^{\rm s}$,
Dec=-02$\arcdeg$37$\arcmin$25.9$\arcsec$. $\sigma$ Ori itself is the bright object near center, and
greater extinction is seen in the 2008 field than that from 2007.}
\label{fields}
\end{figure}

We compiled a list of likely and candidate $\sigma$ Orionis cluster members from 
\citet{1999ApJ...521..671B}, \citet{2001ApJ...556..830B}, \citet{2001A&A...377L...9B}, 
\citet{2003A&A...404..171B}, \citet{2004AN....325..705B}, \citet{2004A&A...424..857C}, 
\citet{2004AJ....128.2316S}, \citet{2004A&A...419..249S}, \citet{2005MNRAS.356.1583B}, 
\citet{2005MNRAS.356...89K}, \citet{2006A&A...446..501F}, \citet{2007A&A...470..903C}, 
\citet{2007ApJ...662.1067H}, \citet{2008A&A...478..667C}, \citet{2008ApJ...688..362L}, and 
\citet{2009A&A...505.1115L}, including available signatures of youth and kinematic measurements. Our 
observations target two fields (as shown in Fig.\ \ref{fields}) selected to avoid bright stars such as 
$\sigma$ Ori AB itself, while maximizing both the density of confirmed or suspected low-mass cluster members 
and number of objects with previously observed variability. We cross-correlated the positions of objects in 
our fields with the above-mentioned sources to assemble a final list of confirmed and likely members 
appearing in our imaging data, which is provided in Table~1.

\footnotetext[3]{http://archive.stsci.edu/cgi-bin/dss\_form}
\footnotetext[4]{http://irsa.ipac.caltech.edu/data/IRIS/}

\section{Data Acquisition \& Reduction}
A field centered on RA=05$^{\rm h}$38$^{\rm m}$00.6$^{\rm s}$ and Dec=-02$\arcdeg$43$\arcmin$46.3$\arcsec$ in 
the $\sigma$ Orionis cluster was observed for 12 consecutive nights from 2007 December 27 to 2008 January 7 
with the CTIO 1.0-m telescope and Y4KCam detector. A second field at RA=05$^{\rm h}$39$^{\rm m}$31.2$^{\rm 
s}$, Dec=-02$\arcdeg$37$\arcmin$25.9$\arcsec$ was observed from 14 to 24 Dec 2008. During this second run, two 
repeat observations per night were also obtained of the first field, such that long-term photometric trends 
might be investigated. Skies were clear and photometric for the majority of observations, with little moon and 
seeing from 0.9$\arcsec$--1.8$\arcsec$. The CCD consists of a 4064$\times$4064 chip with 15~$\mu$m pixels, 
corresponding to a scale of 0.289$\arcsec$~pixel$^{-1}$ and an approximately 20$\arcmin\times$20$\arcmin$ 
field of view. Because readout occurs in quadrants, bias levels vary in the four regions. This effect 
unfortunately cannot be completely calibrated out, because both the mean bias level across the amplifiers as 
well as the two-dimensional spatial dependence are highly time variable, as seen in the behavior of the 
overscan region and bias images. Our photometry is largely unaffected by this issue since sky subtraction 
takes into account local bias levels around our targets. However, we have masked out data in the central 20 
columns and rows of the CCD where rapid spatial variation in the bias between different quadrants prevents 
proper background extraction. The amplifiers have gains from 1.33 to 1.42 electron ADU$^{-1}$ and readout 
noise $\sim$7 electron pixel$^{-1}$.

The observations targeted 153 candidate very-low-mass $\sigma$ Ori members, including some 15 
spectroscopically confirmed young brown dwarfs (see Table~1). Our goal of acquiring high-precision time series 
photometry on these objects required accumulation of as much signal as possible while maintaining an observing 
cadence well under the $\sim$1-hour time scales of interest for short-period signals. Theoretically, the 
shortest detectable sinusoidal period is twice the cadence; we elaborate on this relationship in \S 5. In 
practice, exposure times are limited by contamination from large numbers of cosmic ray hits and diffraction 
spikes from saturation of numerous nearby bright stars when count levels reach 50,000 ADU. As a compromise 
between these competing effects, we initially chose an exposure time of 360 seconds in the Cousins $I$ band, 
where the optical spectral energy distribution of brown dwarfs nears its maximum. During the 2008 
observations, we increased integrations to 600 seconds for slightly improved signal-to-noise. Due to the 
consistent night-to-night observing conditions, these set-ups did not need to be adjusted throughout the runs. 
With a detector read-out time of 90 seconds in the unbinned mode, the resulting cadences were 7.5 and 11.5 
minutes per photometric data point in the 2007 and 2008 run, respectively. The corresponding total observation 
times were 72 and 60 hours, resulting in 523 and 338 data points.

Careful calibration procedures were followed to ensure that the ultimate photometry was restricted mainly by 
source and sky background noise inherent to the measurements. Sets of bias images and dome flats were acquired 
daily. Since dome flat field images taken with the CTIO 1.0-m telescope are known to deviate from the true 
pixel sensitivity distribution by up to 10\% toward the corners of the detector, we only used sky flat fields. 
Twilight sky flats were obtained at the beginning and end of each night in the $I$ band. Uniform bright sky 
illumination and detector response can be achieved with exposures of at least 10 seconds (to mitigate shutter 
shading effects) and less than a few minutes (to avoid the appearance of many stars in the flat field). 
Conditions allowed for four consecutive sky flats with flux levels averaging 30,000 counts, providing a good 
representation of pixel sensitivity variations within the linearity limit of the CCD. We checked that the 
combination of all eight twilight flats per night should contribute an uncertainty of less than 0.002 
magnitudes per pixel to the photometry, sufficient for our precision requirements. For two nights when thin 
cirrus prevented uniform twilight exposures, we incorporated observations from adjoining nights into the 
composite flat field after confirming that the detector sensitivity did not change significantly over 24-hour 
time scales. In a few cases, new dust did appear on the CCD window midway through the night and its 
corresponding ``donut'' could not be adequately removed from the images. Affected areas were noted and 
confirmed not to lie in close proximity to any of our photometric targets or potential reference stars. We 
ensured that the pointing remained stable by choosing the same guide star from night to night and centering it 
in the same pixel of the guide camera.

We cleaned the images of cosmic rays with the IRAF {\em cosmicrays}\ utility. This detects and replaces sharp, 
non-stellar sources appearing more than five standard deviations above the background. Rare cosmic ray hits 
coincident with the stars and brown dwarfs are not removed in this way and must be identified separately in 
the later light curves. Standard reductions including subtraction of biases and flatfielding were carried out 
with the IRAF {\em imred} package. Images were split into quadrants, and each corrected with a high-order fit 
to its individual overscan, to account for highly variable bias structure at the edge between the bottom and 
top amplifiers. Quadrants were subsequently trimmed and pasted back together to form a seamless image. 
Residual two-dimensional bias structure was removed by subtracting a master frame of 20 median-combined zero 
images.

Because the $I$ band\footnote[1]{Filter profiles are available here: 
http://www.astronomy.ohio-state.edu/Y4KCam/Filters/y4kcam\_Ic.txt} extends well beyond 8000$\AA$ and the 
typical CCD thickness is 20~$\mu$m or less, our images suffer from fringing, in which long-wavelength emission 
from OH night sky lines reflects multiple times within the CCD to create a complicated interference pattern 
superimposed on the images. An SDSS i filter, which better suppresses sky emission, was unavailable at the 
time of our observations. The fringing effect is additive and fixed with respect to detector position, but its 
strength varies throughout the night, depending on sky conditions. For the Y4KCam, we find that its amplitude 
typically fluctuates on scales of 30-50$\arcsec$, with amplitudes reaching 2\% with respect to the background. 
While guiding generally keeps stars on the same pixel, steep gradients in the fringe pattern and an 
unexplained 4--5 pixel drift in $x$ position throughout the night could affect background subtraction for 
aperture photometry, introducing artificial variability on the same levels as potential rapid rotation or 
pulsation signatures. Hence we developed a procedure to effectively model and subtract the fringing from all 
images. Throughout the first run, we took 360-second exposures of sparsely populated areas of sky, amassing a 
total of 68 ``fringe'' flat fields. To isolate the fringe pattern in these images, it is important to extract 
the two-dimensional continuum sky background as well as stellar point sources. We generated object masks for 
each field, eliminating images with highly saturated stars. Because of varying bias levels in the different 
quadrants, we modeled the background to second order, allowing the fit to vary in each of the four regions. 
This piecewise background was then subtracted from each image, leaving a fringe pattern with mean value zero. 
A high signal-to-noise master fringe frame devoid of stars and background was created by median combining the 
individual fringe images, incorporating the object masks. To defringe an image, it is necessary to subtract 
the fringe frame scaled by the value determined to best reproduce the time-dependent fringe amplitude. The 
IRAF task {\em rmfringe} performed this process by iterating to minimize the difference between scaled fringe 
flat field and each background-subtracted, object-masked image. After a first round of fringe subtraction from 
the fringe fields themselves, we repeated these steps but instead used the processed images from the previous 
iteration to determine the sky background. This resulted in a slightly more accurate master fringe frame.

To defringe the two science fields in $\sigma$ Ori, we followed the same procedures, subtracting the scaled 
master fringe frame from the science images in two iterations. The second round again included input sky 
background as determined from the first round fringe-subtracted images. Since no fringe field exposures were 
taken during the 2008 run, we used the same 2007 master frame for this data, resulting in slightly higher 
residuals. We found that these steps effectively removed fringes in some 95\% of images if liberal object 
masking was applied, especially in the northeast corner of the field where stray light from a bright nearby 
star reflected into the detector field of view. The remaining 5\% of images were corrected by manual 
defringing. Fringe subtraction was successful in removing background variations down to the 0.1\% level, 
suitable for our photometric purposes. Images were then aligned to the same $x$-$y$ coordinates with a small 
flux-preserving shift using the IRAF script IMAL2 provided by \citet{2001phot.work...85D}. This script takes 
as input a number of bright reference stars across an image, determines their centers using the IRAF {\em 
imcentroid} task, and outputs the mean shift in $x$ and $y$. It then uses the IRAF {\em imshift} task to 
perform the shift calculated for each image.

\section{Photometry}

The aim of our monitoring campaign is to obtain light curves with as high a cadence and precision as possible, 
thereby providing sensitivity to variability below the 0.01 magnitude level on sub-hour time scales. 
Optimizing signal-to-noise (S/N) ratios on the low-mass cluster targets in our fields is particularly 
challenging with a 1-meter telescope, as the selected 6 to 10-minute exposure times result in S/N=100 only on 
the brighter brown dwarfs in the sample. These exposures also lead to moderate numbers of cosmic ray hits as 
well as slightly non-symmetric psf shapes resulting from accumulated guiding errors. Consequently, we paid 
special attention to our photometric analysis procedures and tested several different routines to identify the 
one providing the best S/N performance.

\subsection{Aperture Photometry}

Since our fields are not particularly crowded, we expect aperture photometry to outperform point spread 
function (psf) fitting. We employed the IRAF script VAPHOT \citep[based on {\em phot};][]{2001phot.work...85D} 
to calculate instrumental magnitudes with apertures optimized to provide the best signal-to-noise ratios as a 
function of stellar flux, sky background, and seeing. Photometry of bright objects typically benefits from 
large apertures since the flux signal dominates over the background, while for faint objects smaller apertures 
are needed because photometric precision is sky-limited, as discussed by \citet{1989PASP..101..616H}. 
Moreover, the optimal aperture size scales approximately with seeing, such that it is nearly constant when 
expressed as a multiple of the psf size. VAPHOT makes use of these properties to perform high-precision 
differential photometry without the need for multiple trials of different aperture sizes or aperture 
corrections. The program dynamically determines the best apertures for all desired photometric targets on a 
single input frame with seeing representative of the average for the entire run. The ratio of the calculated 
aperture sizes to the full width at half maximum (FWHM) of the psf is then fixed, and aperture sizes in all 
other frames are scaled relative to those determined for the chosen ``typical'' frame. All measurements on an 
object should thereby recover the same fraction of its total flux from frame to frame and night to night, in 
the limit that the psf is circularly symmetric. In reality, the psf is not perfectly symmetric, and this 
assumption introduces the need for a small correction to the measured fluxes. We have not applied such a 
correction here but discuss a method that we have used to reduce the error using image subtraction photometry 
in $\S$4.2.

Aperture photometry with the scaled aperture sizes was then carried out with the IRAF {\em phot} task, 
including redetermination of the object centroids before aperture placement. Typical aperture radii were 10.5 
pixels ($\sim 3\arcsec$) for bright stars and 7 pixels ($\sim 2\arcsec$) for faint targets such as BDs. We do 
not perform aperture corrections since this introduces additional errors and our instrumental magnitudes 
differ from their flux-corrected counterparts by the same constant value, a situation entirely suitable for 
differential photometry. We have measured the sky background around each object within an annulus extending 
between 4.5 to 6 times the FWHM.

The primary difficulty we have encountered in producing high-precision photometry with VAPHOT is the implicit 
assumption of a psf fixed in both size across the image and in shape from night to night. The psf size across 
the Y4KCam detector is in fact known to vary by up to 25\% from the center to corner\footnote[2]{See 
http://www.lowell.edu/users/massey/obins/y4kcamred.html for details.}. As provided, VAPHOT determines the 
seeing FWHM in each image by fitting a gaussian profile to a single bright star specified by the user. This 
value is then used to scale the apertures for {\em all} other objects in the field. We altered the script to 
instead output an average psf of several bright stars across the field. In addition, we found that the 
calculated optimal apertures for all but the faintest targets were too small, in that the aperture scaling 
based on psf size estimates introduced significant noise on night-to-night time scales. Doubling the aperture 
sizes for targets with $I<18$ reduced RMS spreads over the entire observing duration by more than 50\% in most 
cases. Therefore, we adopted the larger aperture sizes for all object in the brighter half of our sample. 
These improvements confirm that neglecting spatial variations and non-gaussian shapes in the point spread 
function introduces substantial artificial variability in photometry with relatively small apertures.

Differential photometry was carried out with a suite of reference stars for which peak flux remained below the 
detector saturation and linearity limits on all nights. In each of the two fields, we selected an initial set 
of 10--20 bright (all {\em I}$\sim$13) reference stars, summed the fluxes in each image, and converted to a 
magnitude. Tests of several weighting schemes, such as the one suggested by \citet{2001MNRAS.326..553S} did 
not produce substantially different results. Differential magnitudes relative to this ensemble magnitude were 
computed for each of the reference stars in turn, with that particular star removed from the ensemble. We 
computed the light curve RMS values, and objects with variability visible by eye or RMS more than one standard 
deviation above the average RMS for that magnitude were removed from the ensemble. The process was repeated 
with the new subset of reference stars until no outliers remained. The final ensembles consisted of 4--6 
reference stars, with spreads of $\sim$0.002 magnitudes over the course of the entire observing run. Based on 
this reference, differential light curves were generated for all objects in the field with signal below the 
saturation limit but at least five times the background.

A number of the light curves displayed significant zero-point changes on time scales of one or more days. 
These variations appeared even among some of the brightest targets but did not seem to occur systematically 
across all objects. We suspect that slow changes in the pointing and thus object mapping in $x$-$y$ pixel 
coordinates and other parameters such as seeing and airmass affect the photometry in a position-dependent way. 
To investigate associated trends in the light curves, we fit object magnitudes linearly as a function of psf 
FWHM and ellipticity, sky counts, object $x$ and $y$ position, relative centroid position, as well as airmass. 
The fit to most light curves was only weakly dependent on these parameters. Out of concern for unnecessary 
addition of noise to the data, we did not remove these low-level trends.

An additional consideration for the photometry is potential differences in color between the late-type objects 
in our sample and the brighter stars in the reference ensemble. To first order, extinction effects due to 
changing airmass cancel out in differential photometry. However, second-order color terms can introduce 
significant trends in the light curves if target objects are substantially redder than the reference ensemble 
\citep[e.g.,][]{1991PASP..103..221Y}. Atmospheric extinction is weaker at longer wavelengths, and this can 
emerge as a gradual brightening of differential light curves for fainter, redder objects as airmass decreases. 
No such behavior is visible in the light curves of faint cluster members in our sample, and the absence of 
significant airmass-flux correlations confirms this finding. We suspect that the lack of obvious trends is due 
to the relatively weak dependence of extinction on wavelength beyond $\sim$ 7000$\AA$, as indicated by the 
small $I$-band color-dependent extinction coefficient determined later in $\S$4.3. Variable extinction due to 
changing atmospheric conditions could also produce artificial offsets in the object brightness, whereby the 
differential magnitudes would correlate with reference ensemble magnitude rather than airmass. Again, we fit 
the light curves for this effect, but did not detect significant trends and hence did not apply any 
corrections to the data.

The major sources of random error in the light curves are photon shot noise and sky background noise. We 
estimate based on the relation given by \citet{1967AJ.....72..747Y} that atmospheric scintillation effects 
will introduce brightness fluctuations of less than 5$\times 10^{-4}$ magnitudes for the observational set-up 
here and hence should be negligible. To assess the quality of our light curves, we extracted photometry on all 
$\sim$3200 point sources identified in the fields and removed severely saturated objects from the sample. On 
time scales of less than one night, the floor of the distribution is well accounted for by photon and sky 
noise, plus an additional allowance of $\sim$0.002--0.0025 magnitudes in systematic error. The adopted 
uncertainty for our unbinned data range from 0.002 magnitudes for the bright reference stars, to just over 
0.01 for the brown dwarfs near $I$=17, and 0.1 at the faint end where targets reach $I$=21. On the longer time 
scales corresponding to the observing duration, RMS light curve fluctuations are increased by up to 50\% over 
these values because of night-to-night systematic effects.

\subsection{Image Subtraction Photometry}

Several concerns prompted us to perform an independent test of our results with a different set of photometric 
reduction procedures. For a few of the target brown dwarfs, flux from faint sources near our object apertures 
may have interfered with proper sky subtraction during aperture photometry. In addition, night-to-night 
variations in the mean magnitude of many sources suggests that spatial and temporal psf variations as well as 
slightly non-circular psf shape may be significant enough to alter the photometric zero point. Comparison 
tests of psf fitting photometry and image subtraction \citep[e.g.,][]{2002AJ....123.3460M} have shown that the 
latter method can result in significantly smaller light curve scatter. Therefore, we opted to employ the 
method of differential image analysis \citep{1998ApJ...503..325A,2002AJ....123.3460M} to produce a separate 
photometric dataset with reduced sensitivity to crowding and other psf effects. The Hotpants package 
\citep{2004ApJ...611..418B} compares the fluxes of objects in every exposure to their counterparts in a 
selected reference image, thereby enabling a differential brightness measurement. Images are first accurately 
aligned to a common grid. A high-quality stacked reference image is then convolved with a time-dependent 
kernel which is mathematically optimized to reproduce the psf (size and shape) in all individual images. The 
science images are then subtracted from the convolved reference to reveal residuals possibly indicative of 
variability.

We found that subtraction from the reference template produced relatively clean images, with background 
consistent with the levels expected from noise properties of the input images. By specifying spatial 
variations of the background and psf kernel, we are able to obtain subtracted images devoid of systematic 
effects. Systematic residual flux is detectable above the background only in the brightest stars, where it 
appears in saturation-related peaks or a circular pattern with alternating positive and negative flux on 
either side. As pointed out by \citet{1998ApJ...503..325A}, the latter pattern is likely the effect of 
small-scale atmospheric turbulence, which causes offsets of the psf centers even in well-aligned frames. We 
measured the residual flux in each subtracted image by performing nearly the same aperture photometry 
routines, as described in $\S$4.1. Inputs for aperture placement and size were determined from the convolved, 
unsubtracted images. To convert the measurements to differential magnitudes, we also measured fluxes of each 
star in the reference template, again using the same optimal aperture sizes determined by VAPHOT for the more 
standard photometry discussed in $\S$4.1. Magnitudes were then computed relative to the reference frame. For a 
selection of variables in which the signal dominated noise, we confirmed that the image subtraction routine 
produced the same light curves as the photometry performed on un-subtracted images, to within the photometric 
uncertainties. This technique is a hybrid version of the variable-aperture and image subtraction methods, the 
second of which typically involves an aperture correction even to compute the differential magnitude. Our 
approach thus eliminates important systematic noise contributions and should perform significantly better than 
either method alone.

\begin{figure}
\begin{center}
\includegraphics[scale=0.45]{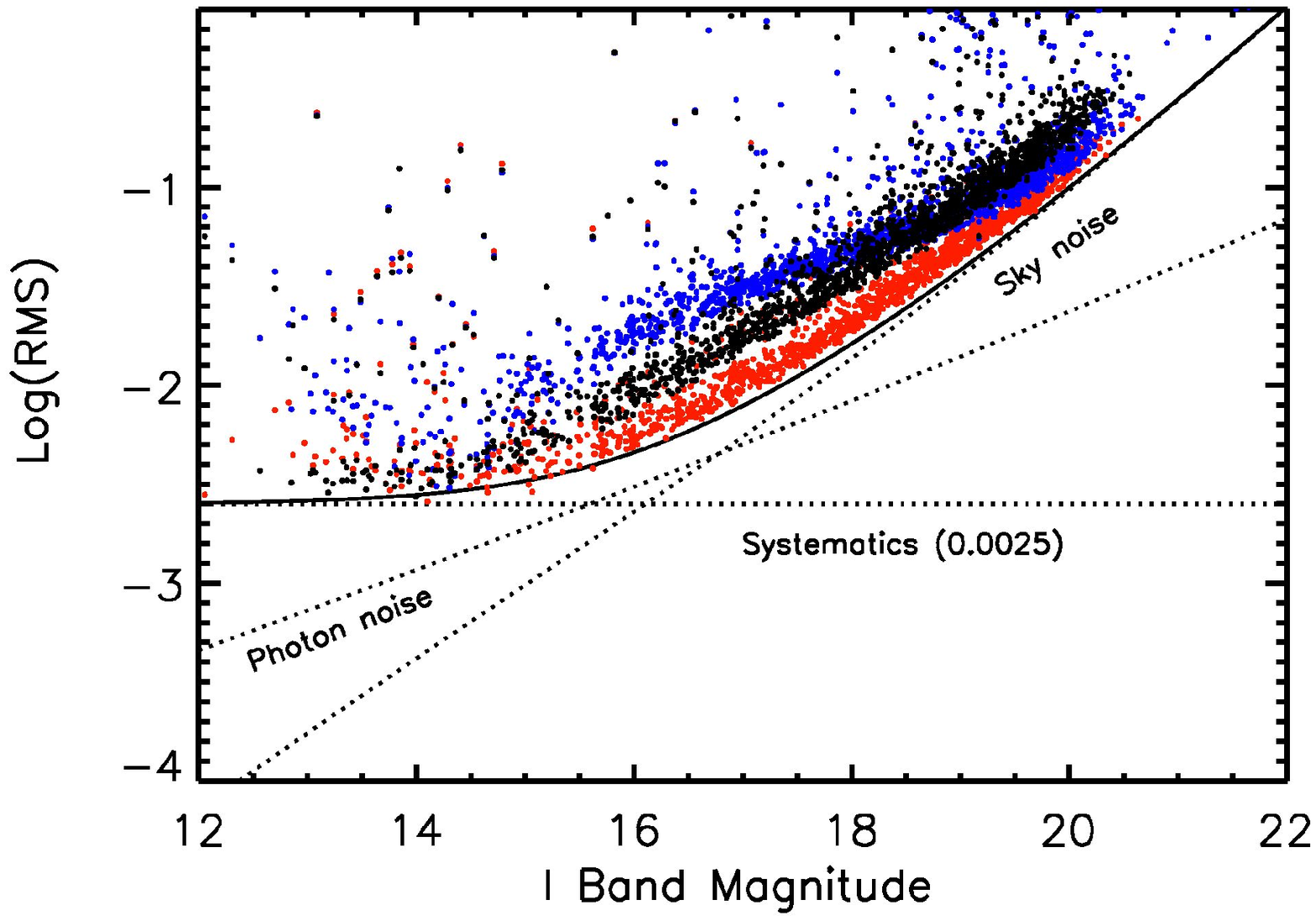}
\includegraphics[scale=0.45]{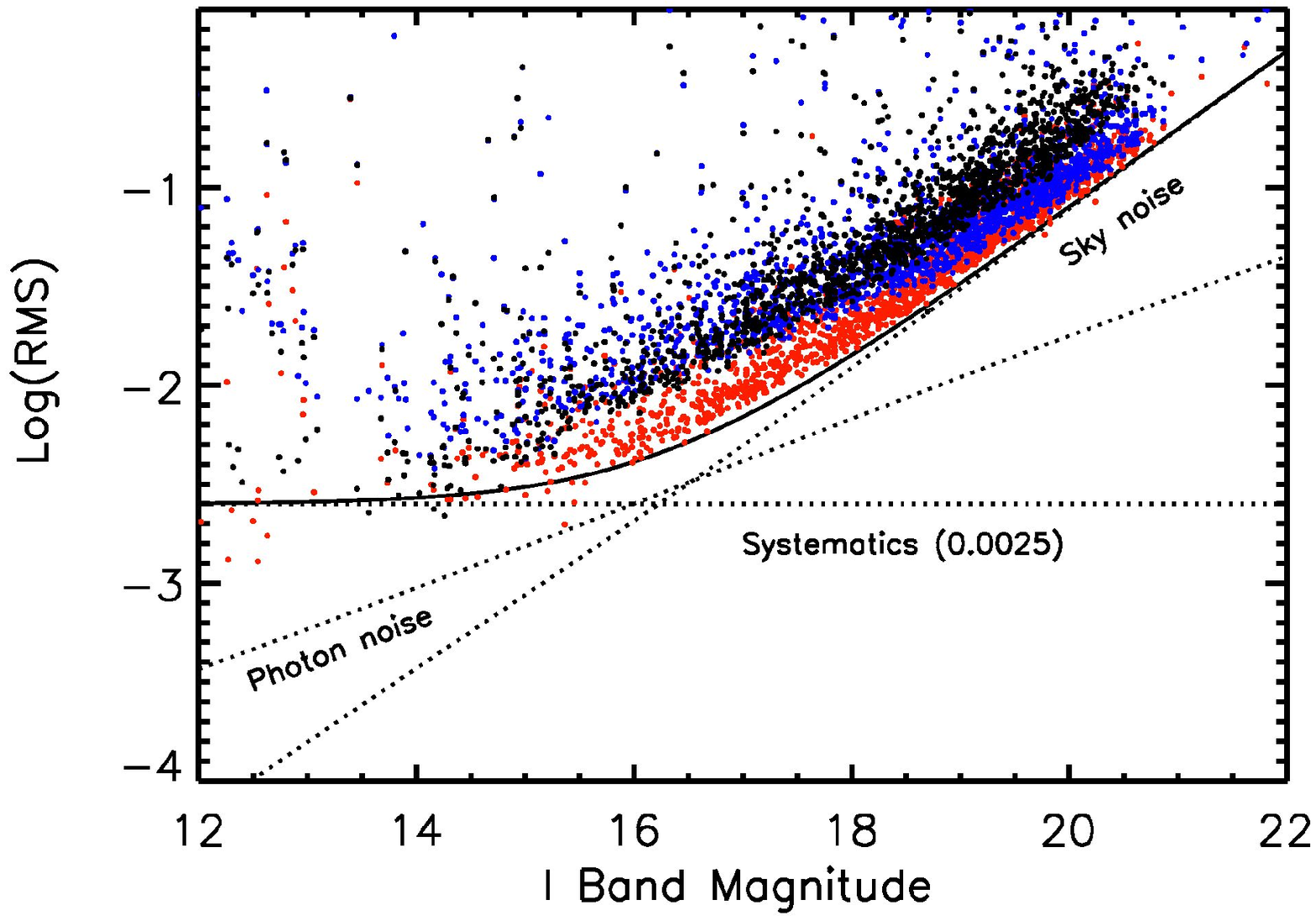}
\end{center}
\caption[]{Spread of photometry over the duration of each observing
run, as a function of magnitude for three methods of variable-aperture photometry. The 2007 field is
represented on the top, while the 2008 field is on the bottom. Blue dots represent photometry with the
calculated optimal apertures, black dots are the same photometry with double-sized apertures, and red
dots are the result of image subtraction followed by photometry with optimal-sized apertures. While   
the first two methods exhibit systematic errors particularly in the middle magnitude range, the
trends for image subtraction photometry in both fields are well described by a combination of photon noise,
sky background, and a small systematic contribution. Larger deviations at the bright end are due in
part to CCD non-linearity. Points lying significantly above the trend signify variable objects or
erroneous photometry (e.g., bad pixel or saturation effects) that was later removed.}
\label{sigrms}
\end{figure}

We expect the photon and sky noise components of the image subtraction light curves to be similar to those 
derived from standard optimal aperture photometry. But since image subtraction photometry involves 
measurements on {\em residuals} (with at least an order of magnitude less flux, even for variable objects) 
resulting from the image subtraction optimization process, the light curves should be much less sensitive to 
errors in psf and aperture size. To test this assumption, we plot in Fig.\ \ref{sigrms} the RMS light curve 
spread as a function of magnitude over the duration of each observing run for the different photometry 
methods. We find that while doubling the aperture sizes (as explained in $\S$4.1) offers improvement in 
photometric precision in the standard optimal aperture method, image subtraction photometry indeed 
significantly outperforms both of these approaches. To assess each method in comparison with the expected 
uncertainties, we have estimated the poisson and sky noise components, based on the variable aperture size as 
a function of magnitude as well as the mean sky background value over all nights of each run. Apart from the 
brightest 3\% of objects which are affected by our neglect of CCD non-linearity ($I\lesssim$14), the 
combination of image subtraction and optimal aperture selection produces light curves consistent with the 
analytically determined photon and sky noise floors plus a 0.002--0.0025 magnitude systematic uncertainty over 
the entirety of each run. These curves are shown in Fig.\ \ref{sigrms}; they pass slightly below, as opposed 
to {\em through} the data distribution because of small systematics evidently unaccounted for in the sky 
background. Based on this assessment, we have adopted as our final dataset the image subtraction results for 
targets with $I>14$, and light curves from standard aperture photometry with double-sized apertures for 
$I<14$.

\vspace{1cm}
\subsection{Absolute Photometry \& Colors}

Because of the precision requirements of our observations, it was not efficient to observe standard fields 
frequently or collect multi-color data. Telescope motion compromises object pixel placement and thus 
introduces flat-fielding error effects.  Filter changes are also associated with focus shifts and small 
position increments which often degrade data quality. However, standard magnitudes and color information can 
be very useful in distinguishing between the intrinsic properties of different variable sources. As a 
compromise, we obtained one or two $R$-band exposures of each $I$-band field every night. To derive the 
Cousins $R$ and $I$ magnitudes, we also observed a spatially dense Stetson photometric standard field in 
NGC~2818 at several different airmasses and performed aperture-corrected photometry on over 500 stars with 
available Stetson $R$ and $I$ magnitudes \citep{2000PASP..112..925S}. The conversions from the CTIO filter 
(``r'' and ``i'') magnitudes was determined by fitting the following linear trends across a wide range of 
magnitudes and colors, as well as several airmass values ($X$): $$I=i+(\epsilon_I+k_I^{'}X)(R-I)+k_IX+Z_I$$ 
$$R=r+(\epsilon_R+k_R^{'}X)(R-I)+k_RX+Z_R$$ $$R-I=\epsilon_{RI}(r-i)+Z_{RI},$$ where $\epsilon$ is an 
extinction coefficient and $k$ denotes an airmass coefficient. Aperture-corrected photometry of these sources 
resulted in an $R$-band zero point $Z_R=22.908$, $I$-band zero point of $Z_I=22.140$, and small airmass 
coefficients ($k_I\sim -0.06$; $k_I^{'}\sim0.002$) consistent with typical values for CTIO. Based on these 
conversions, we derived average Cousins $R$ and $I$ magnitudes for all targets in the field within the 
linearity limit corresponding to $I\sim$12.5. Since the airmass during our observations was restricted to be 
less than 2 while the $R-I$ values of our targets covered a range of $\sim$2.0, the small value of the 
color-dependent extinction coefficient ($k_I^{'}$) suggests that we are justified in neglecting the 
flux-airmass trends described in $\S$4.1. These secondary color effects should contribute at most 0.004 
magnitudes of variation to the light curves-- generally far less than other sources of noise and variability, 
and therefore difficult to remove without compromising the data.

The majority of objects in our cluster sample were also detected in the 2MASS survey, which provides $J$, $H$, 
and $K_s$-band data. We cross-referenced the positions of likely cluster members to identify all 2MASS sources 
in our sample. Since young VLMSs and BDs have very red colors, all but the faintest (e.g., $I>20$) have 
$J/H/K_s$ detections. Table~2 contains a compilation of our own absolute photometry of confirmed and candidate 
$\sigma$ Orionis members, along with the corresponding 2MASS magnitudes. For objects covered in prior 
photometric surveys, our $I$ and $R$ values are in good agreement with those reported previously. For example, 
photometric data for the 59 objects in our fields observed by \citet{2004AJ....128.2316S} show an average 
offset of 0.025$\pm$0.10 magnitudes in the $I$ band and 0.035$\pm$0.20 magnitudes in the $R$ band when 
compared to our values. The scatter is consistent with that expected from both the listed uncertainties and 
intrinsic variability.

\section{Periodic Variability Detection}

\begin{figure}
\begin{center}
\includegraphics[scale=0.5]{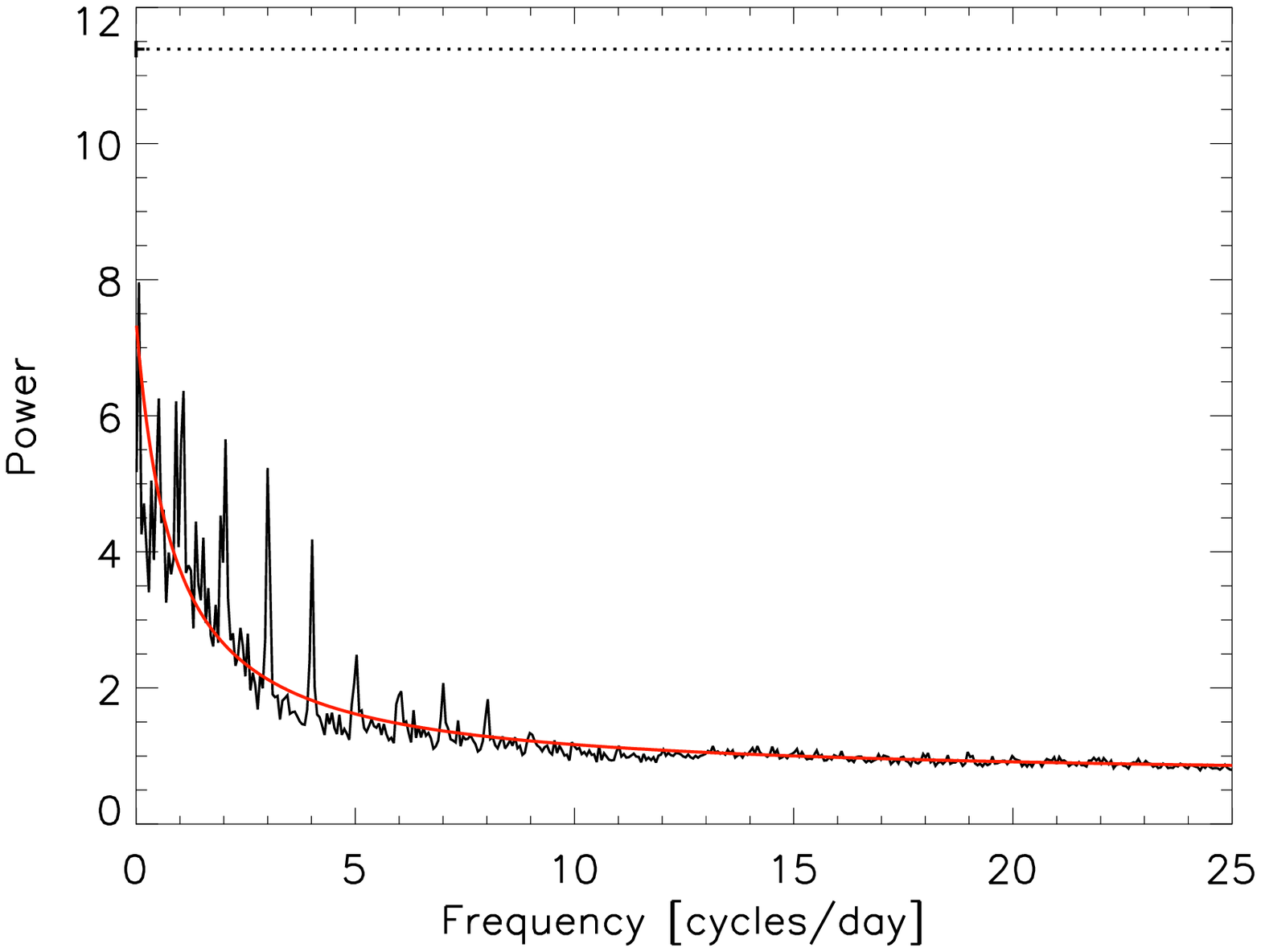}
\includegraphics[scale=0.5]{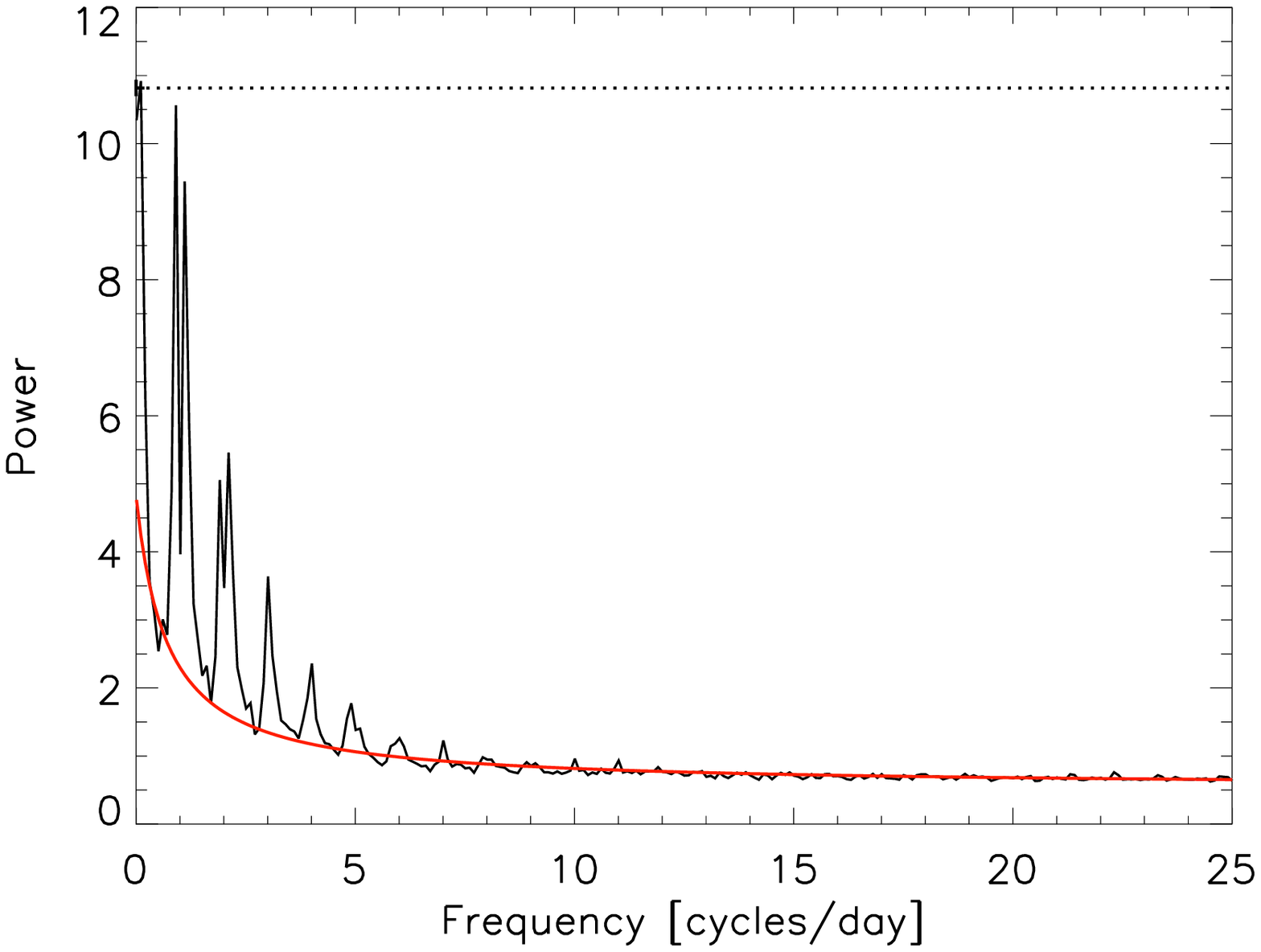}
\end{center}
\caption[]{Average Lomb-Scargle periodogram for the ensemble of 2007 (top) and 2008 (bottom) data.
Dashed lines show the analytically determined 99\% detection limit, as estimated with the
Lomb-Scargle formalism. Red curves indicate our fit to the noise as a function of frequency, disregarding the
systematic peaks at integer values. The roughly constant noise floor continues out to the Nyqyist limit
at $\sim$65 (2008) and $\sim$100 c~d$^{-1}$ (2007).}
\label{avperiod}
\end{figure}

A major focus of our photometric campaign is the detection of variability on short time scales (i.e., 1--10 
hours). It is in this regime that observations of surprisingly fast-rotating VLMSs and BDs have been reported 
and the new phenomenon of deuterium-burning pulsation has also been proposed \citep{2005A&A...432L..57P}. 
Rotating magnetic spots on young low-mass stars typically manifest themselves at a level of a few percent in 
light curves, whereas amplitudes of the pulsation effect are thus far unconstrained by existing theory 
\citep{2005A&A...432L..57P}. Therefore it is crucial to probe the data for potentially weak signals, with 
careful attention to the noise limit, which is generally frequency-dependent. In designing the observational 
set-up, we selected cadences to provide sensitivity to these short periods. Since our data are very evenly 
spaced, modulo daytime gaps (we were fortunate in that nighttime weather was completely pristine), the Nyquist 
limit stipulates that signals may be detected up to half the sampling frequency-- corresponding to 15-minute 
time scales in the 2007 observations, and 23-minute time scales for those from 2008. Because of the long time 
baseline for each run, we are also sensitive to periodicities up to the total observing run duration (12 and 
11 days for the respective runs). However, since most types of photometric errors produce correlated (``red'') 
noise on night-to-night time scales, the minimum detectable variability level at low frequencies is generally 
a factor of a few higher than amplitudes observable at higher frequencies (shorter time scales; see Fig.\ 
\ref{avperiod}).

Prior surveys of the region around $\sigma$ Ori have generated a fairly large sample of low-mass cluster 
objects in which to search for variability (e.g., Table~1). Nevertheless, the census may not be 100\% complete 
in our selected regions. To include young VLMSs and BDs that may have escaped previous identification via 
color-magnitude diagrams, we have produced light curves for all $\sim$3200 unsaturated point sources in the 
two fields. To avoid biases in variability classification, all subsequent analysis was performed without 
regard to the objects' membership status. In this way, we can identify new $\sigma$ Ori candidates as well as 
potentially interesting field stars that happen to lie in the field of view. We have searched for 
periodicities before performing a more generic variability search ($\S$6) to limit the number of variables 
contaminating our analysis of photometric uncertainty as a function of magnitude.

\subsection{Periodogram Analysis} 

As an initial test for periodic variability in the data, we produced Lomb-Scargle periodograms 
\citep{1982ApJ...263..835S} for all light curves. False alarm probabilities (FAP) for detected peaks were 
determined from the prescription of \citet{1986ApJ...302..757H}, which is valid even for datasets with 
non-uniform time spacing.  They estimated FAPs based on large simulations of data with added gaussian noise, 
and their result depends on the number of independent frequencies, which they denote $N_i$.  The formula for 
the parameter $N_i$ is a function of the total number of data points and has been shown to significantly 
overestimate FAPs for small datasets \citep{2007A&A...467.1353R}. This issue is not of great concern to the 
current study, given the 300-500 points from each run.  However, the test must still be used with caution, 
since it assumes all noise sources are white. In reality, the frequency-dependent red noise contributes 
significantly to the light curve RMS on $\sim$ 1-day and longer time scales. Consequently, FAPs can be 
severely underestimated at low frequency and somewhat overestimated at high frequency. The results of the 
Lomb-Scargle test are nevertheless suitable for eliminating targets with no detected variability from the 
sample. With a selection criterion of FAP$<$1\%, we assembled an initial set of possible periodic variables 
for additional analysis.

The collection of Lomb-Scargle periodograms for all targets-- variable or not-- is also a useful tool for 
identifying systematic effects in the data that may cause certain frequencies to consistently appear at 
artificially high probability. This effect is often seen when color-airmass effects are not taken into account 
in the light curves, resulting in trends that mimic intra-night variability. Because of the very uniform 
sampling of our datasets, we expect most of these spurious frequencies to occur at or near multiples of 1 
cycle per day (cd$^{-1}$). To quantitatively map out these values, we constructed a histogram from all 
frequencies corresponding to peaks significant at the 99\% level in the Lomb-Scargle periodogram. This 
diagnostic plot confirms that there are indeed pile-ups near integer frequencies, and we discarded potential 
variability detections corresponding to periodogram peaks occurring only at these values.

As an additional way to identify suspicious frequencies and examine the typical variability power distribution 
in frequency-amplitude space, we also generated a mean periodogram from all $\sim$1500 objects in each field, 
as seen in Fig.\ \ref{avperiod}. This plot clearly displays not only the mathematical clustering of 
``significant'' peaks around integer frequencies but also the steep increase in the noise floor toward low 
frequencies. We attribute this latter effect to red noise and fit it with an exponential of form 
$P=a_0+a_1/(f+a_2)$, where $P$ is power, $f$ is frequency, and $a_0$, $a_1$, and $a_2$ are constant fitting 
parameters such that power declines to match the white noise baseline at $\sim 15$cd$^{-1}$ \citep[e.g., 
``$1/f$'' noise; ][]{1978ComAp...7..103P}. The model for this $1/f$ component was incorporated into our 
computation of detection limits ($\S$5.2).

After removing from consideration targets with either no detectable variability or periodogram peaks only near 
integer frequency values, we performed additional analysis on the remaining light curves. All exhibited one or 
more peaks at the 99\% significance level in the periodogram. To further probe these signals, we employed the 
program Period04 \citep{2005CoAst.146...53L}, which computes a fourier transform \citep{1975Ap&SS..36..137D} 
of the light curve and may also be applied to time series with gaps. Results are similar to the Lomb-Scargle 
periodogram, but the program oversamples frequencies by a factor of 20 and contains an extended analysis 
package to calculate phases, subtract out signals, and search for periodicities at lower levels. Our input 
light curves were shifted to zero mean and cleaned of outliers at more than 4 standard deviations. Period04 
includes an option to assign weights to each data point, such that deviant points do not overly influence the 
determination of the periodogram.  However, based on our assessment of light curve RMS as a function of 
magnitude we conclude that uncertainties are difficult to determine on a point-to-point basis. We believe the 
approach of neglecting weights but removing clear outliers is therefore sufficient to accurately identify the 
frequencies of variability in the sample.

For each light curve, we used Period04 to identify the largest peak in the periodogram and extract a 
preliminary amplitude and phase for each epoch of observation. We then used the program to perform a 
non-linear least-squares fit for frequency, amplitude, and phase. A corresponding sinusoid was then subtracted 
from the light curve (this procedure is known as ``prewhitening'') and a new periodogram was produced. We 
examined the residuals to determine whether they contained further significant frequencies or were consistent 
noise. If another suspected peak appeared, the data were once again prewhitened and the original light curve 
subjected to a multiperiodic least-squares fit \citep{1998CoAst.111....1S,2005CoAst.146...53L}. We repeated 
the process until all significant fourier components were extracted from the data. While significant harmonics 
appeared in cases where periodic variability was not completely sinusoidal, in no case did we identify 
multiple unassociated periods in a single object.

The statistical significance of identified peaks is difficult to determine directly but can be estimated from 
the noise properties of the periodogram. One criterion for detection of a signal to better than 99.9\% 
certainty proposed by \citep{1993A&A...271..482B} requires S/N$>$4 in the amplitude spectrum \citep[see 
also][]{1997A&A...328..544K}. For individual periodograms, noise levels were computed from the prewhitened 
periodogram as a running mean over boxes of 10 cd$^{-1}$ in frequency. We confirmed that no peaks remained at 
more than four times the noise baseline. As an additional check that all significant periodic components were 
removed from the data, we examined the light curve residuals and compared them to the typical RMS of 
non-variable objects with similar magnitudes (as shown in Fig.\ \ref{sigrms}). The values were generally 
consistent with the noise in the non-variable targets.

Errors for the derived frequencies and amplitudes can be computed analytically in terms of the average light 
curve noise and number of data points \citet{1999A&A...349..225B}, but this approach is known to underestimate 
the true uncertainties. The least-squares fit also provides an error matrix, but neither of these methods 
fully account for the properties of noise in the frequency domain. We have therefore opted to run a set of 500 
Monte Carlo simulations with Period04 for each object displaying periodic variability. The detected signals 
are extracted, and remaining noise data points are randomly rearranged such that the original timestamps are 
preserved. The identification of periodogram peaks and least-squares fit to the light curve is then carried 
out as before for each simulated light curve. The distribution of frequencies and amplitudes returned by these 
simulations then determine our uncertainties. Since the distributions are not strictly gaussian, we estimate 
1--$\sigma$ uncertainties based on the values enclosing 68\% of the simulated data. For signals that are near 
the detection limit, the simulations take into account the possibility that noise causes an alias to be 
selected instead of the true peak. This effect is included in our uncertainties listed in Table~3, which are 
provided at the 3--$\sigma$ level.

\subsection{Detection Limits}

Knowledge of our sensitivity to light curve periodicities as a function of both amplitude and frequency is 
crucial to determining whether lack of variability in some objects is related to detection techniques or real 
physical properties. In the presence of pure white noise, the signal-to-noise ratio for detection of a 
periodic signal in a periodogram scales as $A\sqrt{N}$/(2$\sigma$), where $A$ is the amplitude, $N$ is the 
total number of data points, and $\sigma$ is the photometric uncertainty. Therefore, for long time series it 
is possible to detect signals with amplitudes well below the level of the uncertainties in light curves. For 
example, data from our 12-night CTIO observations in 2007 reach a noise level of 0.001 magnitudes in the 
periodogram for objects near $I$=17, making detections as low as $\sim$ 0.004 magnitudes (e.g., S/N=4) 
possible. Red noise diminishes our ability to distinguish signals below about 5-10 cd$^{-1}$, or periods longer 
than a few hours. But across most of the frequency spectrum, sensitivity to periodicities is nearly uniform 
since the time sampling for both runs was uninterrupted, apart from the consistent daily gaps. We find the 
mean periodogram to be entirely adequate in eliminating the anomalous peaks, and because of our relatively 
uniform sampling do not find any deviations other than multiples of one cycle per day.

\begin{figure}
\begin{center}
\includegraphics[scale=0.45]{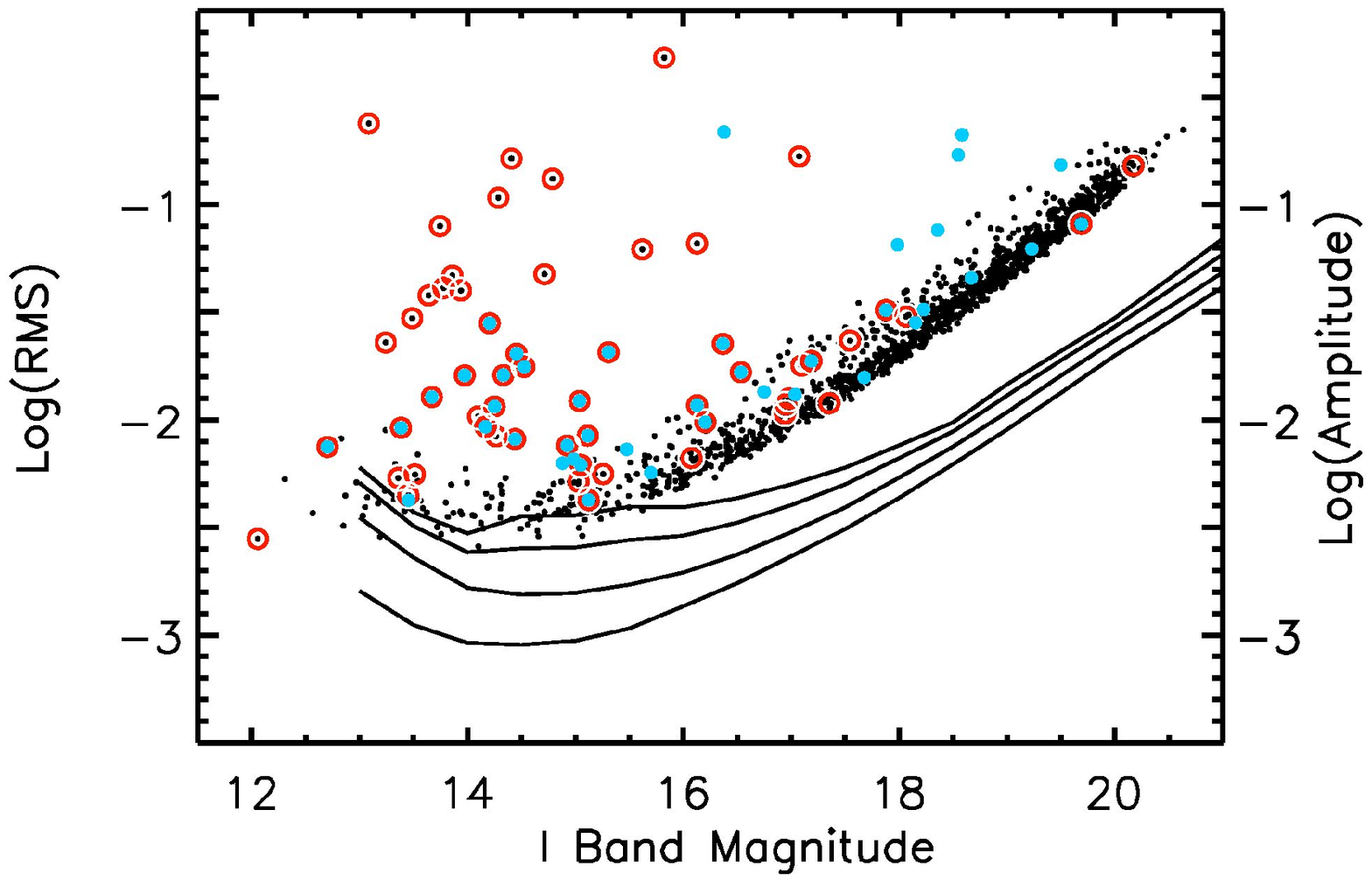}
\includegraphics[scale=0.45]{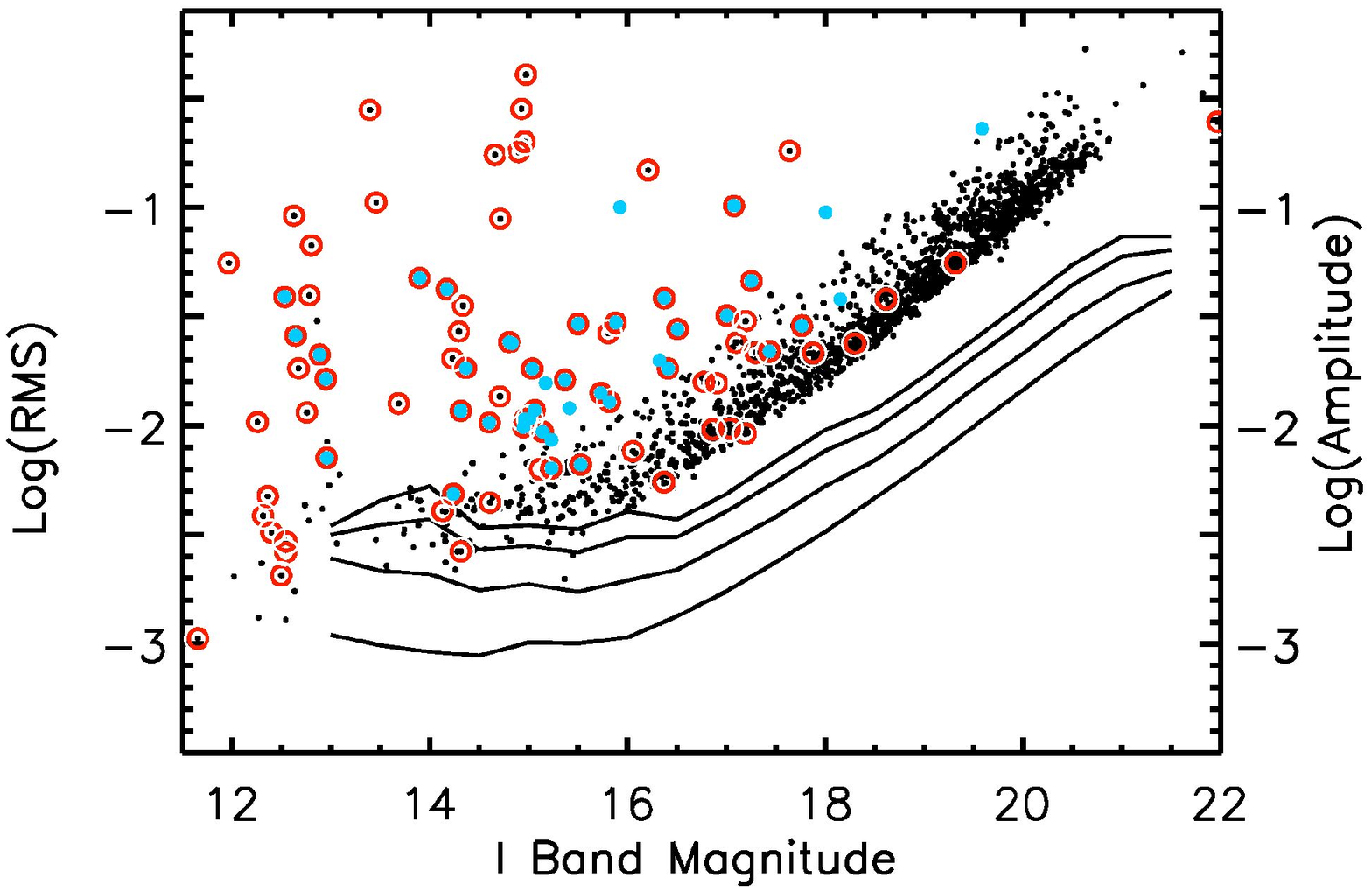}
\end{center}
\caption[]{Relative to the left axis, spread in photometry as a function of magnitude (top: 2007 data; bottom:
2008 data). Detected periodic variables are marked as blue dots, while confirmed and likely cluster
members appear as red circles. Relative to the right axis, we plot the 99\% sensitivity {\em amplitude} limit
to periodic variability on four different time scales. From the top curve to bottom, these correspond to
periods of 10 days, 0.8 days, 3.2 hours, and 1.6 hours. The 2007 field contains 1493 data points, while that
from 2008 has 1683. Fewer objects appear at the bright end in the 2007 field because of variations in the   
underlying distribution of stellar magnitudes and also slightly different saturation limit.}
\label{detlim}
\end{figure}

Nevertheless, we must also determine the frequency dependence of our sensitivity to periodic signals, {\em in 
the presence of red noise}. We therefore measure the mean noise level at four characteristic frequencies (0.1, 
1.2, 7.4, and 15.2 cd$^{-1}$; corresponding to periods of 10 days, 0.8 days, 3.2 hours, and 1.6 hours) at 
intervals of 0.5 magnitudes. The mean noise levels are determined by generating periodograms for all objects 
not displaying variability (as measured by an RMS within 1--$\sigma$ of the median for that magnitude). We 
then measure the power in the periodograms at each of the four frequencies, and average together the values in 
0.5-magnitude bins. Since we expect to be able to detect periodic amplitudes at four times the noise level, we 
have plotted these results, multiplied by a factor of 4.0, in Fig.\ \ref{detlim}. These values represent the 
minimum amplitude detectable in a periodic variable, as a function of period and magnitude.

In some cases, objects displayed signs of variability that were too weak to confirm. Those with unexpectedly 
high residual RMS but no obvious periodogram peaks were set aside for further analysis as part of the 
aperiodic variability group (\S 6). For targets with a possible peak in the periodogram just below the S/N$>$4 
criterion, we analyzed the light curves produced by both image subtraction and standard aperture photometry; 
because of the slightly different processing, occasionally a low-level signal appeared with one method but not 
the others. For the particularly faint BDs with photometry was subject to large sky background noise, we 
required the peak to pass several tests for detection. First, when the putative signal is subtracted from the 
light curve, any other high-amplitude structure in its immediate vicinity (e.g, within $\sim 5$ cd$^{-1}$) 
must also disappear. Peaks that prove difficult to remove cleanly are typical of noise. Furthermore, we look 
for signals with one distinct peak, as opposed to two or more of roughly equal height separated by 
$\sim$1~cd$^{-1}$. Multiple peaks this close are not probable given the types of variability expected in VLMSs 
and BDs (e.g., one peak corresponding to the rotation period, and one or more additional peaks due to rotation 
of a binary companion or pulsation, for which overtones should be separated by at least 5~cd$^{-1}$).

\begin{figure*}
\begin{center}
\epsscale{0.85}
\plotone{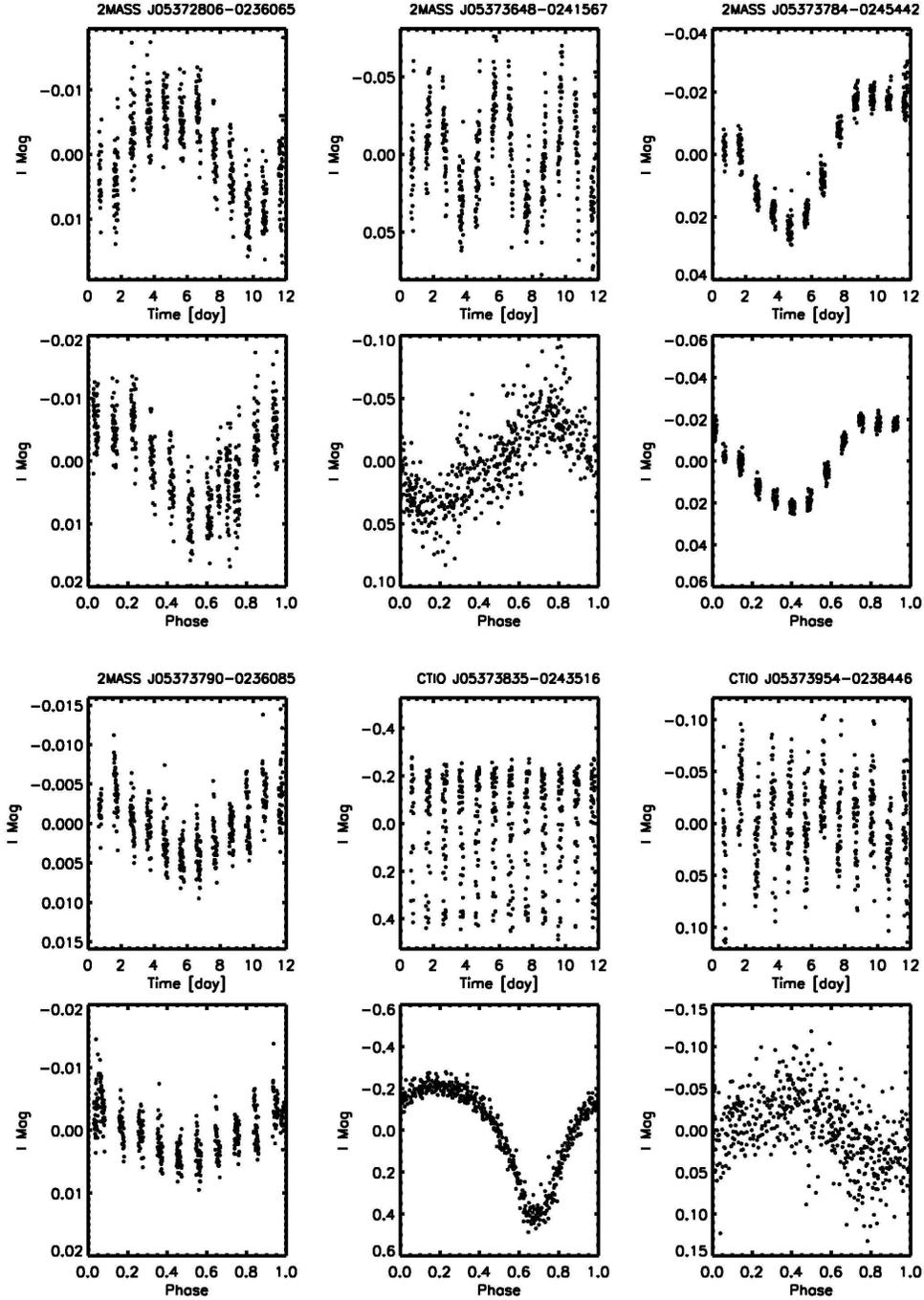}
\end{center}
\caption[]{Differential light curves with detected periodic variability, in order of right ascension.
First and third rows show the original light curve, while those in the second and fourth rows 
are phased to the detected period. There are also a few that are not likely cluster members; membership  
status is listed in Table~3. Only the first panel of the figure is depicted here; Fig.\ 5 in its 
entirety is available in the online version of the Journal as well as 
\url{www.astro.caltech.edu/~amc/SigmaOri.html}.}
\label{lightcurves}
\end{figure*}

The final sample of periodic variables contains 84 objects with clear variability by all criteria. Phased 
light curves for these targets are presented in Fig.\ \ref{lightcurves}, and their measured properties are 
listed in Table~3. The majority are VLMSs with roughly sinusoidal variability. However, the shapes of 19 are 
more characteristic of traditional pulsators or eclipsing binaries, and their blue colors are indicative of 
locations in the background field. For completeness, these are included in Table~3 as well. We have also 
identified a small number of objects with possible but questionable periodic variability. In these cases, the 
RMS of the residual light curves remains significantly larger than the expected noise level after subtraction 
of the detected signal. Objects in this small sample may consist of either undulating noise levels or other 
sources of non-periodic variability and are noted as unknown variable type in Table~3.

\section{Aperiodic variability detection}

Past monitoring campaigns have revealed not only well-behaved periodic variability among low-mass young 
cluster members, but also sporadic, aperiodic brightness fluctuations likely indicative of accretion or 
time-variable disk extinction. While the light curves are a challenge to analyze quantitatively, their 
features offer clues into the mechanisms behind star-disk interaction.  To fully mine our data for variables 
of all types, we have subjected the light curves to a battery of statistical tests in addition to the 
periodogram analysis. We examine the RMS magnitude spread for light curves of all objects in each of the two 
observed fields, as shown in Fig.\ \ref{detlim}. Such plots are standard tools for not only assessing the 
photometric performance, but also identifying outliers whose light curve RMS greatly exceeds the expected 
precision and hence suggests underlying variability. While the overall spread in light curves is well modeled 
by a combination of poisson errors, sky background, and a small systematic uncertainty ($\sim$0.002 
magnitudes), many outliers that were not identified through the periodogram analysis are obvious in Fig.\ 
\ref{detlim}-- indicating variability of a more erratic sort.

\subsection{Chi-squared analysis} 

\begin{figure}
\begin{center}
\includegraphics[scale=0.45]{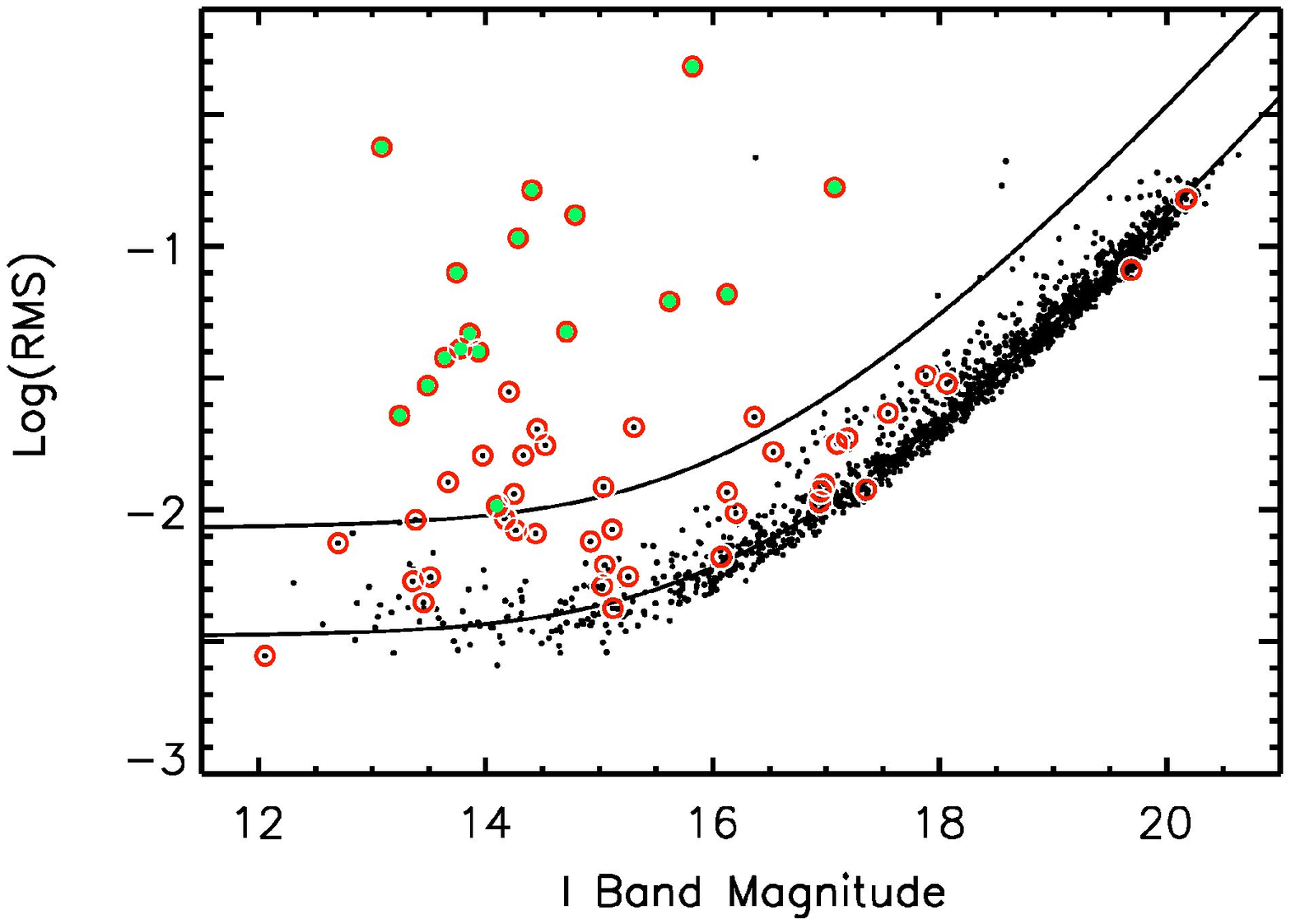}
\includegraphics[scale=0.45]{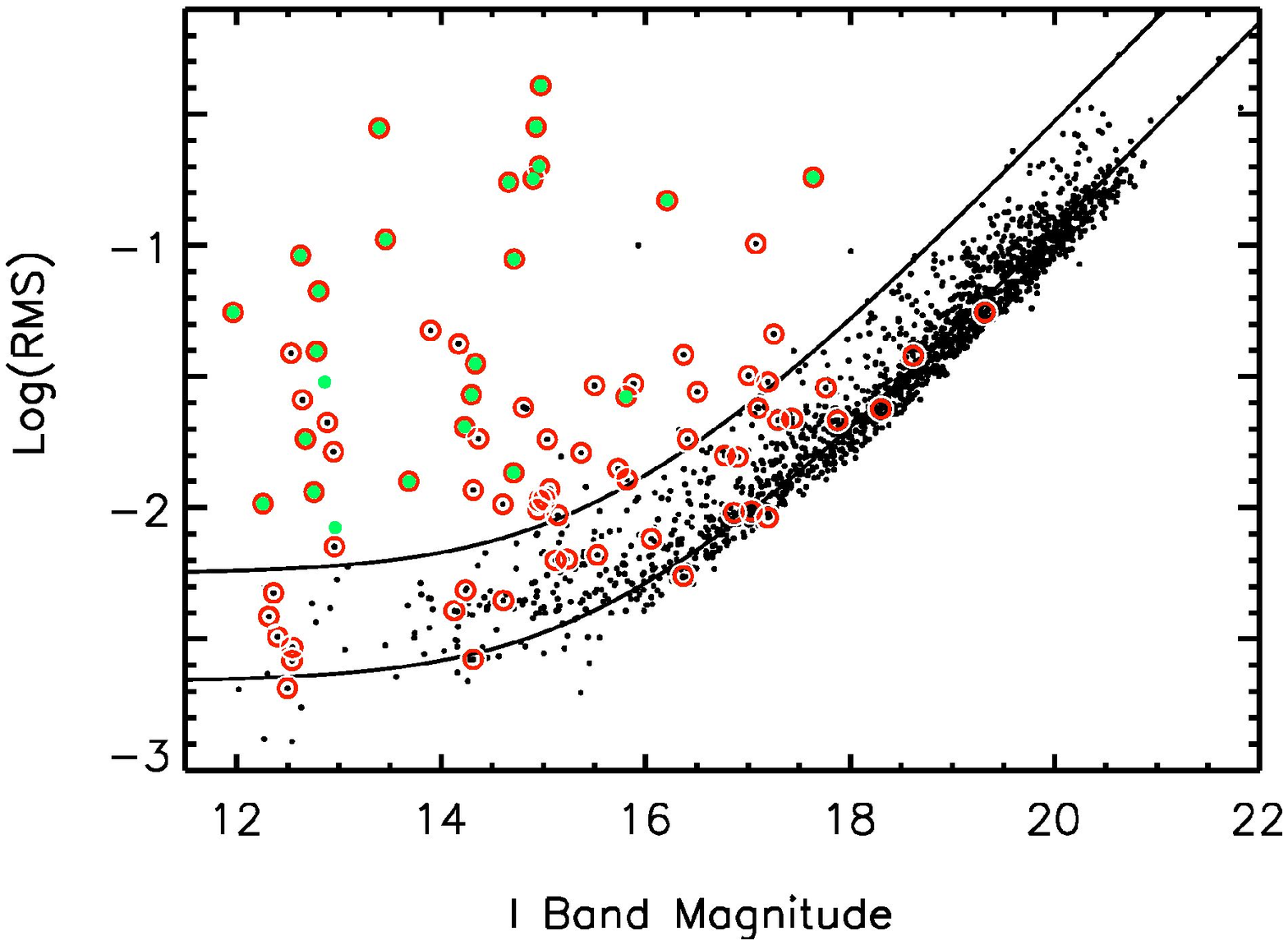}
\end{center}
\caption[]{Same as Figure 4, except now showing aperiodic variables in
green. We plot the estimated total contributions from poisson, sky, and systematic
noise, shifted upward by 0.12 dex so as to match the median of the data (solid line). The curve   
corresponding to 99\% probability of variability detection via the $\chi^2$ test appears above this.}
\label{apermagrms}
\end{figure}

To distinguish between true variables and photometric errors, we disregarded targets with photometry 
clearly affected by bad pixels, saturation, or close proximity to neighboring stars, as the large RMS values 
are due to measurement issues rather than intrinsic variability. We subjected the remaining group of objects 
with inexplicably large RMS to a reduced Chi-squared criterion: if the photometric uncertainty of an 
individual data point $x_i$ is $\sigma_i$, then for a light curve with mean 0 and N total points, we have: 
$$\chi^2=\Sigma\frac{x_i^2}{\sigma_i^2(N-1)}.$$ In addition, the measured standard deviation of the light 
curve, $\sigma$, is given by: $$\sigma^2=\Sigma\frac{x_i^2}{(N-1)}.$$ If the individual photometric 
uncertainties are well represented by some typical value dependent on the object magnitude $m$, e.g., 
$\sigma_i\sim\sigma_{\rm typ}(m)$, then we see that the reduced $\chi^2$ criterion translates to a requirement 
on the standard deviation: $$\chi^2=\frac{\sigma^2}{\sigma_{\rm typ}(m)^2}.$$ To detect aperiodic variables 
with an estimated 99\% certainty, we select only light curves with $\chi^2>6.6$, or equivalently, a spread of 
more than 2.58 times the photometric uncertainty. These values are approximate, since the noise is not 
strictly gaussian, as assumed by the statistics. We estimated typical photometric uncertainties by performing 
a median fit as follows to the RMS as a function of magnitude using the combined poisson, sky, and systematic 
noise model: The values of all three noise sources were fixed (as a function of magnitude) according to the 
noise model components derived in $\S$4.2.  A constant was then added to the model and adjusted such that half 
of the RMS light curve values lay above the model, and half lay below. The detected periodic variables as well 
as all 3-$\sigma$ outliers were rejected, and the fitting process was iterated until the median-fit function 
did not change. The variability detection cut-off was then taken to be the median fit, raised by a factor of 
2.58. These curves are superimposed on the data in Fig.\ \ref{apermagrms}.

Like the periodic variability search, the excess RMS analysis was conducted on all objects with available 
photometry, irrespective of cluster membership status.  After selection of probable variables via the $\chi^2$ 
criterion, we overplotted in Fig.\ \ref{apermagrms} those confirmed or likely to be members. It is evident 
that the vast majority of high-amplitude variables in our fields are known $\sigma$ Ori members, and the 
remainder are therefore good candidates. Objects exhibiting large RMS light curve spreads but not shown as 
variables (green dots) in Fig.\ \ref{apermagrms} were already found to be periodic (e.g., $\S$5) and displayed 
instead in Fig.\ \ref{detlim}.  Quite a few of the identified {\em periodic} variables lie below the $\chi^2$ 
detection threshold, indicating the power of the periodogram for identification of variability isolated to 
specific frequencies. In addition to the $\chi^2$ test, we probed all light curves for variability by 
calculating the single-band Stetson index \citep[e.g.,][]{1996PASP..108..851S}, which is a measure of the 
degree of correlation between successive data points. The distribution of Stetson index as a function of 
magnitude was fairly tight, such that the number of variables selected was relatively insensitive to the 
threshold chosen for variability detection. While this test confirmed all cases of aperiodic variability 
uncovered with the $\chi^2$ criterion and a number of the previously identified periodic variables, it did not 
reveal any additional variable objects. This result may reflect a large typical intrinsic light curve scatter 
for the aperiodic variables in our sample.
 
\begin{figure*}
\begin{center}
\epsscale{0.85} 
\plotone{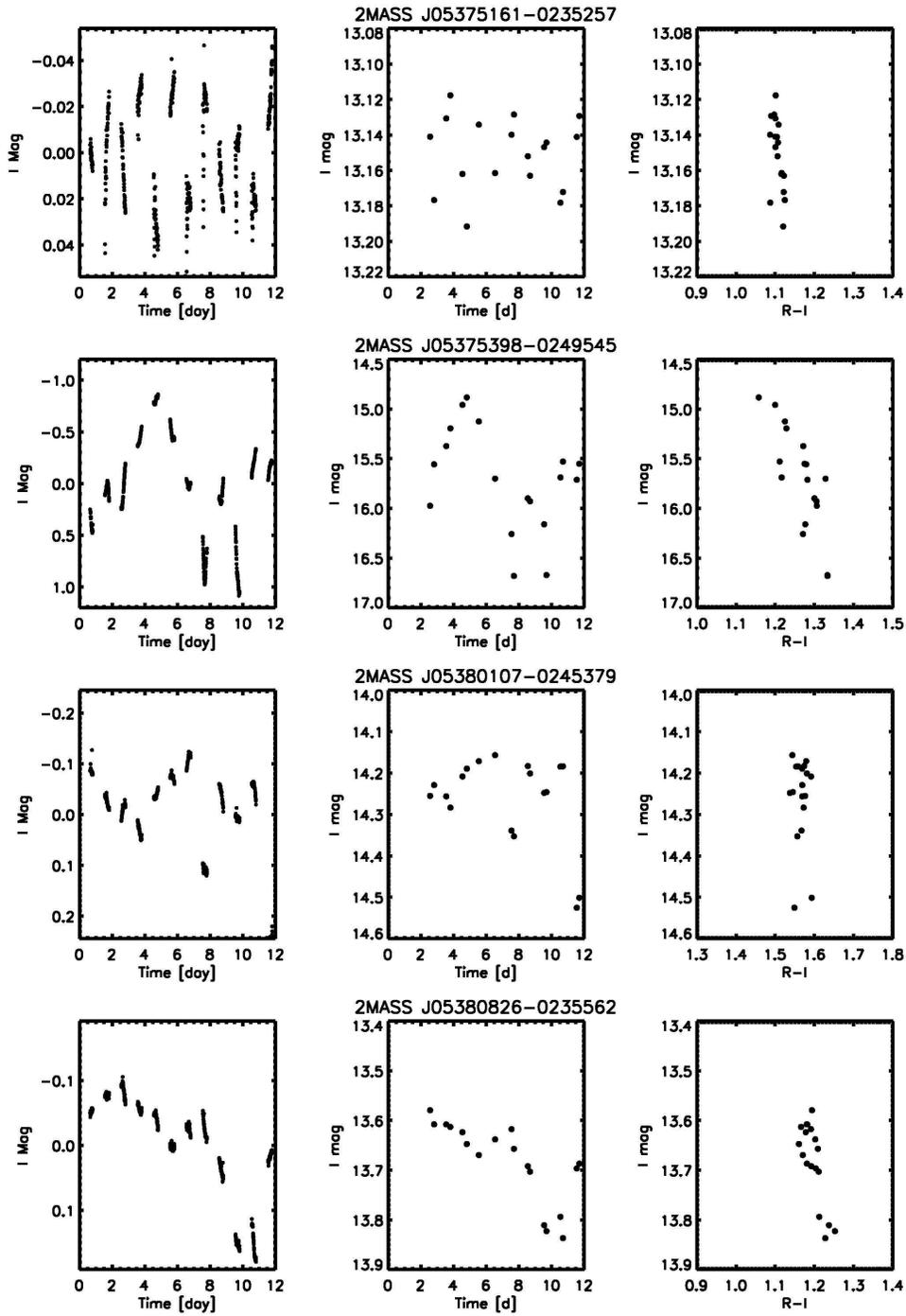}
\end{center}
\caption[]{Light curves selected as aperiodic based on large $\chi^2$ values and lack of periodicities. 
Objects are arranged in order of right ascension, and membership information is available in Table~4. The left 
column displays the full $I$-band light curves, while the middle shows the same data at the reduced cadence
corresponding to the $R$-band observations. The right column shows $R$-$I$ color trends. Figure~5
in its entirety is available in the online version of the Journal as well as
\url{www.astro.caltech.edu/~amc/SigmaOri.html}.}
\label{aperlightcurves}
\end{figure*}

In total, we identified 42 aperiodic variables, as listed in Table~4 and shown in Fig.\ 
\ref{aperlightcurves}. In order to explore the relationship between erratic variability and the presence of 
disks and accretion, we have noted the objects in Table 1 with observed infrared or near-infrared excess, 
and also provide the H$\alpha$ equivalent width where available in Table~4; in $\S$7.4 we discuss the 
correspondence between these quantities.

\subsection{Sensitivity to combined aperiodic and periodic variability}

In $\S$5.2 we simulated our sensitivity to photometric periodicities at different frequencies by assuming that 
the underlying light curves are well represented by a combination of simple noise sources (white and red) and 
a single sinusoidal signal. However, the large number of aperiodic variables detected via the $\chi^2$ test 
indicates that many light curves are in fact dominated by other types of variability, such as that associated 
with accretion. In these cases, we may not be able to detect periodicities superimposed on the 
larger-amplitude erratic fluctuations. We have investigated this reduction in sensitivity by injecting 
sinusoids of various frequency and amplitude into the light curves of a large subset of our aperiodic 
variables. The sample includes objects with RMS ranging from 0.01 to 0.3 magnitudes and $I$-band brightnesses 
from 12.0 to 17.5 magnitudes. We then attempted to recover the injected signals in the periodograms. The 
erratic nature of these light curves produces a steep trend in the frequency domain similar to the red noise 
from correlated photometric errors, but reaching higher amplitudes.

Since detection of periodic variability is frequency dependent, we have performed signal recovery tests in 
three regimes: frequencies less than 1 cd$^{-1}$ (e.g., periods greater than 1 day), frequencies between 1 and 
3 cd$^{-1}$, and frequencies greater than 3 cd$^{-1}$. These domains were chosen based on the typical 
exponential shape that we find for periodograms in our aperiodic variable sample. Our tests indicate that the 
periodogram noise levels for these objects are well correlated with the RMS spread in their light curves, 
regardless of brightness. This RMS ranges from 0.01 to 0.4 (see Table~3) and should not be confused with the 
photometric noise level, which is typically much smaller. Amplitudes of the injected signals ranged from 
25--400\% of the RMS for the two lower frequency regimes and 5--50\% of the RMS for the high frequency regime.

Most of the injected signals appeared clearly in the periodogram, but the decision as to whether they were 
``detectable'' depended on the surrounding noise level. For frequencies less than 1 cd$^{-1}$, the mean 
periodogram noise is approximately the light curve RMS divided by 2.2 (e.g., $\sim$0.45$\times$RMS), whereas 
for frequencies from 1 to 3 cd$^{-1}$, this decreases to the RMS divided by 2.9 (e.g., $\sim$0.34$\times$RMS). 
Noise in the periodograms of aperiodic variables decreases drastically toward higher frequencies or short 
periods, and consequently for frequencies beyond 3 cd$^{-1}$, the mean periodogram noise level decreases to 
RMS/23 (e.g., $\sim$0.04$\times$RMS). Detectability of a periodic signal requires an amplitude of at least 4.0 
times the periodogram noise level. Therefore, our ability to detect periodic signals superimposed on aperiodic 
variability requires periodic amplitudes larger than $\sim$1.8$\times$RMS, $\sim$1.36$\times$RMS, and 
$\sim$0.16$\times$RMS in the three respective frequency ranges. Based on a median {\em periodic} variability 
amplitude of 0.02 magnitudes, we then expect to detect both aperiodic and periodic variability in cases where 
the period is less than eight hours (e.g., frequency $>$3 cd$^{-1}$) and the RMS of aperiodic variability is 
less than 0.13 magnitudes. It may also be possible to detect periodicities with longer periods, but only if 
the RMS of aperiodic variability is near 0.01-- an uncommon occurrence, according to Table~3. We conclude that 
it is a challenge to identify both periodic and aperiodic variability in individual objects because of the 
different characteristic amplitudes of these phenomena.

\section{Variability in the Context of Stellar and Circumstellar Properties}

\begin{figure}
\begin{center}
\includegraphics[scale=0.45]{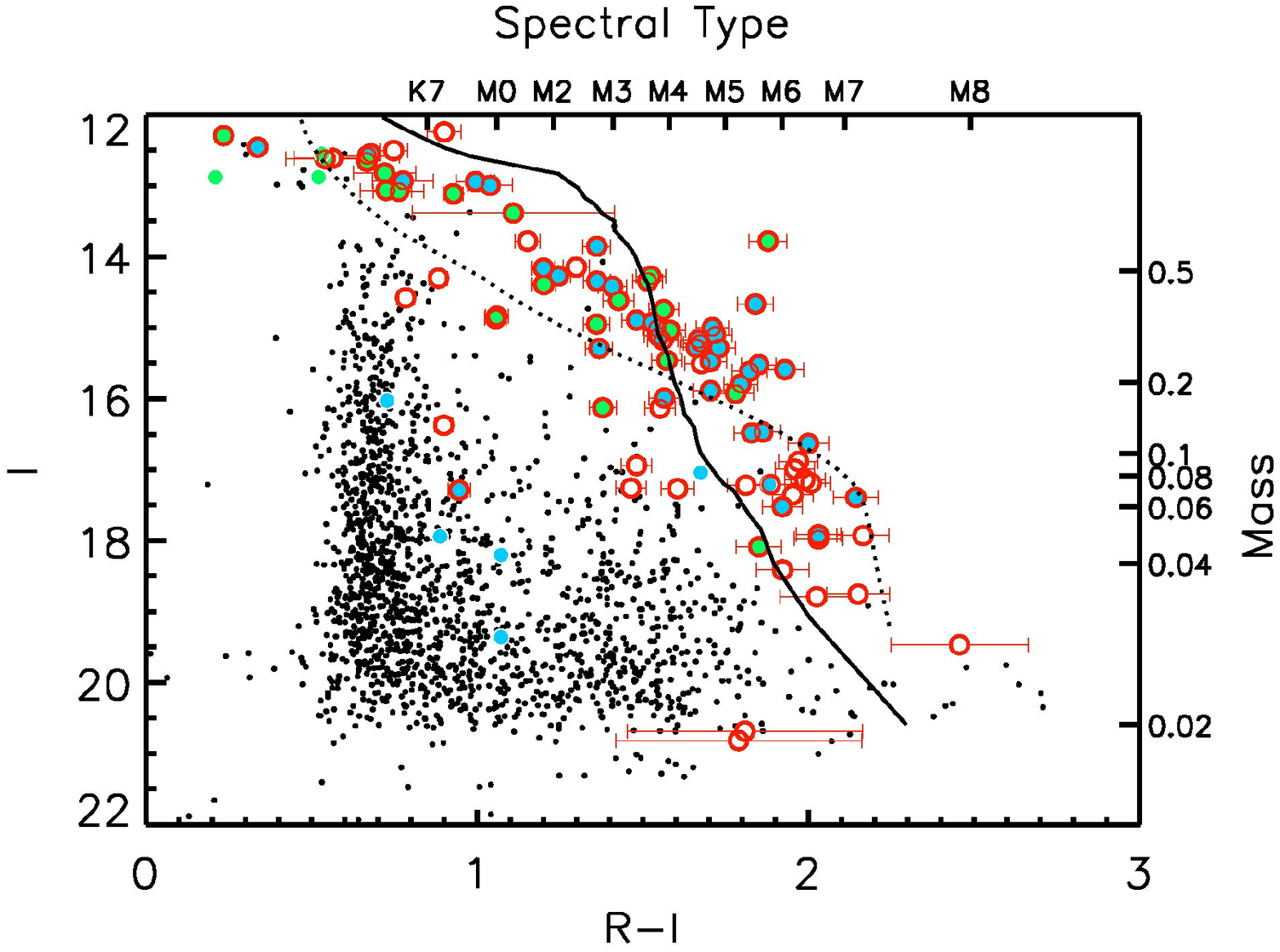}
\includegraphics[scale=0.45]{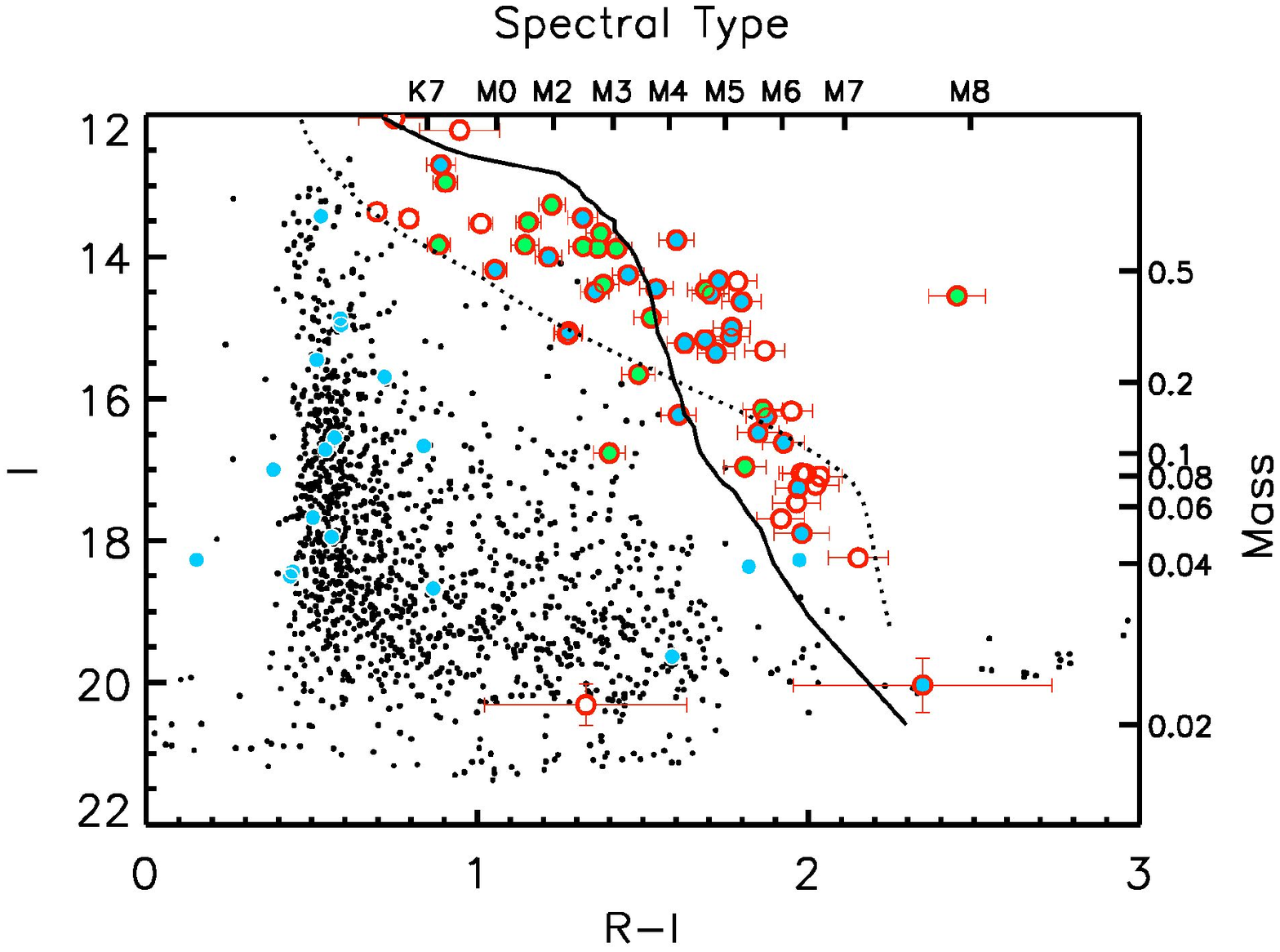}
\end{center}
\caption[]{$R$ and $I$ color-magnitude diagrams for all objects with photometry (black dots) derived from
our fields (top: 2008; bottom: 2007). Red circles are confirmed and candidate $\sigma$ Ori cluster members,
while cyan filled circles are detected periodic variables and green filled circles are aperiodic variables.
We have overplotted 3-Myr isochrones from \citet{1998A&A...337..403B} (solid curve) and
\citet{1997MmSAI..68..807D} (dotted curve) to illustrate the theoretically
predicted sequence for young cluster members. Masses are from \citet{1998A&A...337..403B}, but those from
\citet{1997MmSAI..68..807D} are similar. Spectral types shown were derived from the empirical relationship
between $R$-$I$ and spectral type among objects in our data and a few from $\sigma$~Ori datasets in the
literature. The two fields exhibit different average reddening due to spatial variations in extinction.}
\label{ricolmag}
\end{figure}

We have identified 126 variables in our fields, including at least 107 suspected $\sigma$ Ori members (101 of 
these are previously proposed members and six are candidate members newly identified here). In Fig.\ 
\ref{ricolmag} we present $R$-$I$ versus $I$ optical color-magnitude diagrams derived from our photometric 
data (\S 4.3) and overplotted with 3 Myr theoretical isochrones from \citet{1998A&A...337..403B} and 
\citet{1997MmSAI..68..807D}, incorporating a conversion to photospheric colors using color-temperature and 
bolometric-correction-temperature relationships, along with a distance of 440~pc \citep{2008AJ....135.1616S}.  
The vast majority of the variables fall above the main sequence and along a possible young cluster sequence. 
This finding confirms that single-band photometric monitoring is an efficient way to identify 
pre-main-sequence low-mass stars and brown dwarfs, and thus an effective technique in fields where the 
pre-main-sequence stars do not stand out in color-magnitude diagrams as distinct from the field stars.

The light curves and their temporal properties offer insights into the origin and prevalence of brightness 
variations, which we discuss in $\S$7.1. Yet we can also make use of the rich array of data from previous 
spectroscopic studies (e.g., Table 1) as well as the $Spitzer$ mission to analyze variability from several 
additional angles. In the forthcoming sections, we assess the correlations of variability with stellar and 
circumstellar properties. The $R$-$I$ photometry available from our work provides not only information on the 
relationship between brightness and color changes ($\S$7.5), but also a means to investigate the 
mass-dependent properties of young stars and brown dwarfs ($\S$7.3). In addition, we employ mid-infrared data 
to connect variability with the presence of disks around these objects ($\S$7.4).

\subsection{Overall variability properties}
\subsubsection{Variability classification and persistence}

Characterization of variability can illuminate our understanding of the physical processes that take place on 
and around few Myr old low-mass stars. We have identified several types of variability among our sample of 126 
variables, including irregular variability and various forms of periodic variability such as spot modulated 
stellar rotation, pulsations, and full or partial eclipse signatures, as listed in Table~3.  Among 147 
previously known or suspected cluster members included in our photometry, the overall variability fraction is 
69\%, with irregulars (27\%) and periodic objects (42\%) comprising this cluster sample. Furthermore, we 
uncovered 25 variables with no prior membership information, most of whose light curves resemble eclipsing 
binaries or short-period pulsators. However, six have colors consistent with membership in $\sigma$ Ori and 
light curves consistent with either spot modulation or accretion. Since these six objects encompass a range of 
brightnesses, it is not clear as to why they they were missed in previous surveys. The new candidates are 
included in Table~1 and noted in Tables 3 and 4 as possible members. Just under half (44\%) of objects in the 
remaining 31\% of our sample for which no variability is detected have strong evidence for $\sigma$~Ori 
membership based on Table~1. Hence we conclude that at least 15\% of young cluster members may not display 
obvious brightness fluctuations on time scales up to two weeks.

Among the 41 $\sigma$ Ori members in our fields previously identified as variable objects (35 aperiodic and 6 
periodic; see Appendix A), we confirm variability in 33 (30 aperiodic and 3 periodic); this suggests that the 
variability mechanisms are long-term rather than sporadic phenomena. In the subset for which we do not 
redetect variability, there are no particular biases toward long or short time scale. We suspect that the 
combination of low numbers of data points, uneven time sampling, and underestimated uncertainties could have 
contributed to previous false detections in some cases. However, it is also possible that the variability 
mechanism itself turned off during the time of our observations.

In addition to comparing our variability detections with those of other works, we can use our own repeat 
observations of the 2007 field to glean further information about the time scales on which various types of 
variability operate. While the small number of data points per light curve (23, or two per night taken in 
2008) precludes detailed comparison of variability properties from one year to the next, we can nevertheless 
identify objects with high-amplitude variability persisting on this longer time scale. Of the 17 aperiodic 
variables found in our 2007 field, we re-detect {\em all} of them again in 2008, based on the $\chi^2$ 
analysis described in $\S$6.1. In addition, 22, or over 80\%, of our 27 periodic variables identified as 
likely $\sigma$~Ori members in the 2007 field display significant variability at a similar period (the 
majority agreed to within 5\%) in 2008.

We can estimate a minimum characteristic timescale, $T$, on which the various types of variability operate by 
considering the set of all objects with repeat observations separated by at least one year. In total, there 
are 52 aperiodic variables that were either observed in both 2007 and 2008 by us, or identified by another 
group and observed later by us. Of these, 47 displayed aperiodic variability during both sets of observations. 
We suppose that for a typical duration of accretion (or other source of aperiodic variability) $T$, the 
probability that variability will persist one year after its initial detection is $p\sim e^{-1/T}$. Taking 
this probability to equal 47/52, we find the typical characteristic for aperiodic variability time scale to be 
$T\sim10$~years. A similar result is obtained using a binomial distribution to describe the probabilities for 
the outcomes of measuring variability.

Likewise, we can perform the same analysis for the periodic variables. In this case, 25 of 33 objects 
exhibited variability at roughly the same period during repeat observations over one year apart. The 
corresponding time scale for persistence of periodic variability is then at least $\sim$4~years.  Based on 
these results, we conclude that the types of variability present among these young cluster sources are 
long-lived in comparison to the objects' rotation periods ($\sim$1--10 days) as well as the intra-night time 
scale of abrupt light variations seen in aperiodic objects.

\subsubsection{Variability demographics across time scale and brightness}

In addition to visual classification of light curves, we can also consider variability properties in the time 
and magnitude domains. In doing so, it is important to understand any selection or other effects that may mask 
certain kinds of variability from being observed. The observing set-up imposes practical constraints on 
variability detection through photometric cadence, precision, interruptions, and total duration. These details 
translate into a maximum detectable amplitude for periodic variables and sets the range of detectable periods. 
The demographics of variability present additional considerations for our ability to classify light curve 
behavior. Some fraction of young stars and brown dwarfs may not have magnetic spots, or their surface features 
may be too small to induce observable variability and potentially infer a rotation period. Other objects may 
have multiple sources of variability (e.g., spots, accretion, circumstellar variability) that are difficult to 
separate from each other. In what follows, we carefully consider the connection between these effects and the 
variability trends that we have uncovered.

In the time domain, our observations are sensitive to photometric periods between $\sim$20 minutes and 
$\sim$12 days, as discussed in $\S$5.  While we do encounter periodic variability close to the longest 
possible time scales, we detect no periodicities on the shortest time scales-- less than 7 hours (e.g., Fig.\ 
\ref{colpd}). If this effect is the result of our photometric sensitivity, then it should be explained by the 
detection limits determined in ($\S$5 and \S 6). Instead, we find (Fig.\ \ref{detlim}) that we are {\em more} 
sensitive to short periods and could recover signals down to 0.001 magnitude amplitudes for objects brighter 
than $I$=16, or signals with 0.01 magnitude amplitudes out to $I\sim$19 or 20. Another possibility is that we 
are somehow missing periodic variability in cases where the light curves are dominated by aperiodic behavior. 
In $\S$6.2 we concluded that we are likely to identify both types of variability in a single object only if 
the time scale for the periodic component is less than 8 hours and the light curve RMS is below $\sim$0.13 
magnitudes. A number of the detected aperiodic variables do indeed have RMS values that satisfy this criterion 
(Table~4). Hence while detection limits may explain our failure to identify combinations of aperiodic 
variability and longer time scale periodicity in single targets, they do not account for the dearth of 
short-period variables. We conclude that the lack of periodic variability on time scales under 7 hours is a 
real physical effect.

Changes in variability properties as a function of magnitude can also shed light on the properties of young 
stars and brown dwarfs. To estimate the correspondence between mass, $I$-band magnitude, and $R$-$I$ color, we 
have overlaid 3 Myr theoretical isochrones from \citet{1998A&A...337..403B} and \citet{1997MmSAI..68..807D} on 
our data in Fig.\ \ref{ricolmag}. Since reddening is low in $\sigma$~Ori, the observed $R$-$I$ values are 
close to the intrinsic photospheric colors. Although mass predictions are fairly uncertain at these ages 
\citep{2002A&A...382..563B}, the two models agree well with each other and we have adopted the mass values of 
\citet{1998A&A...337..403B}. These estimates indicate that our dataset encompasses objects with masses from 
approximately 0.02 to 1.0~$M_{\odot}$.  The substellar limit, at $\sim$0.08~$M_{\odot}$, lies near $I$=17 or 
spectral type M6. The spectral types shown in Fig.\ \ref{ricolmag} were adopted directly from the objects in 
our $\sigma$~Orionis sample with available spectroscopy (Table~1).

We find variables of all types spanning the entire range of magnitudes, but Fig.\ \ref{ricolmag} displays a 
subtle decrease in variable cluster members at the faint end, which might be explained by the decline in 
photometric sensitivity. For the subclass of variables identified as aperiodic, we note that the brightest 
objects have light curve RMS values from 0.03 to 0.2. Based on the detection limits described in $\S$6, we 
lose sensitivity to this type of variability around an $I$ magnitude of 18.0. For objects brighter than this 
limit, we find that aperiodic variables seem to populate the entire range of magnitudes, including a portion 
of the brown dwarf regime. Attributing aperiodic variability to accretion and its associated hot spots or 
fluctuating dust extinction levels, we do not find significant evidence for physical changes in these effects 
across the substellar boundary.

Magnitude trends in {\em periodic} objects are slightly more difficult to determine, as they are dependent on 
period as well as the potential presence of aperiodic variability at larger amplitude. Our detection limits 
(Fig.\ \ref{detlim}) indicate that we are sensitive to amplitudes of $\sim$0.01 magnitudes out to 
$I\sim$18.5-19.5, depending on period. Thus we should be able to detect whether the properties of periodic 
variability are similar from the stellar through the brown dwarf regime. If we divide our sample into 
``bright'' ($I<17$) and ``faint'' ($I>17$) groups, we find the fraction of periodically variable faint objects 
to be 34$\pm$10\%. Compared to the number of targets that are periodically variable at brighter magnitudes 
(46$\pm$6\%), there appears to be a reduction in the fraction of variable members for faint magnitudes and 
thus lower mass.  The significance level of this finding is difficult to assess since cluster membership 
status is not secure for many of the fainter objects. However, if we restrict our estimate to {\em confirmed} 
(e.g., via spectroscopy or infrared excess) cluster members, the periodic variability fractions are similar to 
those of uncertain cluster members: 45$\pm$7\% for objects with $I<17$, and 26$\pm$12\% for those with $I>17$. 
The majority of periodically variable cluster members display roughly sinusoidal light curves consistent with 
rotational modulation of stellar spots. Therefore the apparent reduction in periodic variables toward fainter 
magnitudes suggests a difference in the photospheric properties of young brown dwarfs, as compared to the 
higher mass stars.

\subsection{Origin of periodic variability}

The periodic variability in our cluster sample is most likely due to spot modulation of the light curves. On 
time scales of 0.3--12 days and with amplitudes of 0.003--0.12 mag, the periods of the brightness changes 
among known and suspected cluster members are too long to be explained by the pulsation theory 
\citep{2005A&A...432L..57P}. We would have detected the shorter periods predicted by the theory if they had 
amplitudes of $\sim$0.001 (bright sample; $I<$16) to 0.01 magnitudes (faint sample; $I\sim$20). Further, the 
roughly sinusoidal shapes of the periodic variables are not consistent with other varieties of pulsators or a 
population of eclipsing systems, apart from the 19 field objects listed in Table~3. Instead, the time scales 
and amplitudes are compatible with modulation of spots that may be either cooler than the photosphere, as in 
active chromosphere models, or hotter than the photosphere, as in accretion column models 
\citep{2001AJ....121.3160C,2009MNRAS.398..873S}. Comparison of theoretical spot models with multi-color 
photometric data has shown that both scenarios can produce larger amplitude light curves at shorter wavelength 
\citep[e.g.,][]{2009A&A...508.1313F}. Although we have a small sample of $R$-band data points for each target, 
the color data are not extensive enough to allow for detailed modeling. In either case we assume that the 
periodicities extracted from our analysis can be attributed to rotational modulation of surface 
inhomogeneities and directly adopted as rotation periods.

\subsection{Rotation rates in $\sigma$ Orionis}
\subsubsection{Distribution with color/mass}

For ``higher'' mass ($>$0.3--0.4 $M_{\odot}$) stars in the ONC, NGC~2264, and IC~348 clusters derived periods 
have in some cases revealed double-peaked distributions, with two groups clustered near 1-2 and 8-10 days 
\citep{2002A&A...396..513H,2005A&A...430.1005L,2006ApJ...649..862C}. For other young cluster datasets, the 
distribution is not bimodal but peaks near 3--5 days \citep{2007ApJ...671..605C,2008MNRAS.384..675I}. In 
contrast, our $\sigma$ Ori sample extends well into the brown dwarf regime and the corresponding periods 
cluster at short time scales, 1-2 days, with a uniform or exponentially decreasing tail extending out to and 
perhaps beyond 10 days. Only a few objects in the sample have periods in the 8-10 day range. Since the dataset 
includes a representative sampling of the $\sigma$~Ori IMF between $\sim$0.02 and 1.0~$M_\odot$ it is 
possible to search for trends in the period distribution along the color and magnitude axes.

\begin{figure}
\begin{center}
\includegraphics[scale=0.5]{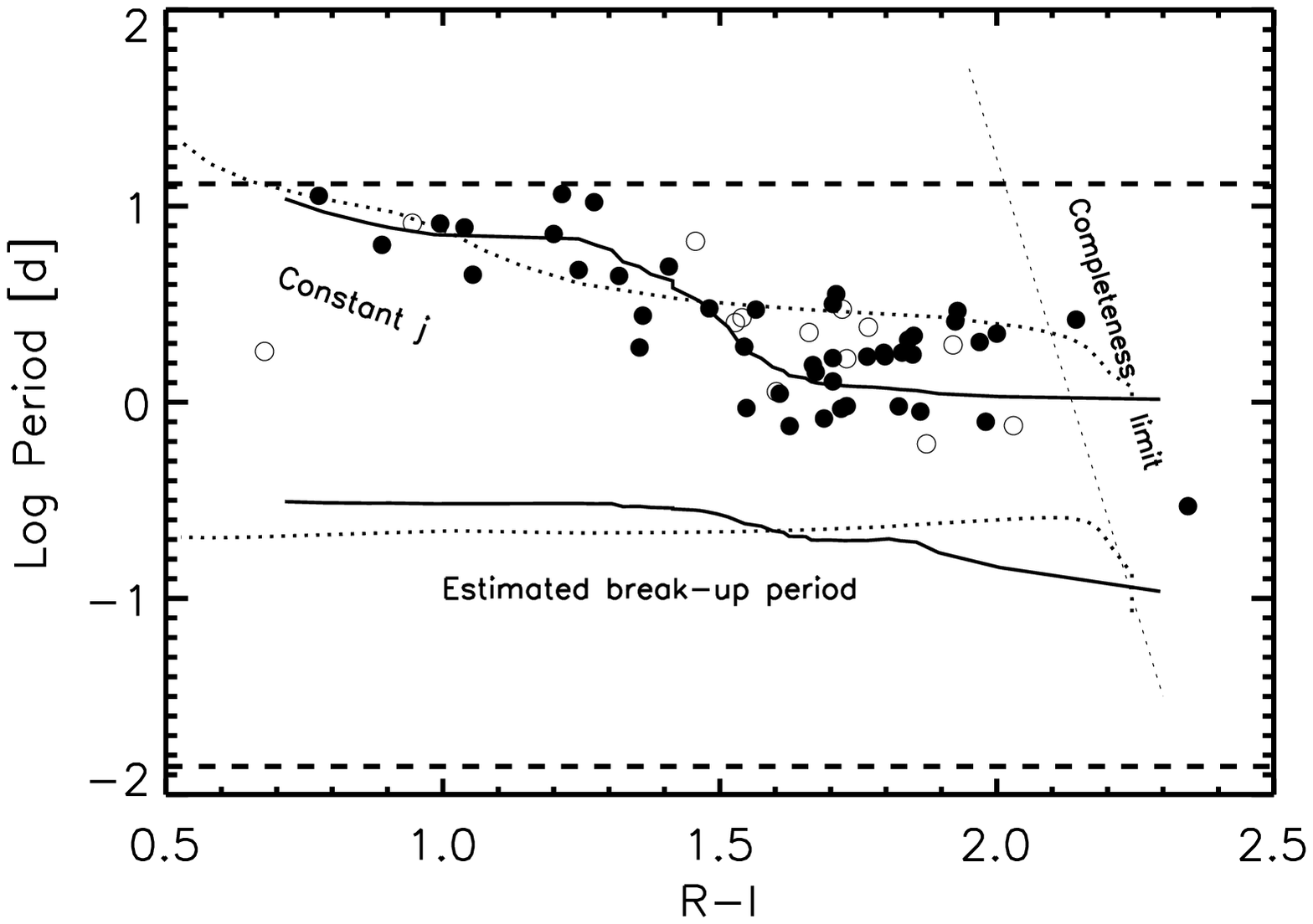}
\includegraphics[scale=0.5]{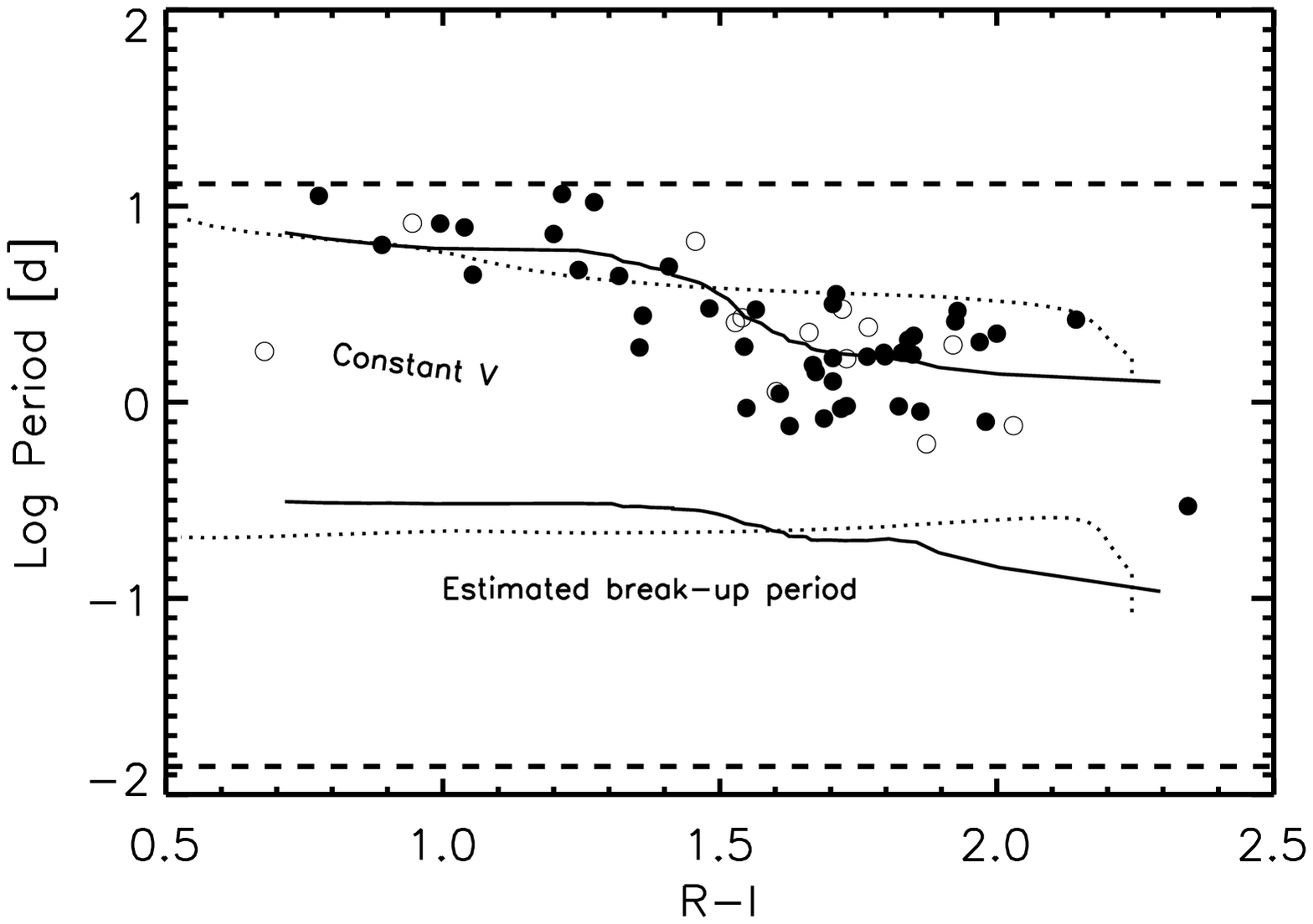}
\end{center}
\caption[]{Period of variables versus their $R-I$ color. Variables without obvious periods are not
included, nor are those periodic variables having colors inconsistent with
cluster membership. Objects with infrared excesses indicative of disks ($\S$7.4) are marked as open
circles, whereas objects without evidence of a disk are filled circles. In the top diagram, we have    
overplotted models of constant specific angular momentum ($j$) derived from radii provided by the 3~Myr
isochrones of \citet{1998A&A...337..403B} (solid curve) and \citet{1997MmSAI..68..807D} (dotted). The
dotted line at the right side represents the completion limit redward of which we cannot detect periodic
signals of amplitude less than 0.007 magnitudes. In the bottom diagram, we overplot models of
constant angular velocity from the same isochrones. In both plots, we show estimated break-up
periods derived from mass and radii predicted by the same theoretical models.}
\label{colpd} 
\end{figure}

\begin{figure}
\begin{center}
\includegraphics[scale=0.5]{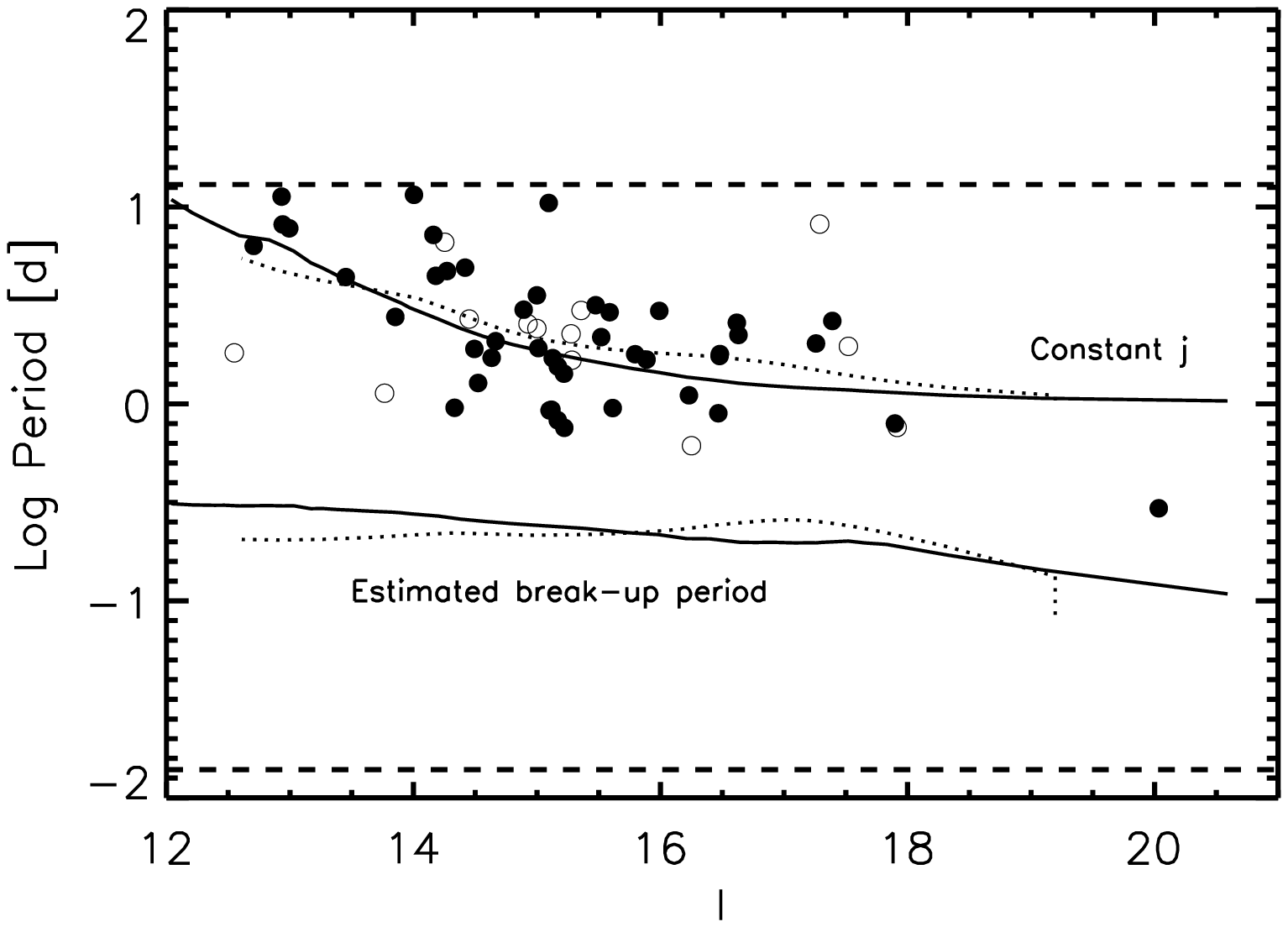}
\includegraphics[scale=0.5]{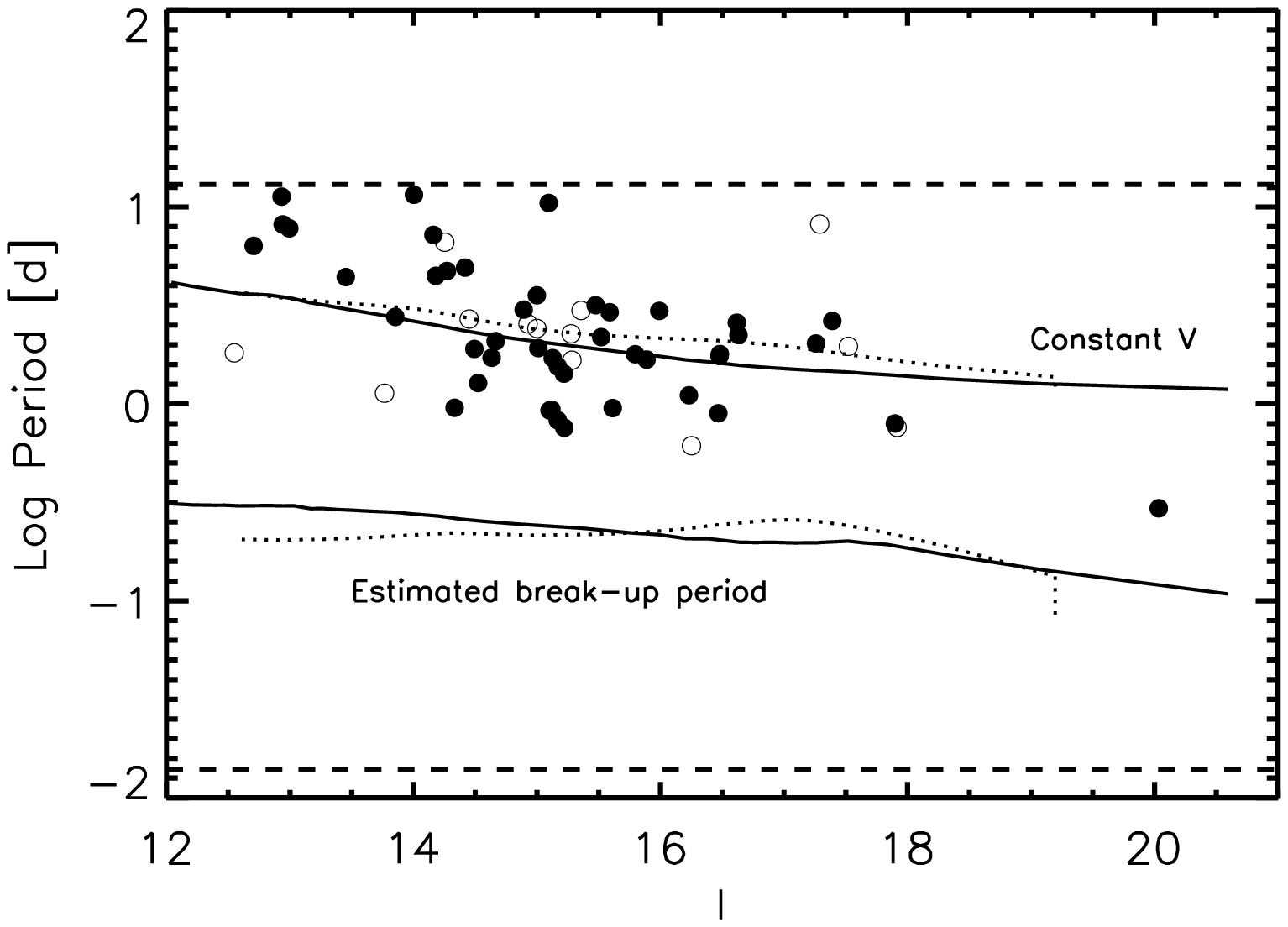}
\end{center}
\caption[]{Period of variables versus their $I$-band magnitude. The sample, as well as the symbols and
curves, are the same as in Fig.\ \ref{colpd}. Likewise in the bottom diagram, we show models of
constant angular velocity.}
\label{ipd}
\end{figure}

In Figs.\ \ref{colpd} and \ref{ipd}, we present the period as a function of $R$-$I$ and $I$, both of which 
serve as a proxy for mass since extinction is low. Included are only those periodically variable objects with 
solid or likely cluster membership status based on colors and spectroscopic data available in the literature 
(Tables 1 and 3). In this way, contamination by periodicities of field variables should be negligible. Apart 
from one or two outliers, there is a significant decrease in period with progressively redder color or fainter 
magnitude, implying that within this mass range, lower mass objects rotate faster than the higher mass ones. 
Taking the substellar boundary to be near spectral type M6 or $R$-$I\sim 1.9$ and $I\sim 16.5$ (see Fig.\ 
\ref{ricolmag}), there are nine brown dwarfs in the rotation sample with periods ranging from $\sim$7 hours to 
$\sim$3 days. On the other hand, the higher mass stars with $R$-$I<1.3$ or $I<14.3$ and 
$M\gtrsim 0.45 M_{\odot}$ have periods larger than 4.5 days, with the exception of one object. The 
correlation of period with mass is statistically significant at the $10^{-6}$--10$^{-5}$ level, depending on 
whether the test is run on period and color or period and magnitude. Masses estimated from photometry are 
dependent on the theoretical model used, and the values presented here are derived from 
\citet{1998A&A...337..403B}, based on $I$-band magnitude and an age of 3~Myr. Previous works have used 
cut-offs between young ``low'' and ``high'' mass stars of spectral type M2.5 and masses of either 0.25 or 
0.4~$M_\odot$ depending on the theoretical model \citep[e.g.,][]{2007prpl.conf..297H}. We adopt a slightly 
higher value of 0.45~$M_{\odot}$ corresponding to $I$=14.3 and find that 78\% of our sample falls in the 
low-mass end.

An intriguing aspect of our data is that several regions of the color-period and magnitude-period diagrams are 
nearly devoid of data points. Only one cluster member appears with a rotation period less 14 hours. This 
finding cannot be a result of our detection limits, as our sensitivity {\em increases} on shorter time scales 
($\S$5.2). To test whether a short-period cut-off might be explained by the maximum allowed rotational 
velocities, we have estimated the periods required for break-up as a function of mass, using masses and radii 
from the 3~Myr models of \citet{1998A&A...337..403B} and \citet{1997MmSAI..68..807D}. Break-up is assumed to 
occur when the centrifugal force from rotation exceeds self gravity; the results of these computations are 
shown in Figs.\ \ref{colpd} and \ref{ipd}. The break-up periods increase slowly with mass and range from 2 to 
7 hours, and thus there is a significant gap between the break-up curve and the observed rotation data. 
Consequently, some physical mechanism seems to limit the rotation speed of most low-mass objects to at most 
40\% of break-up, and even slower speeds at higher mass.

In addition to a lack of variability on few-hour time scales, we also find a dearth of periodic variables in 
two other regions of the period-color and period-magnitude diagrams:\ from Figs.\ \ref{colpd} and \ref{ipd}, 
we see that only two blue objects (e.g., $R$-$I\lesssim$1.3, $I\lesssim$14.3, or spectral type earlier than 
M2.5) rotate with periods faster than 3 days, and only one of the redder objects (e.g., $R$-$I>1.5$, 
$I\gtrsim$15, or spectral types later than M3.5) rotates with a period greater than 3.2 days. It is these two 
largely empty regions that conspire to create the pattern of increasing period with mass. To confirm that this 
trend is not a data selection effect, we have explored several scenarios that might prevent detection of 
rotation periods in the two regions.

As emphasized previously, our sensitivity to periodic signals increases on shorter time scales down to 20 
minutes; hence this does not explain the gap in period detections at the bright end. However, detection also 
depends on variability amplitude. In Fig.\ \ref{detlim}, we have shown that we are sensitive to amplitudes of 
$\gtrsim$0.001 magnitudes for the brightest ($I<$16) and bluest objects.  The entire sample of periodic 
variables associated with rotation has a mean amplitude of 0.02 magnitudes, with a standard deviation of 0.013 
magnitudes. Thus we expect only a small fraction of periodic variables to display amplitudes less than 0.007 
magnitudes. To determine whether a population of ``missing'' blue objects with such low amplitudes could 
explain the deficit of data points in the lower left portion of the color-period diagram, we examined the 
periodograms of all cluster members with $R$-$I<$1.3 and no detected variability. In the majority of these 
objects, we are able to rule out the presence of periodicities with amplitudes greater than 0.007 magnitudes. 
For those members that display aperiodic variability, identification of underlying periodicities is nearly 
impossible (see $\S$6.2). However, we see no reason that the light curves of aperiodic objects would contain 
periodic variability with preferentially short period, unless there is some additional spin-up due to ongoing 
accretion. Thus we tentatively conclude that there is a real deficit of $\sigma$~Ori members blueward of 
$R$-$I$=1.3 and $I$=14.3 with periods less than 3 days.

The second empty region of the color-period diagram, where $R$-$I\gtrsim$1.5 or $I\gtrsim$15, displays an 
apparent boundary at periods over $\sim$4 days. It is tempting to identify this as a physical trend, but not 
immediately clear whether it could simply reflect our diminished sensitivity to longer periods at faint 
magnitudes. To find the locus of colors, magnitudes, and periods for which we could detect periodic 
variability amplitudes as low as 0.007 magnitudes, we averaged all periodograms of non-variable field objects 
in 0.5-magnitude bins. For each bin, we fit an exponential curve to the mean periodogram, as in Fig.\ 
\ref{avperiod}. To detect a signal of amplitude 0.007, the noise level must be approximately 1/4 of this, or 
0.0018 magnitudes. The point at which the exponential fit reaches this value was then taken to be the minimum 
frequency required for a detection. We then converted this frequency to period, and employed an empirical 
isochrone fit to Fig.\ 8 to translate the $I$-band magnitude of each bin to an $R$-$I$ value. The resulting 
set of data points from all magnitude bins forms a locus on the color-period diagram which declines steeply 
with color, as shown by the completeness limit line in Fig.\ \ref{colpd}. Redward of this relation, we cannot 
uncover signals of amplitude less than 0.007 magnitudes and thus the periodic sample may not be complete. The 
locus crosses our maximum detectable period, $\sim$12 days, at $R$-$I\sim$2.0 and reaches a period of 1 day 
between $R$-$I$=2.1 and 2.15. While several data points fall redward of this line (these detections had higher 
amplitudes), a large swath of the empty region still lies on the blue side and cannot be explained by the 
completeness limit. As with the other gap in the color-period and magnitude-period diagrams, a survey of the 
periodograms of non-variable objects shows no evidence of overlooked periodicities with amplitudes greater 
than 0.007 magnitudes. It is once again possible that we may be missing periods in objects that are accreting 
and display high-amplitude erratic variability or have very small surface spots, but we cannot explain why 
these effects would only occur for certain combinations of colors and periods. Consequently, the trend of 
increasing period with decreasing color seen in Figs.\ \ref{colpd} and \ref{ipd} appear to reflect a physical 
correlation between rotation and mass.

\begin{figure}
\begin{center}  
\includegraphics[scale=0.5]{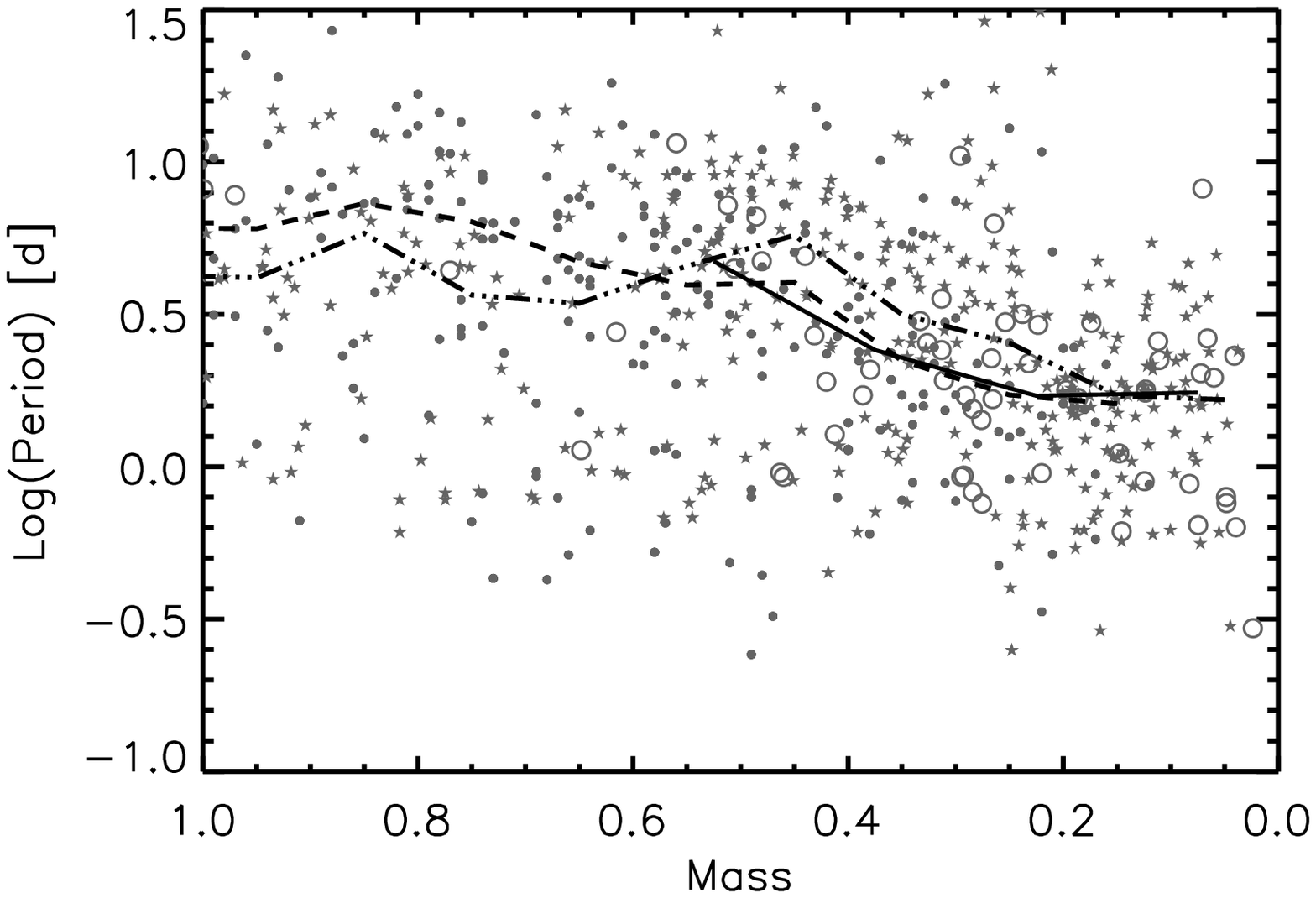}  
\end{center}
\caption[]{Period of variables in our $\sigma$~Orionis sample (open circles), NGC~2264
\citep[stars; ][]{2005A&A...430.1005L}, and NGC 2362 \citep[small circles; ][]{2008MNRAS.384..675I} versus
estimated mass based on $I$-band magnitude and the theoretical models of \citet{1998A&A...337..403B}. 
Curves show the the median period in 0.1~$M_\odot$ bins (or 0.15 $M_\odot$ for our sparser data): a 
dash-dotted line for \citet{2005A&A...430.1005L}, dashed line for \citet{2008MNRAS.384..675I}, and a solid 
line for our own data, which stops at $\sim$0.55$M_\odot$.}
\label{masspd}
\end{figure}

To explore whether the gaps found in our period-color and period-magnitude diagrams are a general feature of 
young star and brown dwarf rotation, we have compared our data to the period-mass distributions of the similar 
age clusters NGC~2264 \citep[$\sim$2~Myr;][]{2005A&A...430.1005L,2009IAUS..258..363I} and NGC~2362 
\citep[$\sim$5~Myr;][]{2008MNRAS.384..675I}. We in fact find quite a few objects with periods from 1--3 days 
across all masses. Nevertheless, there does appear to be a relative deficit of fast rotators at higher mass, 
as well as slow rotators at lower mass, similar to $\sigma$~Orionis. To compare rotation data from the three 
clusters more quantitatively, we have plotted them together in Fig.\ \ref{masspd}. $I$-band magnitudes from 
each set have been transformed to masses using the models of \citet{1998A&A...337..403B}, as well as cluster 
distances and $I$-band extinctions. Although there are inherent uncertainties to the theoretical models at 
this age, the systematic errors should be similar for each cluster. Superimposed on the data in Fig.\ 
\ref{masspd} are median fits to each set of periods and masses, which are remarkably similar for each of the 
three clusters, particularly for masses below 0.4~$M_\odot$. In addition, the rotation distributions in all 
three clusters appear to transition to longer periods above this mass (which is model dependent and 
corresponds roughly to $I\sim$14.5 for $\sigma$~Orionis. A Kolmogorov-Smirnov test reveals no significant 
differences between the three period distributions from the brown dwarf regime up to 0.5~$M_\odot$ where our 
own data peter out.

\subsubsection{Connection to internal structure and surface physics}

The measured periods and amplitudes can inform us about the angular momentum and magnetic field properties of 
very low mass stars and brown dwarfs. The fact that rotation period seems to be connected with color or 
magnitude, and hence mass, implies that a physical conservation law may be at work.

Light curve period, $P$, is related to specific angular momentum, $j$, via $j\propto R^2/P$. If specific 
angular momentum from the natal cluster gas is conserved among $\sigma$~Ori members, then we expect periods to 
scale as $R^2$. The actual radii of our sample objects are unknown, but theoretical models predict their 
values with significant uncertainty due to lack of information about initial conditions, opacity, and 
treatment of convection \citep{2002A&A...382..563B}. We have used the 3~Myr isochrones of 
\citet{1998A&A...337..403B} and \citet{1997MmSAI..68..807D} to estimate $R^2$ as a function of mass. 
Converting masses to $R$-$I$ and $I$ as in Fig.\ \ref{ricolmag}, determination of a relationship between 
period and color requires the selection of a scaling constant to represent fixed specific angular momentum. 
Since the moments of inertia of young, low-mass objects are not well known, we have simply used one end of the 
{\em observed} color-period relation to anchor the calculated constant angular momentum function. We present 
the results in Figs.\ \ref{colpd} and \ref{ipd} (top panels) for data from both \citet{1998A&A...337..403B} 
and \citet{1997MmSAI..68..807D}; both curves fit the color-period data surprisingly well. In particular, the 
model derived from the \citet{1998A&A...337..403B} isochrone can be adjusted so as to pass through the center 
of the data, reproducing the ``gaps'' seen in the lower left and upper right quadrants of the color-period 
diagram.

If young ($\sim$3--5~Myr) stars maintain constant angular velocity rather than angular momentum, we would 
expect periods to scale as $R$ instead of $R^2$. Although there is reason to believe that individual stars may 
{\em evolve} at constant angular momentum \citep[e.g.,]{2004AJ....127.1029R}, we have adopted this model 
primarily to illustrate how much freedom there is in fitting the data. We generated a constant angular 
velocity curve in the same way as we did for specific angular momentum and once again anchored one end to the 
observational data. As shown at the bottom of Figs.\ \ref{colpd} and \ref{ipd}, this function fits the 
observed periods and colors almost as well as the $R^2$ model, although two curves derived from the 
\citet{1997MmSAI..68..807D} isochrone are a bit flatter than the data. So while there certainly seems to be a 
trend in periods with color and magnitude, it is not tight enough to conclusively determine its cause. In 
addition, a single outlier (2MASS J05391883-0230531) at $R$-$I$=0.7 and a clear period of 1.8 days confounds 
the idea.

\begin{figure}
\begin{center}
\includegraphics[scale=0.6]{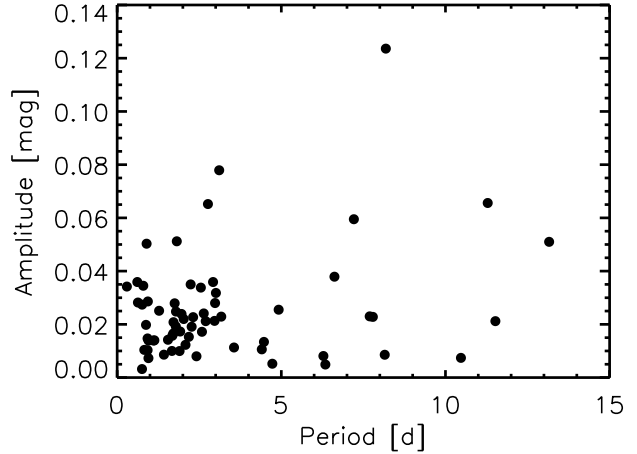}
%\plotone{f12.eps}
\end{center}
\caption[]{Periods and amplitudes of variable $\sigma$ Orionis members. Most error bars are smaller than 
the size of the points.}
\label{peramp}
\end{figure}

While observed period may tell us something about physical properties of the variability mechanisms in the 
very-low-mass regime, light curve amplitude can also offer valuable information. This parameter is related to 
surface spot coverage and contrast. In Fig.\ \ref{peramp}, we show amplitude as a function of period for the 
sample of variables with good $\sigma$~Ori membership information. Short-period rotators appear slightly more 
likely to have amplitudes below 0.04 magnitudes than those with periods greater than 5 days, but it is 
difficult to sort out observational biases from this effect. Although different spot configurations may 
produce the same brightness patterns, we estimate a typical spot coverage of at least $\sim$2\% based on the 
median 0.02-magnitude light curve amplitudes, assuming black spots. If, on the other hand, the temperature 
contrast between spots and the surrounding photosphere is closer to 80\% (e.g., $T_{\rm spot}/T_{\rm phot}$), 
then coverage increases to $\sim$10\%. Such contrasts and amplitudes are characteristic of either cool or hot 
spot covering fractions in young star samples \citep{2009A&A...508.1313F}. Since amplitude does not appear to 
be correlated with period or color, we suggest that the mechanism producing the spots does not vary 
appreciably with rotation and possibly mass. Furthermore, because the majority of our objects are expected to 
be fully convective, the lack of correlation between spot coverage and other parameters may be indicative of 
uniform magnetic properties across the low-mass regime.

\subsection{The relationship between variability and circumstellar disks}

Disks around young stars can be readily identified through thermal emission from circumstellar dust, manifest 
as infrared excess, or from gaseous emission lines attributed to accretion and outflow processes close to the 
star and seen spectroscopically.  In this section we investigate the correlation between optical photometric 
variability and the evidence for circumstellar dust and gas.

We cross-referenced our photometric sample with that of \citet{2008ApJ...688..362L}, which provides $Spitzer$ 
Infrared Array Camera (IRAC; 3.6-8.0 $\mu$m) photometry derived from the observations of 
\citet{2007ApJ...662.1067H}. We find that 133 of 153 confirmed or candidate $\sigma$~Ori members in our time 
series dataset have $Spitzer$ photometry, including 57 of 65 cluster periodic variables. IRAC photometry 
enables nearly unambiguous identification of unevolved disks in this cluster, as noted by 
\citet{2008ApJ...688..362L}. The $\sigma$ Orionis observations are unique among nearby young cluster 
observations with $Spitzer$ in that they were designed to search for disks around low-mass brown dwarfs and 
even planetary-mass objects; hence they are particularly deep. This gives us an unprecedented opportunity to 
study the relationship between variability, rotation, and presence of disks in the very low mass regime, 
potentially illuminating the reason why young cluster rotation period distributions have been reported to 
change around $\sim$0.25 or 0.4$M_{\odot}$ \citep{2006ApJ...646..297R,2007ApJ...671..605C}, and why the 
rotation periods in our own dataset appear to undergo a transition near $R$-$I$=1.3 ($\sim$0.45$M_{\odot}$; as 
discussed in $\S$7.3.1).

\subsubsection{Disk selection criteria}

\begin{figure}
\begin{center}
\includegraphics[scale=0.45]{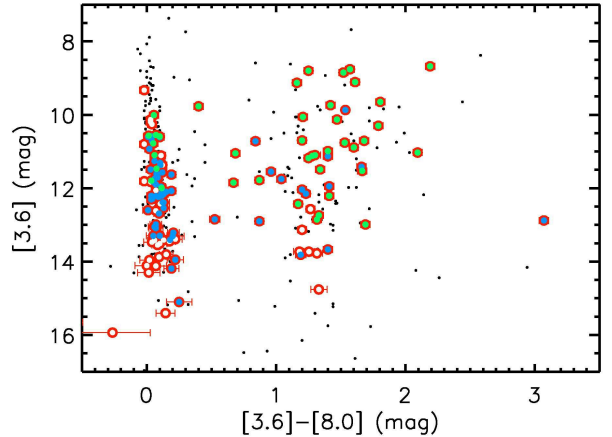}
%\plotone{f13.eps}
\end{center}
\caption[]{{\em Spitzer} photometry of likely $\sigma$ Ori members (dots) from \citet{2008ApJ...688..362L}.
Those found in our photometric sample are marked with red circles. Aperiodic variables detected in our
photometry are overplotted as filled green circles, while periodic variables in our sample are marked by
filled blue circles. The nearly vertical cluster of objects near [3.6]-[8.0]=0 is the sequence of colors 
and magnitudes pertaining to bare photospheres.}
\label{spitcolmag}
\end{figure}

We display in Fig.\ \ref{spitcolmag} the distribution of $Spitzer$/IRAC 3.6-8.0$\mu$m colors for all objects 
in our data with available infrared photometry. As seen in the figure, the sample splits relatively cleanly 
into two groups, with the narrower blue sequence near [3.6]-[8.0]=0 representing bare photosphere colors. The 
cloud of objects with [3.6]-[8.0] colors between 1 and 2 is indicative of infrared excesses signifying the 
presence of a dusty disk. While the sequence of {\em photospheric} colors is fairly well defined, several 
ambiguous objects lie between 0.3 and 0.7 magnitudes. We have therefore chosen a somewhat conservative disk 
selection criteria of [3.6]-[8.0]$>0.7$ \citep[e.g.,][]{2007ApJ...671..605C} so as to omit these objects from 
the disk sample. In total, we identify 47 likely $\sigma$~Ori members with both photometry from our campaign 
and {\em Spitzer} colors indicative of disks. The resulting disk fraction in our sample is roughly 35$\pm$5\%. 
We find that our disk identification is entirely consistent with that of \citet{2007ApJ...662.1067H} and 
\citet{2007A&A...470..903C} (based on the same $Spitzer$ data), apart from one newly-identified disk-bearing 
object, 2MASS J05375398-0249545, which has a [3.6]-[8.0] color of 1.3. The full listing of disk 
classifications is provided in Table~1.

\begin{figure}
\begin{center}
\leavevmode   
\includegraphics[scale=0.5]{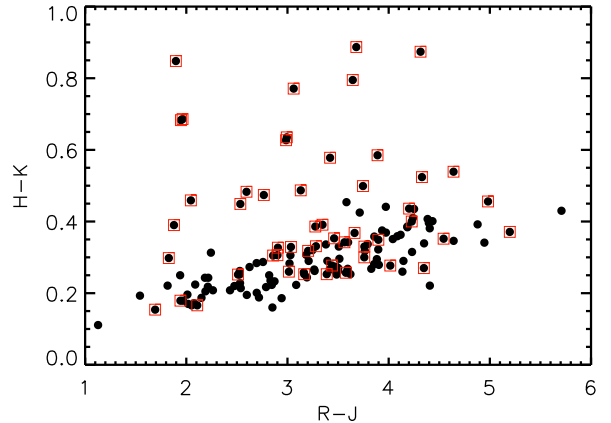}
%\plotone{f14.eps}
\end{center}
\caption[]{$R$-$J$ and $H$-$K$ colors for $\sigma$~Ori cluster members in our sample. Disk
identification at these wavelengths is possible for objects whose $H$-$K$ colors significantly exceed
the trend in photospheric colors visible along thee bottom of the diagram. Targets for which {\em
Spitzer} infrared data implies the presence of a disk are surrounded by red squares. Fewer than half of
disk-bearing members would have been selected based on the near-infrared method.}
\label{rjvshk}
\end{figure}

Previous works exploring connections between variability and the presence of disks often have relied on colors 
at shorter wavelengths to infer the presence of circumstellar dust. To test the suitability of this method, we 
produced another color-magnitude diagram using $R$-$J$ and $H$-$K$ colors, as seen in Fig.\ \ref{rjvshk}. Here 
the {\em Spitzer}-identified disk-bearing objects are highlighted by red squares. While there are a number of 
targets with sufficiently large $H$-$K$ to confirm a dust excess, many others that {\em do} have disks based 
on the $Spitzer$ data cannot be distinguished from the sequence of photospheric colors with $H$-$K$ ranging 
from 0.2 to 0.4.

\subsubsection{Variability-disk connection}

In Fig.\ \ref{spitcolmag} we have distinguished variable objects from the non-variables in the $Spitzer$/IRAC 
color-magnitude diagram. Not all of our photometric targets in $\sigma$~Ori are included in the $Spitzer$ 
sample due to varying spatial coverage. Of the 133 that are, we identified 97 as variables (e.g., Tables~3 and 
4). The majority of objects with clear periodicities have no evidence for a disk (43 of 57), while a subset of 
13 do show clear infrared excess. The disk fraction among period variables is thus $\sim$23$\pm$6\%, somewhat 
lower than the overall disk fraction. However, this measurement may be biased by the fact that we cannot 
measure periods in disk-bearing objects that are undergoing relatively high amplitude accretion events. Four 
objects fall in the ambiguous category with [3.6]-[8.0] colors between 0.3 and 0.7. One of these (2MASS 
J05390808-0228447; [3.6]-[8.0]=0.53) has a clear periodicity with period 1.7 days and amplitude 0.02 
magnitudes, similar to other variables that lack infrared excesses. The remaining three (2MASS 
J05390760-0232391, 2MASS J05390878-0231115, 2MASS J05392677-0242583) exhibit much more erratic and higher 
amplitude (RMS$\sim$0.2--0.3 magnitudes) variability.

In general, we can associate disks with the majority of aperiodic variables in our sample and lack of a disk 
with most of the periodic variables. This outcome is no surprise, since the aperiodic variability is likely 
due to accretion, which requires a disk. Likewise, since the variability in most of these disk-bearing objects 
is relatively high amplitude ($\sim$0.1 magnitudes RMS on average), we do not expect to detect many periodic 
variables among this sample, for the reasons outlined in $\S$6.2. But a number of objects do not fit these 
scenarios. Nine $\sigma$~Ori members display aperiodic variability but no sign of infrared excess in the 
$Spitzer$ data; the additional three objects highlighted above have only weak signs of an excess. On the other 
hand, 13 $\sigma$~Ori members with clear-cut infrared excesses display periodic variability with only low-level 
erratic behavior suggestive of accretion. In a few cases where signal-to-noise is particularly high (e.g., 
2MASS J05391883-0230531 and 2MASS J05381866-0251388), it is possible to see that the phased light curve is a 
combination of a nearly perfect sinusoid and a small additional ``blip'' that may be ascribed to transitory 
accretion.

Since the $Spitzer$ data enables us to conclude only that an object is surrounded by warm {\em dust}, the 
association between an infrared excess and accretion (i.e., infall of {\em gas}), is imperfect. This may 
explain why a small fraction of objects identified as having disks do not exhibit aperiodic variability, if 
the gas supply in these systems has already diminished. Likewise, we conjecture that those targets displaying 
aperiodic variability but no infrared excess probably still have a gas component of a disk, whereas the dust 
is reduced or changed to the point of being undetectable at 8.0$\mu$m and shortward. In the following 
sections, we explore in more detail the connections between each type of variability and the presence or 
absence of a disk.

\subsubsection{Relationship between disks and periodic variability due to rotation}

\begin{figure}
\begin{center}
\leavevmode   
\includegraphics[scale=0.5]{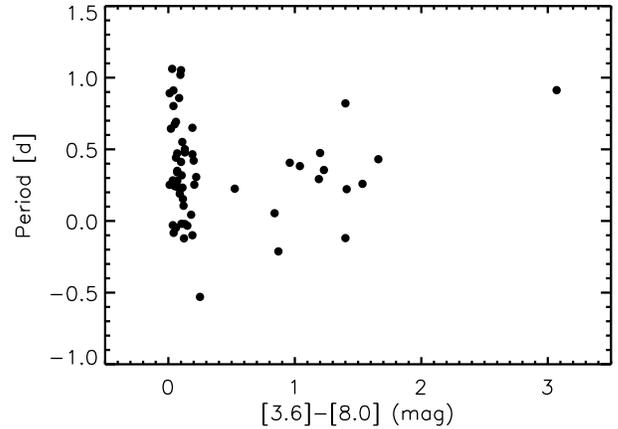}
%\plotone{f16.eps}
\end{center}
\caption[]{$Spitzer$ [3.6]-[8.0] color versus rotation period for our periodic $\sigma$~Ori members.}
\label{spitpd}
\end{figure}

The connection between stellar rotation period and disk presence has long been a subject of speculation. Disks 
have been invoked as a mechanism to remove angular momentum from young stars, in order to explain the slow 
rotation rates seen at older ages, as compared to models of spin-up associated with radial contraction 
\citep{2007IAUS..243..231B}. But while some studies have claimed a correlation between rotation rate and disk 
presence \citep[e.g.,][]{2006ApJ...646..297R,2007ApJ...671..605C}, others have refuted the so-called 
disk-locking theory \citep{1991ApJ...370L..39K,2004AJ....127.2228M}, particularly in the low-mass regime. To 
investigate the disk-rotation connection with our own data, we have examined the subset of 57 objects 
identified with both periodic variability and $Spitzer$ [3.6]-[8.0] data. Among these periodic variables, only 
13 fall in the disk sample with infrared color excesses. Unfortunately for the majority of disk-bearing 
objects, we cannot photometrically measure most of their rotation rates because of the prominent 
high-amplitude aperiodic variability. But we can nevertheless plot the periodic sample against $Spitzer$ 
[3.6]-[8.0] color to discern any large differences between the rotation rates of objects with and without 
disks, as shown in Fig.\ \ref{spitpd}. The sequence of likely diskless objects at [3.6]-[8.0]$\sim$0.0 
contains a large spread of photometric periods from 8 hours to over 10 days. The objects with disks do have a 
slightly lower mean period, but this could be a selection effect. If there is a mass dependence for rotation 
or accretion properties, then this diagram may not indicate the true distribution of rotation periods. For 
example, if low-mass stars rotate faster but accrete for longer, then we may not be detecting a number of 
short rotation periods through the larger-amplitude fluctuations due to accretion in the light curves. In 
addition, the fraction of disk-bearing objects appears to increase from $\sim$40\% of low-mass stars 
(0.1-0.5~$M_\odot$) to $\sim$60\% of brown dwarfs in $\sigma$~Orionis \citep{2008ApJ...688..362L}.

To circumvent the possible mass biases from our data, we have highlighted the disk-bearing objects among the 
rotation sample in Figs.\ \ref{colpd} and \ref{ipd}; these are indicated by open circles. The inclusion of 
color information in addition to periods and disk presence enables us to examine the effect of the mass 
distribution underlying our sample. We have seen from this diagram that the rotation periods have a marked and 
significant trend toward longer time scale at bluer color (and hence higher mass), as discussed in $\S$7.3.1. 
This correlation appears relatively independent of whether an object possesses a disk. To statistically test 
for differences between the rotation periods of objects with disks and without disks, we have plotted 
histograms of each distribution. We restrict both samples to $R$-$I>1.3$ since there are only two disk-bearing 
stars blueward of this boundary, and rotation rates of the diskless stars might be biased by mass.  Using a 
two-sided Kolmogorov-Smirnov test \citep{1992nrfa.book.....P}, we find that any differences between the 
rotation rate distributions of disk-bearing and diskless objects are not statistically significant, at the 7\% 
level (i.e., $p$=0.93). Even if we expand the analysis to include stars with $R$-$I<1.3$, there remain no 
differences, at the 35\% level $p$=0.65). With the caveat that the statistics are based on small numbers, we 
conclude that the disk-locking paradigm is largely inconsistent with our observations. The distribution of 
rotation periods instead appears to be set primarily by mass and additionally by a possible a third parameter.

\subsubsection{Relationship between disks and aperiodic variability}

In this section, we explore more directly a linkage between aperiodic variability, accretion, and disks. 
Erratic light curve variations in young stars have long been tied to spectroscopic signatures of accretion 
(Joy 1942), although they can have several origins \citep{1994ASPC...62...35H}. In particular, classical T 
Tauri stars, classified by their broad H$\alpha$ emission lines, undergo larger brightness fluctuations than 
the periodic variations more often seen in weak-lined stars \citet{1994ASPC...62...35H}. The fact that most of 
our disk-bearing objects display variability that is both higher amplitude and more erratic supports this 
picture.

We can study the relationship between accretion and disk presence more directly by examining the available 
spectroscopy for our detected aperiodic variables. We have listed in Table~4 the H$\alpha$ pseudo-equivalent 
widths (pEW) where available from previous work. The value of this parameter is typically used to distinguish 
between H$\alpha$ emission that is chromospheric in nature, as compared to emission created in an accretion 
column and hence indicative of a disk. An equivalent width greater than 5--15 $\AA$ is typically chosen to 
identify accretors. We adopt here the criteria of \citet{2003A&A...404..171B}, in which the H$\alpha$ pEW 
boundary between accretors and chromospheric emitters varies with spectral type. The value varies from 7 to 
11~$\AA$ across the M spectral type range typical of our sample. We find that 13 of our 17 aperiodic variables 
with H$\alpha$ pEW measurements from the literature have values consistent with accretion. The remaining four 
objects have fairly low RMS spread in their light curves that may indicate a different source for the 
variability.

Two of our targets with the largest H$\alpha$ pEW values are brown dwarfs, based on their faint $I$-band 
magnitudes: 2MASS J05382543-0242412 and 2MASS J05385542-0241208. The photometric data alone suggests that they 
are substellar accretors, because of the high-amplitude variability and lack of detectable periodicities. The 
former object was studied in detail by \citet[; see note in Appendix A]{2006A&A...445..143C}, but the latter 
was heretofore unknown as a variable, although it was noted as having a broad H$\alpha$ emission line with an 
equivalent width of 190$\AA$ and other T~Tauri-like spectroscopic features by \citet{2008A&A...491..515C}.

To tie together the variability features, accretion indicators, and disk presence, we have compared the values 
of light curve RMS, H$\alpha$ pEW, and $Spitzer$ [3.6]-[8.0] color for our aperiodic variables. We detect no 
correlation between RMS and H$\alpha$ pEW, suggesting that the mechanism producing variability is somehow 
decorrelated with the strength of accretion. However, it must be noted that our photometry was taken well 
after (years, in many cases) the spectroscopic data. If either light curve amplitude or H$\alpha$ emission is 
highly time-variable, non-simultaneity of the observations may explain this finding. In addition, we have 
examined the relationship between these parameters and the infrared excess. Large H$\alpha$ pEW ($>10\AA$) 
compares well with infrared excess as a predictor of disk presence in that all but one target with values 
greater than $10\AA$ also have [3.6]-[8.0]$>$1.0. But once again, we do not see any noteworthy trends in RMS 
or H$\alpha$ with [3.6]-[8.0] color among targets identified as having disks.

\begin{figure}
\begin{center}
\includegraphics[scale=0.5]{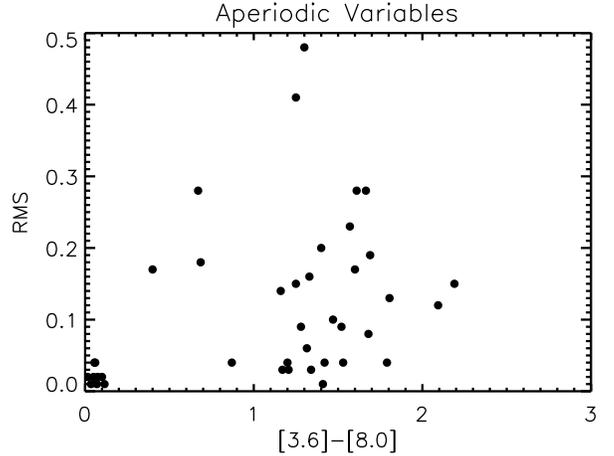}
%\plotone{f15.eps}
\end{center}
\caption[]{$Spitzer$ [3.6]-[8.0] color versus light curve RMS value for our aperiodic variables.}
\label{spitrms}
\end{figure}

There is a curious small population of objects, though, with RMS values ($\sim$0.01-0.03 magnitudes) much 
lower than the other aperiodic variables and whose H$\alpha$ pEW and [3.6]-[8.0] values suggest {\em absence} 
of accretion or an associated disk. In addition to having light curves in which variability is clearly obvious 
by eye, these objects have $\chi^2$ values high enough that their status as variables is not in doubt. All but 
one have $\chi^2>4.5$, or less than $10^{-5}$ probability that the light curve trends arose by chance; the 
remaining object (2MASS J05383922-0253084) has a $\chi^2$ value of 2.85, or an estimated 0.4\% probability 
that its light curve behavior is explained by noise. We show in Fig.\ \ref{spitrms} the RMS and infrared 
color. The subset of nine low-RMS objects is seen as a cluster in the lower left corner and is clearly 
differentiated from the larger cloud of points with colors indicative of disks. Not all of these objects have 
available H$\alpha$ pEW values, but for those that do we find they are all low, between 0 and 10$\AA$.

In summary, both H$\alpha$ emission and [3.6]-[8.0] color are good indicators of disk presence. Light Curve 
RMS is only a moderate indicator, since we encounter a number of disk-bearing objects with only low-level or 
periodic photometric variability. Of 47 targets identified with disks via $Spitzer$ data, we find 19 (40\%) 
have aperiodic variability with RMS values above 0.05 magnitudes. The distinct advantage of photometric 
monitoring thus appears to be the ability to identify aperiodic variables for which the other indicators do 
{\em not} suggest a disk or accretion. The variability in these cases is difficult to reproduce without 
invoking some sort of circumstellar material, since its erratic and short-time-scale nature suggests a dynamic 
process as opposed to thermal or magnetic phenomena associated with the stellar surface. We suggest that this 
small population of objects does in fact have residual disks undetectable at $Spitzer$/IRAC wavelengths, with 
possible accretion or dust occultation as the source of low-level variability.

\subsection{Peculiar variables} 

\begin{figure*}
\begin{center}
\includegraphics[scale=0.6]{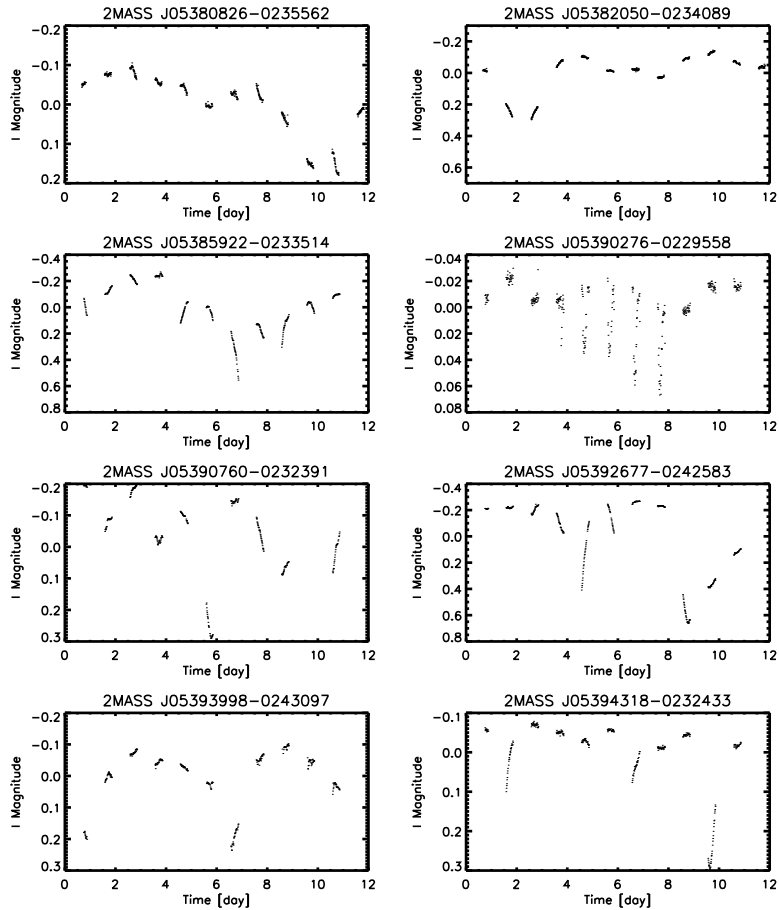}
%\plotone{f17.eps}
\end{center}
\caption[]{Aperiodic light curves with one or more unusually pronounced brightness dips.}
\label{oddvars}
\end{figure*}

While over 40\% of our detected variables are clearly periodic (Table 2; Fig.\ \ref{lightcurves}), some 27\% 
are highly stochastic (Tables 3; Fig.\ \ref{aperlightcurves}).  As discussed above, the former are associated 
with stellar rotation and the latter with processes associated with disk accretion.  A number of intriguing 
objects among the stochastic class appear to have repeating patterns that are not, however, identified as 
periodic, the most prominent eight of which are shown in Fig.\ \ref{oddvars}. They tend to display 
large-amplitude ($\sim$0.2-0.5 mag) dips of short duration (less than one day to a few days) in their light 
curves, preceded and followed by lower amplitude and longer time scale fluctuations.  In some cases the fading 
can take up to a week. A few objects (2MASS~J05382050-0234089 and 2MASS~J05390276-0229558) display brightness 
dips with symmetric ingress and egress suggestive of some sort of occulting body; other brightness dips are 
rapid enough that we have only observed a portion of the event. Among all of the aperiodic $\sigma$~Ori light 
curves we identify approximately 20\% of the sample that undergo fading events.

Stars displaying such distinct fading episodes may represent a low-mass analog of the UX Ori class (UXORs), in 
which brightness decreases of up to several magnitudes appear and persist for up to tens of days. The 
phenomenon has also been referred to as ``Type III'' pre-main-sequence variability 
\citep{1994ASPC...62...35H}. While it is typically associated with objects of spectral type K0 and earlier, it 
has been identified in the form of quasi-periodic, deep (i.e., on the order of a magnitude) brightness dips in 
a few T Tauri stars, notably AA~Tau (Bouvier et al.\ 1999). Among the several theories that have been 
suggested to explain the prominent dips seen in these variables, the most common invokes extinction events, in 
which clumpy material in a surrounding disk occults the central object from time to time. As the opacity 
increases the star becomes fainter and redder until scattering dominates and the object becomes bluer as it 
continues to fade. \citet{2000A&A...363..984B} accounted for the recurrence of brightness dips with a model in 
which the occulting region is a high latitude "warp" that periodically obscures the star above the extinction 
of a flared disk that is typical over the rest of the orbit. For the more sporadic fading, another theory is 
that the behavior may be due to variable accretion \citep{1994ASPC...62...35H}.

The diversity of light curve properties for the ``peculiar'' variables discussed here hints at multiple 
origins for the fading events, some of which may be well described by the periodic disk occultation model. 
While all of these objects have been classified as aperiodic based on the lack of one or more discrete peaks 
in the periodogram, most do display signal patterns in the frequency domain that are not consistent with 
either white or red noise. These include five or more peaks or clusters of peaks in the periodogram, 
indicating semi-periodic light curve behavior. We find that two or three objects or $\sim$25-40\% of this 
sample of eight as being quasi-periodic in their short duration fading behavior. This fraction is similar to 
the 28\% estimated by \citet{2010A&A...519A..88A} for periodic ``AA Tau like'' behavior in a comparable set of 
young stars in NGC~2264 determined from consideration of optical wavelength CoRoT data. Examining in detail 
the light curves of 2MASS~J05390276-0229558 and 2MASS~J05394318-0232433, we can estimate eclipse durations, 
depths and frequencies, assuming that the same ``blob'' of material is responsible for each fading event. For 
2MASS~J05390276-0229558, we estimate an eclipse repeat period of $\sim$1 day and duration of $\sim$0.2 day, 
while the light curve of 2MASS~J05394318-0232433 displays dips of period $\sim$4 days and duration of 
$\sim$0.85 day. The stars, which are of similar $I$-band magnitude, have masses of $\sim$0.4~$M_\odot$ and 
radii $\sim$1.2~$R_\odot$, as estimated from the 3~Myr models of \citet{1998A&A...337..403B}. If the material 
is in a circular orbit, then its distance from the star can be deduced based on these stellar parameters along 
with the ratio of the eclipse duration to the repeat period. This rough estimate reveals that the occulting 
material must be extremely close to the star-- within a stellar radius in both cases. In this scenario, the 
light curves may actually be displaying an impending accretion event, in which migrating material merges with 
the central star. If, on the other hand, the fading events are caused by distinct blobs of material, then 
their locations may be much farther out. The depths of the fading events ($\sim$4\% and $\sim$15\%, 
respectively) imply sizes for the material of 0.2--0.4 stellar radii.

The presence of disks around our peculiar variables also sheds light on the origin of brightness fluctuations. 
Based on Spitzer photometry ($\S$7.3) and the analysis of \citet{2007ApJ...662.1067H}, we find that five of 
the eight peculiar variables shown in Fig.\ \ref{oddvars} are Class~II type young stellar objects, surrounded 
by a thick disk but beyond the stage with significant high latitude (envelope) material. A further two objects 
(2MASS~J05392677-0242583 and 2MASS~J05390760-0232391) have weak $Spitzer$ infrared excesses ([3.6]-[8.0] color 
between 0.3 and 0.7). 2MASS~J05392677-0242583 is probably an ``anemic'' disk \citep{2006AJ....131.1574L}, 
while 2MASS~J05390760-0232391 was classified as a transition disk by \citet{2007ApJ...662.1067H} based on its 
large 24$\mu$m excess. The data suggest that both have optically thin inner regions. 2MASS~J05390276-0229558, 
on the other hand, does not appear to have either a disk or any signs of strong H$\alpha$ emission. The fact 
that the intriguing eclipse-like variations seen in its light curve are much lower in amplitude than the other 
peculiar variables may indicate the presence of more consolidated disk material unobservable at $Spitzer$/IRAC 
wavelengths. For the majority of objects mentioned here, we believe the variability can be plausibly 
interpreted as extinction by ``clouds'' or geometric warps of relatively higher opacity than the disk 
atmosphere which produce fading events as the feature passes through our line of sight to the star while the 
disk rotates.

\begin{figure}
\begin{center}
\includegraphics[scale=0.5]{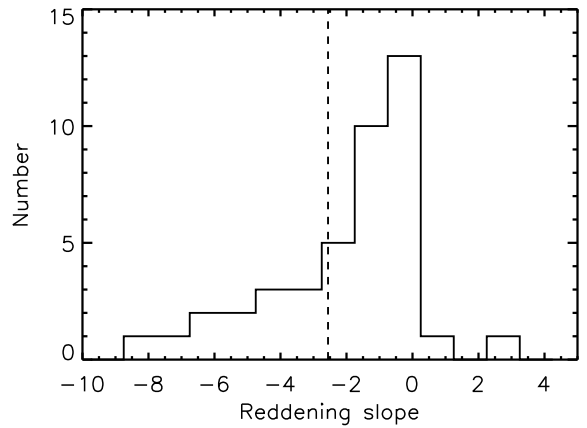}
%\plotone{f18.eps}
\end{center}
\caption[]{Histogram of reddening values derived from the slopes of the $I$ versus $R$-$I$ trends displayed in
Fig.\ \ref{aperlightcurves}. Negative values correspond to increased reddening with decreasing brightness. The
dashed line marks the value for interstellar extinction, -E($R$-$I$)/$A_I$.}
\label{extincts}
\end{figure}

Color data can help further illuminate the source of peculiar variability, since we have not ruled out 
accretion effects. Different trends in color are expected depending on whether the variations are caused by 
extinction, disk scattering, or stellar spots, as explained by \citet{2001AJ....121.3160C} and 
\citet{2009MNRAS.398..873S}. Since we have acquired $R$-band data twice per night for all targets, we can 
examine $R$-$I$ as a function of brightness for all aperiodic variables, and check whether any particular 
pattern stands out for the eight selected peculiar variables. We present in the right panel of Fig.\ 
\ref{aperlightcurves} the available colors and magnitudes. Notably, with only lower cadence data (as 
represented in the middle panel) the richness and coherence of the light curve forms would be hidden. In many 
cases the fading events observed among our aperiodic variables are relatively colorless although both 
significant reddening and significant blueing is observed among the sample. We have measured the slope of 
reddening for all aperiodic variables in Fig.\ \ref{aperlightcurves} by fitting a linear trend to the $I$-band 
magnitude as a function of $R$-$I$. We then negate the result so that slopes less than zero represent 
reddening as an object becomes fainter. The distribution of values is presented in Fig.\ \ref{extincts}. 
Although the color light curves do not have enough points to enable a detailed fit to the various variability 
models, we note that the vast majority of aperiodic variables show either negative or zero slope. For 
comparison, we have also plotted the value expected for pure interstellar extinction. Since the material in 
disks may be substantially different, we do not necessarily expect it to follow the same extinction law. 
Indeed, several of the peculiar variables display much more reddening during their fading episodes. The 
modeling by \citet{2001AJ....121.3160C} and \citet{2009MNRAS.398..873S} showed that hot spots from accretion 
can in fact exhibit steeper reddening slopes than extinction, at least in the near-infrared. This is certainly 
a possible explanation for some of our own sources. Only two objects in our sample, however, exhibit 
variability that may be accounted for by emission or scattering by the circumstellar disk, which is predicted 
to produce relatively {\em blue} fading events \citep{2001AJ....121.3160C}. Intriguingly, 
2MASS~J05390276-0229558, the only peculiar variable with no infrared excess, is one example. The single data 
point caught while this object was at its faint limit shows substantially bluer color than the rest of the 
light curve. We envision a scenario in which material temporarily occulting the star also scatters light 
toward us.

Although we cannot rule out the presence of accretion effects, we conclude that the hypothesis of occultation 
by disk material is qualitatively consistent with both the duration and the color-magnitude behavior of the 
brief fading events seen in the set of eight peculiar variables presented here. Further, as some of the events 
are periodic or semi-periodic, we note that the derived periods are consistent with those expected from an 
inner disk region in co-rotation with a star having typical spin for a Class II T Tauri star (2-10 days).  
Similar features located further out in the disk could be responsible for the non-repeating and/or broader 
fading events.

\section{Discussion}

We have presented photometric monitoring on a collection of low-mass stars and brown dwarfs in 
$\sigma$~Orionis. Extensive vetting of membership via prior spectroscopic information and the relative spatial 
compactness of our fields ($\sim$7~pc across) has ensured that the sample is relatively homogeneous in terms 
of age and initial conditions. In addition, the selection of $\sim$10-minute cadence and time baseline of 
nearly two weeks, along with excellent photometric precision has enabled us to carry out an unprecedented 
analysis of variability in young stars and brown dwarfs, complete to amplitudes below the 1\% level for most 
sources. This combination of cadence and precision has allowed us to probe new areas of variability parameter 
space: those pertaining to short time scale and low-amplitude fluctuations. In the preceding analysis, we have 
explored the general properties of variability in very-low-mass $\sigma$ Orionis members and its connections 
to other stellar parameters. In putting the pieces together, we will now highlight the various phenomena 
encountered and possible connections to physical properties.

\subsection{Variability in young stars and brown dwarfs is persistent - in time and mass}

The sensitivity of our photometric monitoring has given us an unprecedented opportunity to probe for 
variability and explore its trends well into the brown dwarf regime. As discussed in $\S$7.1, we have detected 
variability of various forms in nearly 70\% of our sample, including 85\% of stars with strong evidence for 
cluster membership. The $\sim$15\% of likely cluster members with {\em no} evidence for variability do not 
appear to have any distinguishing characteristics such as belonging to a particular mass range or possession 
of disks. This fraction is similar to the proportion of variables identified as periodic in 2007 but not in 
2008. Using the 2007 field as well as data from other studies, we have also found ($\S$7.1.1) that the 
observed periodic and aperiodic variability is persistent on typical time scales of at least 5--10-years. This 
finding is consistent with studies of other clusters such as IC~348, in which analysis of data acquired by 
different groups retrieve largely the same photometric periods for objects in common 
\citep[e.g.,][]{2006ApJ...649..862C}. \citet{2004A&A...419..249S} also carried out two photometry monitoring 
campaigns in another region of $\sigma$~Ori and identified a number of objects with persistent variablity 
across both datasets. Nevertheless they also suggest evidence for spot evolution based on a subset of targets 
displaying periodicities during only one campaign. While our analysis in $\S$7.1.1 points to long-lived 
accretion and magnetic activity on young, low-mass stars (in comparison to, e.g., the rotation period time 
scale), it is not sensitive to light curve amplitude or phase changes. Thus magnetic spots may come and go, 
but the typical young low-mass star or brown dwarf has one or more spots large enough to be detected in 
photometry at the 0.5\% level for time spans of multiple years.

Also intriguing are variability trends (or lack thereof) with mass, particularly across the substellar 
boundary. \citet{2009ApJ...697..373R} have observed that magnetic field strengths on young brown dwarfs are 
substantially weaker than those in higher mass young stars. As a result, we might expect accretion and spot 
properties to change with mass. We have concluded ($\S$7.1.2) that there is no such evidence for a trend in 
{\em aperiodic} variability. Like several other studies of $\sigma$~Ori 
\citep{2006A&A...445..143C,2004A&A...419..249S} we identify several accreting brown dwarfs based on their 
high-level erratic light curve behavior.  The persistence of T-Tauri-like variability to very low masses may 
reflect more so the presence of disks than the surface magnetic field properties of these objects.  The 
fraction of {\em periodic} variables, on the other hand, does seem to decrease into the brown dwarf regime 
($\S$7.1.2) to an extent not accounted for by our photometric sensitivity. This result is consistent with 
decreasing magnetic field strength in that a lack of spots or decreased coverage would be expected. 
Alternatively, spots may still be present but at much lower temperature contrast.

\subsection{Period correlates with color and magnitude in low-mass $\sigma$~Ori members}

Several previous studies have examined the distribution of rotation periods among stars in a number of young 
clusters. Initially, many of the stellar samples did not include stars with masses less than 
$\sim$0.2~$M_\odot$, and the resulting rotation period exhibited two peaks near 2 and 8 days 
\citep[e.g.,][]{2002A&A...396..513H}. However, extension of rotation studies to lower mass has failed to 
retrieve such a bimodal distribution. \citet{2005A&A...430.1005L} and later \citet{2007ApJ...671..605C} indeed 
observed a change in rotation properties near $R$-$I$=1.3 or spectral type M2--M3, with the redder objects 
rotating faster on average. The disappearance of the long period peak in the rotation distribution when a 
low-mass (or equivalently, red color) cut is applied to the distribution implies that a mass-dependent effect 
is at work. Additional rotational studies incorporating components of the low-mass star population in the IC 
348 cluster \citep{2006ApJ...649..862C}, the northern portion of $\sigma$ Orionis \citep{2004A&A...419..249S}, 
and the ONC \citep{1999AJ....117.2941S,2009A&A...502..883R} have confirmed a trend of increasingly rapid 
rotation toward lower mass.

Although our data includes few periodic objects more massive than $\sim$0.5~$M_{\odot}$, they support the 
conclusion that low mass stars and brown dwarfs have a different period distribution from higher mass but 
similarly aged young stars. The distribution of rotation periods uncovered in our analysis contains few 
objects with 8-10 day periods, but a steady increase in number of objects up to a peak near 1 day. We have 
further explored this phenomenon by plotting periods as a function of photospheric color as well as $I$-band 
magnitude, both of which serve as proxies for mass. The results (Figs.\ \ref{colpd} and \ref{ipd}) and 
statistical tests confirm that there does indeed appear to be a strong trend in rotation with mass.  We have 
ruled out (e.g., $\S$7.3.1) the possibility that biases in our photometric sensitivity and signal detection 
algorithm could produce such a strong correlation of period with color or magnitude.

We observe a transition in rotation periods near $R$-$I$=1.3 (spectral type M2.5), similar to that reported by 
\citet{2005A&A...430.1005L} and \citet{2007ApJ...671..605C}, which they attributed to a possible shift in 
magnetic field properties at low mass. However, we are at a loss to explain such a transition, since low-mass 
stars and brown dwarfs at the age of $\sigma$~Ori should all be fully convective. We have attempted to explain 
the trend of rotation with color (and hence mass) with a much simpler hypothesis of constant angular momentum. 
We consider this to be a ``toy'' model since in reality angular momentum likely adheres to a distribution 
rather than a single value \citep[e.g.,][]{2001AJ....121.1676R}. The internal structure models from 
\citet{1998A&A...337..403B} and \citet{1997MmSAI..68..807D} do provide a reasonable fit to the data, with the 
exception of one prominent outlier at $R$-$I$=0.6. Thus we conclude that it is possible to account for the 
spins of $\sigma$~Ori members with models for mass and radius currently in use, which invoke formation of 
H$_2$ in the atmosphere and increasing importance of electron degeneracy at low mass but do not incorporate 
magnetic fields \citep{1998A&A...337..403B}. Nevertheless, larger numbers of rotation data points and 
additional data incorporating higher mass cluster members is likely required to reach a definitive conclusion 
on the origin of the rotation trend uncovered in our study.

\subsection{A lower limit for the time scale of periodic variability in low-mass $\sigma$~Ori members}

In general, we find no periodic variability at periods less than 7 hours. The cut-off in rotation periods 
around 7-10 days is abrupt and significant, considering that we are fully able to detect periods down to 
$\sim$15 minutes. Within the uncertainties of cluster membership verification, there are approximately 40 
young objects in our sample with masses less than $\sim$0.1$M_{\odot}$. We thus conclude that our data to not 
bear out the predictions for pulsational instabilities \citep{2005A&A...432L..57P}, which call for pulsation 
periods of $\sim$1-4 hours in unstable deuterium-burning objects. If any of our BDs or VLMSs is pulsating, 
then they must be doing so at amplitudes below $\sim$0.01-0.02 magnitudes. Our observations are inconsistent 
with reports of short-period variability in young $\sigma$~Ori brown dwarfs observed by 
\citet{2001A&A...367..218B} and \citet{2003A&A...408..663Z}; details on the failure to redetect periodicities 
in these objects are provided in Appendix~A.

Within the range of $\sim$1-7 hours, we not only do not detect signs of pulsation, but we also see no evidence 
of spot-modulated variability. This result suggests some sort of physical mechanism which limits rotation 
rates. In $\S$7.3.1 we estimated that the break-up period for objects from 0.02 to 0.1 $M_{\odot}$ lies near 
2--7 hours, although there are substantial uncertainties in radius, and hence velocity, at these ages. Based 
on these values, it appears that young BDs rotate at up to, but not beyond, $\sim$40\% of their break-up 
velocity. This result stands in contrast to the observations of \citet{1999AJ....117.2941S} in the younger 
Orion Nebula Cluster, for which a number of low-mass objects were found to rotate at 60--100\% percent of 
break-up speed. And while our data do echo previous suggestions 
\citep[e.g.,][]{2006A&A...445..143C,2009A&A...502..883R} that low-mass objects rotate significantly faster 
than their higher mass counterparts, our fit of constant angular momentum models to the data in Fig.\ 
\ref{colpd} illustrate that magnetic effects need not be invoked to explain the trend.

\subsection{No connection between rotation period and the presence of disk}

Perhaps the most surprising finding to arise from our data is the apparent lack of correlation between the 
derived rotation periods and presence of a circumstellar disk around low-mass stars and brown dwarfs (e.g., 
$M\lesssim 0.5 M_\odot$). $\sigma$ Ori is one of the few clusters for which $Spitzer$/IRAC data is available 
and deep enough to identify disks around even the lowest mass members. Likewise, our photometric monitoring is 
sensitive enough to permit the derivation of rotation periods in all non-accreting objects with spots 
producing brightness deviations greater than 0.007 magnitudes (e.g., $\S$7.3.1). Much attention has been paid 
in previous works to the role of disks in regulating the angular momentum evolution of young stars, and in 
particular the role of disk locking \citep{1991ApJ...370L..39K} in limiting rotation rates. Many measurements 
of rotation periods for stars with and without disks have produced discrepant results in that some studies 
show slower rotation on average for disk-bearing stars and others do not; \citet{2006ApJ...649..862C} provide 
an excellent overview. One issue has been the actual selection of disk candidates. The process has recently 
become much more clear-cut with the advent of $Spitzer$ data, but previous reliance on mainly near-infrared 
data may have muddled the samples, as illustrated in Fig.\ \ref{rjvshk}.

Fortunately in our case we have access to excellent $Spitzer$ data for many of our targets, presenting the 
opportunity to examine for the first time correlations between rotation period and disk presence among 
low-mass members. At the same time, our conclusions are limited by the fact that we can measure rotation 
periods for only 13 (28\%) of the disk-bearing objects. But the spread in rotation periods among these objects 
(as shown in Fig.\ 14) is nevertheless quite wide, encompassing roughly the same range as the diskless 
objects. \citet{2006ApJ...646..297R}'s study of ONC members with $Spitzer$ data revealed significantly slower 
rotation among their disk sample even to low masses, although this result may have been biased by the 
detection limits of their $Spitzer$ data. In contrast, the median rotation periods for both disk-bearing and 
diskless periodic variables in our sample do not differ significantly for either the entire sample or the 
large subsample of objects with $M\lesssim 0.45M_\odot$ (\S 7.4.3), leading us to conclude that any 
disk-locking phenomenon is not prominent in the low-mass regime at the age of $\sigma$~Ori. Since we are 
concerned about mass-dependent effects, we have also highlighted the disk-bearing objects in the period-color 
diagram (Fig.\ \ref{colpd}). Once again, it is clear that these targets do not occupy a region of 
preferentially long or short period, regardless of mass. Instead, we find a substantial spread in rotation 
periods for the disk-bearing sample, independent of both disk presence and other properties. These results 
suggest that the disk may not in fact play the lead role in determining the angular momentum of rates of 
young, very-low-mass stars. They are also consistent with a recent theoretical study by 
\citet{2010ApJ...714..989M} which concluded that other processes like stellar winds must be invoked to explain 
the observed spread in rotation rates.

\subsection{New classes of low-mass star variability}

The sensitivity and cadence of our photometric observations have led to the discovery of several novel types 
of variability among the low-mass young cluster members. We discussed the details ($\S$7.5) of a small set of 
``peculiar'' variables whose abrupt dips in brightness mirror those of the higher mass UX Ori stars, but on 
much shorter time scales. With the recent identification of ``AA-Tau-like'' variables in NGC~2264 
\citep{2010A&A...519A..88A}, this is not an entirely new finding, but it does suggest that the eclipse-like 
brightness dip phenomenon is somewhat common in young clusters. Such variables may have been overlooked in 
previous photometric studies since the fading events only become obvious when data are taken at the 
appropriate fast cadence. Additional multi-color studies of the phenomenon should allow for further evaluation 
of its origin.

We also highlight the subsample of aperiodic variables in our sample whose light curve RMS values are 
particularly low and whose $Spitzer$ infrared data shows no indication of a disk (Fig.\ \ref{spitrms}). 
Although the objects also do not have strong H$\alpha$ emission, the erratic nature of the light curves is 
strongly suggestive of accretion, but perhaps at a lower level than the variables with obvious disks. A 
similar phenomenon was observed in the IC~348 cluster, in which a number of weak T~Tauri stars (i.e., weak 
H$\alpha$) were found to be erratic variables by \citet{2005MNRAS.358..341L}. These results bring into 
question our ability to determine which cluster members are truly surrounded by disk material, which 
ultimately affects the analysis of rotation and possible disk locking. It appears from these light curves that 
a percentage of young objects retain enough gas for accretion beyond the time that we would expect their disks 
to be fully cleared based on infrared observations.

\section{Summary}

We have presented the results of high-precision photometric time series monitoring on two fields in the 
$\sigma$~Orionis cluster, including 153 confirmed and candidate members. Nearly 70\% of the sample displays 
variability, enabling not only the identification of several new candidate cluster members (Tables~3 and 4), 
but also a detailed analysis of the types of variability present and its origins. We have found that the 
majority of periodic variability can be explained by rotational modulation of surface features, with time 
scales too long to be consistent with the pulsation theory of \citep{2005A&A...432L..57P}. The large set of 
rotation rates (for 65 objects in total) spans masses from the brown dwarf regime ($\sim$0.04 $M_\odot$) to 
low-mass stars with $M\lesssim$0.5~$M_\odot$. The inclusion of $R$-$I$ color data led us to identify trends in 
variability as a function of mass. We have measured a robust decline in the fraction of periodic variables 
toward the brown dwarf regime, which may be related to a mass dependence of the surface magnetic field 
structure or strength. We have also presented a clear trend in rotation rates, with BDs rotating significantly 
faster than the low-mass stars; we tentatively connected this finding to the initial angular momentum 
properties of these young stellar objects.

In addition, infrared data from $Spitzer$/IRAC has enabled a search for disks around over 90\% of our targets, 
and the resulting disk fraction is $\sim$35\% for the brown dwarfs and low-mass stars of $\sigma$~Ori. Notably, 
we find no significant connection between the presence of a disk and the rotation periods of cluster members. 
While most of the aperiodic variables in our sample have disks, as would be expected from accretion-induced 
variability, a significant subsample ($\sim$30\%) of those with small $I$-band light curve RMS (e.g., 
$\lesssim$0.04 magnitudes) and masses from 0.3-0.7~$M_\odot$ do not have any evidence for disks or accretion in 
the available infrared and spectroscopic data. To our knowledge, this type of variability has not been reported 
previously, and represents a new class of low-amplitude aperiodic variables which may still be accreting at a low 
level despite dispersal of most of their disk. Finally, the high cadence of our data resulted in the 
identification of an additional intriguing type of variability, involving abrupt dips in brightness, some of 
which appear eclipse-like in nature. We have attributed this phenomenon to occultation by material in the disk. 
Overall, we expect that this dataset will offer a comprehensive library of variability typical in clusters in the 
few-Myr range.

\acknowledgements{This research has been supported by grants to LAH from the NASA Origins and ADP programs. 
AMC would like to thank the CTIO Telescope Operations staff for help in carrying out observations.  
Observation time on SMARTS consortium facilities was awarded through the National Optical Astronomy 
Observatory, operated by the Association of Universities for Research in Astronomy, under contract with the 
National Science Foundation. We are grateful for helpful comments from the referee, Adam Kraus, Krzysztof 
Findeisen, and Nairn Baliber.}

\appendix

\section{Objects with previous reports of variability}

$\sigma$ Orionis is a well-studied cluster, and several previous variability studies have targeted its
brown dwarf and low-mass star population. Despite different cadences and sensitivities, we can use prior
data to assess variability patterns over time scales much longer than the duration of our observing runs.
Repeat detection of a periodicity not only confirms the accuracy of the measurement but also attests to
the long-term stability of the mechanism behind it. However, non-detection of variability can also offer
insights into the physical processes affecting young VLMSs and BDs on relatively short astronomical
time scales. We detail results here on a number of targets in our sample that were put forth as
variables by other authors.

{\bf r053820/SWW124/Mayrit 380287 = 2MASS J05382050-0234089} \citet{2009A&A...505.1115L} report 
variability in this object in the $J$, $H$, and $K$ bands.  The difference in magnitudes over several years 
is 1.0, 0.67, and 0.28 magnitudes, respectively. \citet{2007ApJ...662.1067H} also identified it as a variable 
(see below). In this study, we find significant undulations in the $I$-band light curve (RMS$\sim$0.1 
magnitudes), including an $\sim$0.4-magnitude eclipse-like drop over several days (see $\S$7.5). 

{\bf SWW221/Mayrit 1129222 = 2MASS J05375398-0249545} \citet{2009A&A...505.1115L} detect variability of 
this object at $J$, $H$, and $K$ bands. The brightness in each band differ by 0.4-0.5 magnitudes over a 
baseline of several years. During our shorter campaign we find that the object has an rms variation of 1.95 
magnitudes in the $I$ band-- the largest change among all of our variables.

{\bf Mayrit 458140 = 2MASS J05390458-0241493} \citet{2009A&A...505.1115L} inferred variability in 
this source in the $J$, $H$, and $K$ bands. The change in brightness on time scales of several years is 
$\sim$0.2 magnitudes. We also find up to one magnitude in erratic variations on the two-week time scale in 
the $I$-band, suggesting ongoing accretion.

{\bf S Ori J053855.4-0241208= 2MASS J05385542-0241208} \citet{2009A&A...505.1115L} report changes of 0.29 
and 0.23 magnitudes in the $J$ and $H$ bands, respectively, over several years. We also detect 
variability of aperiodic nature, at an RMS of 0.19 magnitudes in the $I$ band. 
  
{\bf S Ori 2 = 2MASS J05392633-0228376} \citet{2004A&A...419..249S} report this object as variable, 
with an RMS of 0.038 magnitudes. Likewise, we detect it as periodic with amplitude 0.019 magnitudes and period 
2.3 days. After subtracting this signal from the data, we also note slightly non-gaussian residuals possibly 
indicative of additional low-level variability.

{\bf SE77 = 2MASS J05385492-0228583} \citet{2004A&A...419..249S} report this object as variable,
with an RMS of 0.028 magnitudes. We do not detect any variability, down to less than 0.001 magnitudes.

{\bf S Ori J053826.1-024041 = 2MASS J05382623-0240413} \citet{2004A&A...424..857C} detected 
variability on minute to hour time scales with amplitude less than 0.04 magnitudes. We see hints of a potential 
periodicity at amplitude 0.006 magnitudes and period 4.8 days, but it is too weak to confirm (S/N$\sim$4 in the 
periodogram). The RMS spread in our light curve is 0.01 magnitudes.

{\bf S Ori 25 = 2MASS J05390894-0239579} \citet{2004A&A...424..857C} detected periodic variability with a 
period of 40$\pm$8 hours (1.7$\pm$0.3 days) and amplitude 0.15$\pm$0.02 magnitudes. We also find variability, 
but with a period of $\sim$2.6 days, and amplitude $\sim$0.025. The periods could be consistent with each other if 
one of the detections selected an alias of the true value. However, the 0.046-magnitudes RMS of our 
light curve implies strong disagreement between the amplitudes. 

{\bf S Ori 42 = 2MASS J05392341-0240575} \citet{2004A&A...424..857C} detected a brightness change of 
0.11$\pm$0.03 from one set of photometry to the next, on a time scale of $\sim$2 years. We cannot probe 
variability on such long time scales but find an RMS spread of 0.056, in line with uncertainties expected for 
field objects of similar magnitude. We also fail to detect any periodicities down to the 0.02-magnitude level.

{\bf S Ori J054004.5-023642 = 2MASS J05400453-0236421} \citet{2004A&A...424..857C} found variability on 
night-to-night time scales and amplitude 0.073 magnitudes. Likewise, we detect this objects as a variable with a
period $\sim$18 hours and amplitude 0.03 magnitudes.

{\bf S Ori J053948.1-022914 = 2MASS J05394826-0229144} \citet{2004A&A...419..249S} noted this object 
(their \#108) as a variable (although not periodic) with an $I$-band RMS spread of 0.139, as compared to a 
median noise level of $\sim$0.08 magnitudes.  We do not detect any such variability, down to our noise floor of 
$\sim$0.04 magnitudes.

{\bf S Ori J053825.4-024241 = 2MASS J05382543-0242412} 
This brown dwarf and was highlighted by \citet{2006A&A...445..143C} as a substellar accretor, as indicated by 
strong H$\alpha$ and other spectroscopic emission line features. They observed its $I$-band light curve to 
undergo day-to-day variability of $\sim$0.25 magnitudes, with smaller variations on shorter time scales. 
We redetect high-amplitude non-periodic variability with $I$-band RMS 0.55 magnitudes and peak-to-peak 
amplitude 0.16 magnitudes, confirming that this object likely continues to accrete. 

{\bf S Ori 27 = 2MASS J05381741-0240242}
Variability was previously reported by \citet{2004A&A...424..857C}, with a period of 2.8$\pm$0.4 hours. However, 
the source appears to be constant to within the photometric errors of our data; we find no evidence of periodic 
signals with amplitudes greater than several millimagnitudes.

{\bf S Ori 28 = 2MASS J05392319-0246557} Variablity was previously detected \citet{2004A&A...424..857C}, with 
a period of 3.3$\pm$0.6 hours but is not re-detected in this data. For this source, we are sensitive to periodic 
signals down to 0.004 magnitudes at periods less than 8 hours and $\sim$0.01 magnitudes for longer time scales.

{\bf S Ori 31 = 2MASS J05382088-0246132}
Variability was previously detected by \citet{2001A&A...367..218B}, with potential periods of 1.75$\pm$0.13 and 
7.5$\pm$0.6 hours. We do not detect variability on any time scale, but are sensitive down to an amplitude level 
of $\sim$0.004 magnitudes.

{\bf S Ori 45 = 2MASS J05382557-0248370}
Variability was previously detected by \citet{2003A&A...408..663Z}, with 
possible periods of 46.4$\pm$1.5 minutes, 2.56$\pm$0.10 hours, and 3.6$\pm$1.2 hours. \citet{2001A&A...367..218B} 
also reported a tentative detection of periodicity at 0.50$\pm$0.13 hours. We detect variability at a 
longer period of $\sim$7 hours and amplitude 0.03 magnitudes.

\citet{2007ApJ...662.1067H} have extracted a number of objects from the CIDA 
Equatorial Variability Survey \citep{2004AJ....127.1158V}. Twenty-five of these are in our fields, and we redetect 
variability in all but one of them (2MASS J05385317-0243528). These objects, all but five of which display aperiodic variability, have the following identification numbers from \citet{2007ApJ...662.1067H} and 2MASS:
SO848 (2MASS J05390193-0235029), SO1154 (2MASS J05393982-0233159), 
SO1235 (2MASS J05395038-0243307), SO1260 (2MASS J05395362-0233426), SO1361 (2MASS J05400889-0233336), 
SO362 (2MASS J05380826-0235562), SO300 (2MASS J05380107-0245379), SO123 (2MASS J05373784-0245442), 
SO374 (2MASS J05380994-0251377), SO396 (2MASS J05381315-0245509), SO435 (2MASS J05381778-0240500), 
SO462 (2MASS J05382050-0234089), SO482 (2MASS J05382307-0236493),
SO598 (2MASS J05383460-0241087), SO646 (2MASS J05383902-0245321),
SO827 (2MASS J05385922-0233514), SO865 (2MASS J05390357-0246269), 
SO879 (2MASS J05390540-0232303), SO976 (2MASS J05391699-0241171), 
SO1017 (2MASS J05392286-0233330), SO1036 (2MASS J05392519-0238220), SO1057 (2MASS J05392677-0242583),
SO1153 (2MASS J05393982-0231217), SO1182 (2MASS J05394318-0232433).

\bibliographystyle{apj}
\bibliography{sigori}

\begin{thebibliography}{92}
\expandafter\ifx\csname natexlab\endcsname\relax\def\natexlab#1{#1}\fi

\bibitem[{{Alard} \& {Lupton}(1998)}]{1998ApJ...503..325A}
{Alard}, C., \& {Lupton}, R.~H. 1998, \apj, 503, 325

\bibitem[{{Alencar} {et~al.}(2010){Alencar}, {Teixeira}, {Guimar{\~a}es},
  {McGinnis}, {Gameiro}, {Bouvier}, {Aigrain}, {Flaccomio}, \&
  {Favata}}]{2010A&A...519A..88A}
{Alencar}, S.~H.~P., {Teixeira}, P.~S., {Guimar{\~a}es}, M.~M., {McGinnis},
  P.~T., {Gameiro}, J.~F., {Bouvier}, J., {Aigrain}, S., {Flaccomio}, E., \&
  {Favata}, F. 2010, \aap, 519, A88+

\bibitem[{{Andrews} {et~al.}(2004){Andrews}, {Reipurth}, {Bally}, \&
  {Heathcote}}]{2004ApJ...606..353A}
{Andrews}, S.~M., {Reipurth}, B., {Bally}, J., \& {Heathcote}, S.~R. 2004,
  \apj, 606, 353

\bibitem[{{Bailer-Jones} \& {Mundt}(2001)}]{2001A&A...367..218B}
{Bailer-Jones}, C.~A.~L., \& {Mundt}, R. 2001, \aap, 367, 218

\bibitem[{{Baraffe} {et~al.}(1998){Baraffe}, {Chabrier}, {Allard}, \&
  {Hauschildt}}]{1998A&A...337..403B}
{Baraffe}, I., {Chabrier}, G., {Allard}, F., \& {Hauschildt}, P.~H. 1998, \aap,
  337, 403

\bibitem[{{Baraffe} {et~al.}(2002){Baraffe}, {Chabrier}, {Allard}, \&
  {Hauschildt}}]{2002A&A...382..563B}
---. 2002, \aap, 382, 563

\bibitem[{{Barrado y Navascu{\'e}s} {et~al.}(2003){Barrado y Navascu{\'e}s},
  {B{\'e}jar}, {Mundt}, {Mart{\'{\i}}n}, {Rebolo}, {Zapatero Osorio}, \&
  {Bailer-Jones}}]{2003A&A...404..171B}
{Barrado y Navascu{\'e}s}, D., {B{\'e}jar}, V.~J.~S., {Mundt}, R.,
  {Mart{\'{\i}}n}, E.~L., {Rebolo}, R., {Zapatero Osorio}, M.~R., \&
  {Bailer-Jones}, C.~A.~L. 2003, \aap, 404, 171

\bibitem[{{Barrado y Navascu{\'e}s} {et~al.}(2001){Barrado y Navascu{\'e}s},
  {Zapatero Osorio}, {B{\'e}jar}, {Rebolo}, {Mart{\'{\i}}n}, {Mundt}, \&
  {Bailer-Jones}}]{2001A&A...377L...9B}
{Barrado y Navascu{\'e}s}, D., {Zapatero Osorio}, M.~R., {B{\'e}jar}, V.~J.~S.,
  {Rebolo}, R., {Mart{\'{\i}}n}, E.~L., {Mundt}, R., \& {Bailer-Jones},
  C.~A.~L. 2001, \aap, 377, L9

\bibitem[{{Becker} {et~al.}(2004){Becker}, {Wittman}, {Boeshaar},
  {Clocchiatti}, {Dell'Antonio}, {Frail}, {Halpern}, {Margoniner}, {Norman},
  {Tyson}, \& {Schommer}}]{2004ApJ...611..418B}
{Becker}, A.~C., {Wittman}, D.~M., {Boeshaar}, P.~C., {Clocchiatti}, A.,
  {Dell'Antonio}, I.~P., {Frail}, D.~A., {Halpern}, J., {Margoniner}, V.~E.,
  {Norman}, D., {Tyson}, J.~A., \& {Schommer}, R.~A. 2004, \apj, 611, 418

\bibitem[{{B{\'e}jar} {et~al.}(2001){B{\'e}jar}, {Mart{\'{\i}}n}, {Zapatero
  Osorio}, {Rebolo}, {Barrado y Navascu{\'e}s}, {Bailer-Jones}, {Mundt},
  {Baraffe}, {Chabrier}, \& {Allard}}]{2001ApJ...556..830B}
{B{\'e}jar}, V.~J.~S., {Mart{\'{\i}}n}, E.~L., {Zapatero Osorio}, M.~R.,
  {Rebolo}, R., {Barrado y Navascu{\'e}s}, D., {Bailer-Jones}, C.~A.~L.,
  {Mundt}, R., {Baraffe}, I., {Chabrier}, C., \& {Allard}, F. 2001, \apj, 556,
  830

\bibitem[{{B{\'e}jar} {et~al.}(1999){B{\'e}jar}, {Zapatero Osorio}, \&
  {Rebolo}}]{1999ApJ...521..671B}
{B{\'e}jar}, V.~J.~S., {Zapatero Osorio}, M.~R., \& {Rebolo}, R. 1999, \apj,
  521, 671

\bibitem[{{B{\'e}jar} {et~al.}(2004){B{\'e}jar}, {Zapatero Osorio}, \&
  {Rebolo}}]{2004AN....325..705B}
---. 2004, Astronomische Nachrichten, 325, 705

\bibitem[{{Bertout}(2000)}]{2000A&A...363..984B}
{Bertout}, C. 2000, \aap, 363, 984

\bibitem[{{Bouvier}(2007)}]{2007IAUS..243..231B}
{Bouvier}, J. 2007, in IAU Symposium, ed. J.~{Bouvier} \& I.~{Appenzeller},
  Vol. 243, 231--240

\bibitem[{{Bouvier} \& {Bertout}(1989)}]{1989A&A...211...99B}
{Bouvier}, J., \& {Bertout}, C. 1989, \aap, 211, 99

\bibitem[{{Breger} {et~al.}(1999){Breger}, {Handler}, {Garrido}, {Audard},
  {Zima}, {Papar{\'o}}, {Beichbuchner}, {Zhi-Ping}, {Shi-Yang}, {Zong-Li},
  {Ai-Ying}, {Pikall}, {Stankov}, {Guzik}, {Sperl}, {Krzesinski}, {Ogloza},
  {Pajdosz}, {Zola}, {Thomassen}, {Solheim}, {Serkowitsch}, {Reegen}, {Rumpf},
  {Schmalwieser}, \& {Montgomery}}]{1999A&A...349..225B}
{Breger}, M., {Handler}, G., {Garrido}, R., {Audard}, N., {Zima}, W.,
  {Papar{\'o}}, M., {Beichbuchner}, F., {Zhi-Ping}, L., {Shi-Yang}, J.,
  {Zong-Li}, L., {Ai-Ying}, Z., {Pikall}, H., {Stankov}, A., {Guzik}, J.~A.,
  {Sperl}, M., {Krzesinski}, J., {Ogloza}, W., {Pajdosz}, G., {Zola}, S.,
  {Thomassen}, T., {Solheim}, J., {Serkowitsch}, E., {Reegen}, P., {Rumpf}, T.,
  {Schmalwieser}, A., \& {Montgomery}, M.~H. 1999, \aap, 349, 225

\bibitem[{{Breger} {et~al.}(1993){Breger}, {Stich}, {Garrido}, {Martin},
  {Jiang}, {Li}, {Hube}, {Ostermann}, {Paparo}, \&
  {Scheck}}]{1993A&A...271..482B}
{Breger}, M., {Stich}, J., {Garrido}, R., {Martin}, B., {Jiang}, S.~Y., {Li},
  Z.~P., {Hube}, D.~P., {Ostermann}, W., {Paparo}, M., \& {Scheck}, M. 1993,
  \aap, 271, 482

\bibitem[{{Burningham} {et~al.}(2005){Burningham}, {Naylor}, {Littlefair}, \&
  {Jeffries}}]{2005MNRAS.356.1583B}
{Burningham}, B., {Naylor}, T., {Littlefair}, S.~P., \& {Jeffries}, R.~D. 2005,
  \mnras, 356, 1583

\bibitem[{{Caballero}(2008)}]{2008A&A...478..667C}
{Caballero}, J.~A. 2008, \aap, 478, 667

\bibitem[{{Caballero} {et~al.}(2007){Caballero}, {B{\'e}jar}, {Rebolo},
  {Eisl{\"o}ffel}, {Zapatero Osorio}, {Mundt}, {Barrado Y Navascu{\'e}s},
  {Bihain}, {Bailer-Jones}, {Forveille}, \&
  {Mart{\'{\i}}n}}]{2007A&A...470..903C}
{Caballero}, J.~A., {B{\'e}jar}, V.~J.~S., {Rebolo}, R., {Eisl{\"o}ffel}, J.,
  {Zapatero Osorio}, M.~R., {Mundt}, R., {Barrado Y Navascu{\'e}s}, D.,
  {Bihain}, G., {Bailer-Jones}, C.~A.~L., {Forveille}, T., \& {Mart{\'{\i}}n},
  E.~L. 2007, \aap, 470, 903

\bibitem[{{Caballero} {et~al.}(2004){Caballero}, {B{\'e}jar}, {Rebolo}, \&
  {Zapatero Osorio}}]{2004A&A...424..857C}
{Caballero}, J.~A., {B{\'e}jar}, V.~J.~S., {Rebolo}, R., \& {Zapatero Osorio},
  M.~R. 2004, \aap, 424, 857

\bibitem[{{Caballero} {et~al.}(2006){Caballero}, {Mart{\'{\i}}n}, {Zapatero
  Osorio}, {B{\'e}jar}, {Rebolo}, {Pavlenko}, \&
  {Wainscoat}}]{2006A&A...445..143C}
{Caballero}, J.~A., {Mart{\'{\i}}n}, E.~L., {Zapatero Osorio}, M.~R.,
  {B{\'e}jar}, V.~J.~S., {Rebolo}, R., {Pavlenko}, Y., \& {Wainscoat}, R. 2006,
  \aap, 445, 143

\bibitem[{{Caballero} {et~al.}(2008){Caballero}, {Valdivielso},
  {Mart{\'{\i}}n}, {Montes}, {Pascual}, \&
  {P{\'e}rez-Gonz{\'a}lez}}]{2008A&A...491..515C}
{Caballero}, J.~A., {Valdivielso}, L., {Mart{\'{\i}}n}, E.~L., {Montes}, D.,
  {Pascual}, S., \& {P{\'e}rez-Gonz{\'a}lez}, P.~G. 2008, \aap, 491, 515

\bibitem[{{Carpenter} {et~al.}(2001){Carpenter}, {Hillenbrand}, \&
  {Skrutskie}}]{2001AJ....121.3160C}
{Carpenter}, J.~M., {Hillenbrand}, L.~A., \& {Skrutskie}, M.~F. 2001, \aj, 121,
  3160

\bibitem[{{Cieza} \& {Baliber}(2006)}]{2006ApJ...649..862C}
{Cieza}, L., \& {Baliber}, N. 2006, \apj, 649, 862

\bibitem[{{Cieza} \& {Baliber}(2007)}]{2007ApJ...671..605C}
---. 2007, \apj, 671, 605

\bibitem[{{Cohen} {et~al.}(2004){Cohen}, {Herbst}, \&
  {Williams}}]{2004AJ....127.1602C}
{Cohen}, R.~E., {Herbst}, W., \& {Williams}, E.~C. 2004, \aj, 127, 1602

\bibitem[{{D'Antona} \& {Mazzitelli}(1997)}]{1997MmSAI..68..807D}
{D'Antona}, F., \& {Mazzitelli}, I. 1997, Memorie della Societa Astronomica
  Italiana, 68, 807

\bibitem[{{Deeg} \& {Doyle}(2001)}]{2001phot.work...85D}
{Deeg}, H.~J., \& {Doyle}, L.~R. 2001, in Third Workshop on Photometry, ed.
  W.~J. {Borucki} \& L.~E. {Lasher}, 85

\bibitem[{{Deeming}(1975)}]{1975Ap&SS..36..137D}
{Deeming}, T.~J. 1975, \apss, 36, 137

\bibitem[{{Fernandez} \& {Eiroa}(1996)}]{1996A&A...310..143F}
{Fernandez}, M., \& {Eiroa}, C. 1996, \aap, 310, 143

\bibitem[{{Franciosini} {et~al.}(2006){Franciosini}, {Pallavicini}, \&
  {Sanz-Forcada}}]{2006A&A...446..501F}
{Franciosini}, E., {Pallavicini}, R., \& {Sanz-Forcada}, J. 2006, \aap, 446,
  501

\bibitem[{{Frasca} {et~al.}(2009){Frasca}, {Covino}, {Spezzi}, {Alcal{\'a}},
  {Marilli}, {F{\.z}r{\'e}sz}, \& {Gandolfi}}]{2009A&A...508.1313F}
{Frasca}, A., {Covino}, E., {Spezzi}, L., {Alcal{\'a}}, J.~M., {Marilli}, E.,
  {F{\.z}r{\'e}sz}, G., \& {Gandolfi}, D. 2009, \aap, 508, 1313

\bibitem[{{Gonz{\'a}lez Hern{\'a}ndez} {et~al.}(2008){Gonz{\'a}lez
  Hern{\'a}ndez}, {Caballero}, {Rebolo}, {B{\'e}jar}, {Barrado Y
  Navascu{\'e}s}, {Mart{\'{\i}}n}, \& {Zapatero Osorio}}]{2008A&A...490.1135G}
{Gonz{\'a}lez Hern{\'a}ndez}, J.~I., {Caballero}, J.~A., {Rebolo}, R.,
  {B{\'e}jar}, V.~J.~S., {Barrado Y Navascu{\'e}s}, D., {Mart{\'{\i}}n}, E.~L.,
  \& {Zapatero Osorio}, M.~R. 2008, \aap, 490, 1135

\bibitem[{{Herbst}(1994)}]{1994ASPC...62...35H}
{Herbst}, W. 1994, in Astronomical Society of the Pacific Conference Series,
  Vol.~62, The Nature and Evolutionary Status of Herbig Ae/Be Stars, ed.
  {P.~S.~The, M.~R.~Perez, \& E.~P.~J.~van den Heuvel}, 35--+

\bibitem[{{Herbst} {et~al.}(2002){Herbst}, {Bailer-Jones}, {Mundt},
  {Meisenheimer}, \& {Wackermann}}]{2002A&A...396..513H}
{Herbst}, W., {Bailer-Jones}, C.~A.~L., {Mundt}, R., {Meisenheimer}, K., \&
  {Wackermann}, R. 2002, \aap, 396, 513

\bibitem[{{Herbst} {et~al.}(2007){Herbst}, {Eisl{\"o}ffel}, {Mundt}, \&
  {Scholz}}]{2007prpl.conf..297H}
{Herbst}, W., {Eisl{\"o}ffel}, J., {Mundt}, R., \& {Scholz}, A. 2007, in
  Protostars and Planets V, ed. B.~{Reipurth}, D.~{Jewitt}, \& K.~{Keil},
  297--311

\bibitem[{{Hern{\'a}ndez} {et~al.}(2007){Hern{\'a}ndez}, {Hartmann}, {Megeath},
  {Gutermuth}, {Muzerolle}, {Calvet}, {Vivas}, {Brice{\~n}o}, {Allen},
  {Stauffer}, {Young}, \& {Fazio}}]{2007ApJ...662.1067H}
{Hern{\'a}ndez}, J., {Hartmann}, L., {Megeath}, T., {Gutermuth}, R.,
  {Muzerolle}, J., {Calvet}, N., {Vivas}, A.~K., {Brice{\~n}o}, C., {Allen},
  L., {Stauffer}, J., {Young}, E., \& {Fazio}, G. 2007, \apj, 662, 1067

\bibitem[{{Horne} \& {Baliunas}(1986)}]{1986ApJ...302..757H}
{Horne}, J.~H., \& {Baliunas}, S.~L. 1986, \apj, 302, 757

\bibitem[{{Howell}(1989)}]{1989PASP..101..616H}
{Howell}, S.~B. 1989, \pasp, 101, 616

\bibitem[{{Irwin} \& {Bouvier}(2009)}]{2009IAUS..258..363I}
{Irwin}, J., \& {Bouvier}, J. 2009, in IAU Symposium, ed. E.~E. {Mamajek},
  D.~R. {Soderblom}, \& R.~F.~G. {Wyse}, Vol. 258, 363

\bibitem[{{Irwin} {et~al.}(2008){Irwin}, {Hodgkin}, {Aigrain}, {Bouvier},
  {Hebb}, {Irwin}, \& {Moraux}}]{2008MNRAS.384..675I}
{Irwin}, J., {Hodgkin}, S., {Aigrain}, S., {Bouvier}, J., {Hebb}, L., {Irwin},
  M., \& {Moraux}, E. 2008, \mnras, 384, 675

\bibitem[{{Jeffries} {et~al.}(2006){Jeffries}, {Maxted}, {Oliveira}, \&
  {Naylor}}]{2006MNRAS.371L...6J}
{Jeffries}, R.~D., {Maxted}, P.~F.~L., {Oliveira}, J.~M., \& {Naylor}, T. 2006,
  \mnras, 371, L6

\bibitem[{{Joergens} {et~al.}(2003){Joergens}, {Fern{\'a}ndez}, {Carpenter}, \&
  {Neuh{\"a}user}}]{2003ApJ...594..971J}
{Joergens}, V., {Fern{\'a}ndez}, M., {Carpenter}, J.~M., \& {Neuh{\"a}user}, R.
  2003, \apj, 594, 971

\bibitem[{{Joy}(1949)}]{1949ApJ...110..424J}
{Joy}, A.~H. 1949, \apj, 110, 424

\bibitem[{{Kenyon} {et~al.}(2005){Kenyon}, {Jeffries}, {Naylor}, {Oliveira}, \&
  {Maxted}}]{2005MNRAS.356...89K}
{Kenyon}, M.~J., {Jeffries}, R.~D., {Naylor}, T., {Oliveira}, J.~M., \&
  {Maxted}, P.~F.~L. 2005, \mnras, 356, 89

\bibitem[{{Koenigl}(1991)}]{1991ApJ...370L..39K}
{Koenigl}, A. 1991, \apjl, 370, L39

\bibitem[{{Kuschnig} {et~al.}(1997){Kuschnig}, {Weiss}, {Gruber}, {Bely}, \&
  {Jenkner}}]{1997A&A...328..544K}
{Kuschnig}, R., {Weiss}, W.~W., {Gruber}, R., {Bely}, P.~Y., \& {Jenkner}, H.
  1997, \aap, 328, 544

\bibitem[{{Lada} {et~al.}(2006){Lada}, {Muench}, {Luhman}, {Allen}, {Hartmann},
  {Megeath}, {Myers}, {Fazio}, {Wood}, {Muzerolle}, {Rieke}, {Siegler}, \&
  {Young}}]{2006AJ....131.1574L}
{Lada}, C.~J., {Muench}, A.~A., {Luhman}, K.~L., {Allen}, L., {Hartmann}, L.,
  {Megeath}, T., {Myers}, P., {Fazio}, G., {Wood}, K., {Muzerolle}, J.,
  {Rieke}, G., {Siegler}, N., \& {Young}, E. 2006, \aj, 131, 1574

\bibitem[{{Lamm} {et~al.}(2005){Lamm}, {Mundt}, {Bailer-Jones}, \&
  {Herbst}}]{2005A&A...430.1005L}
{Lamm}, M.~H., {Mundt}, R., {Bailer-Jones}, C.~A.~L., \& {Herbst}, W. 2005,
  \aap, 430, 1005

\bibitem[{{Lee}(1968)}]{1968ApJ...152..913L}
{Lee}, T.~A. 1968, \apj, 152, 913

\bibitem[{{Lenz} \& {Breger}(2005)}]{2005CoAst.146...53L}
{Lenz}, P., \& {Breger}, M. 2005, Communications in Asteroseismology, 146, 53

\bibitem[{{Littlefair} {et~al.}(2005){Littlefair}, {Naylor}, {Burningham}, \&
  {Jeffries}}]{2005MNRAS.358..341L}
{Littlefair}, S.~P., {Naylor}, T., {Burningham}, B., \& {Jeffries}, R.~D. 2005,
  \mnras, 358, 341

\bibitem[{{Lodieu} {et~al.}(2009){Lodieu}, {Zapatero Osorio}, {Rebolo},
  {Mart{\'{\i}}n}, \& {Hambly}}]{2009A&A...505.1115L}
{Lodieu}, N., {Zapatero Osorio}, M.~R., {Rebolo}, R., {Mart{\'{\i}}n}, E.~L.,
  \& {Hambly}, N.~C. 2009, \aap, 505, 1115

\bibitem[{{Luhman} {et~al.}(2008){Luhman}, {Hern{\'a}ndez}, {Downes},
  {Hartmann}, \& {Brice{\~n}o}}]{2008ApJ...688..362L}
{Luhman}, K.~L., {Hern{\'a}ndez}, J., {Downes}, J.~J., {Hartmann}, L., \&
  {Brice{\~n}o}, C. 2008, \apj, 688, 362

\bibitem[{{Makidon} {et~al.}(2004){Makidon}, {Rebull}, {Strom}, {Adams}, \&
  {Patten}}]{2004AJ....127.2228M}
{Makidon}, R.~B., {Rebull}, L.~M., {Strom}, S.~E., {Adams}, M.~T., \& {Patten},
  B.~M. 2004, \aj, 127, 2228

\bibitem[{{Matt} {et~al.}(2010){Matt}, {Pinz{\'o}n}, {de la Reza}, \&
  {Greene}}]{2010ApJ...714..989M}
{Matt}, S.~P., {Pinz{\'o}n}, G., {de la Reza}, R., \& {Greene}, T.~P. 2010,
  \apj, 714, 989

\bibitem[{{Maxted} {et~al.}(2008){Maxted}, {Jeffries}, {Oliveira}, {Naylor}, \&
  {Jackson}}]{2008MNRAS.385.2210M}
{Maxted}, P.~F.~L., {Jeffries}, R.~D., {Oliveira}, J.~M., {Naylor}, T., \&
  {Jackson}, R.~J. 2008, \mnras, 385, 2210

\bibitem[{{Mochejska} {et~al.}(2002){Mochejska}, {Stanek}, {Sasselov}, \&
  {Szentgyorgyi}}]{2002AJ....123.3460M}
{Mochejska}, B.~J., {Stanek}, K.~Z., {Sasselov}, D.~D., \& {Szentgyorgyi},
  A.~H. 2002, \aj, 123, 3460

\bibitem[{{Muzerolle} {et~al.}(2003){Muzerolle}, {Hillenbrand}, {Calvet},
  {Brice{\~n}o}, \& {Hartmann}}]{2003ApJ...592..266M}
{Muzerolle}, J., {Hillenbrand}, L., {Calvet}, N., {Brice{\~n}o}, C., \&
  {Hartmann}, L. 2003, \apj, 592, 266

\bibitem[{{Nguyen} {et~al.}(2009){Nguyen}, {Jayawardhana}, {van Kerkwijk},
  {Brandeker}, {Scholz}, \& {Damjanov}}]{2009ApJ...695.1648N}
{Nguyen}, D.~C., {Jayawardhana}, R., {van Kerkwijk}, M.~H., {Brandeker}, A.,
  {Scholz}, A., \& {Damjanov}, I. 2009, \apj, 695, 1648

\bibitem[{{Palla} \& {Baraffe}(2005)}]{2005A&A...432L..57P}
{Palla}, F., \& {Baraffe}, I. 2005, \aap, 432, L57

\bibitem[{{Press}(1978)}]{1978ComAp...7..103P}
{Press}, W.~H. 1978, Comments on Astrophysics, 7, 103

\bibitem[{{Press} {et~al.}(1992){Press}, {Teukolsky}, {Vetterling}, \&
  {Flannery}}]{1992nrfa.book.....P}
{Press}, W.~H., {Teukolsky}, S.~A., {Vetterling}, W.~T., \& {Flannery}, B.~P.
  1992, {Numerical recipes: The art of scientific computing} (Cambridge:
  Cambridge University Press)

\bibitem[{{Rebull}(2001)}]{2001AJ....121.1676R}
{Rebull}, L.~M. 2001, \aj, 121, 1676

\bibitem[{{Rebull} {et~al.}(2006){Rebull}, {Stauffer}, {Megeath}, {Hora}, \&
  {Hartmann}}]{2006ApJ...646..297R}
{Rebull}, L.~M., {Stauffer}, J.~R., {Megeath}, S.~T., {Hora}, J.~L., \&
  {Hartmann}, L. 2006, \apj, 646, 297

\bibitem[{{Rebull} {et~al.}(2004){Rebull}, {Wolff}, \&
  {Strom}}]{2004AJ....127.1029R}
{Rebull}, L.~M., {Wolff}, S.~C., \& {Strom}, S.~E. 2004, \aj, 127, 1029

\bibitem[{{Reegen}(2007)}]{2007A&A...467.1353R}
{Reegen}, P. 2007, \aap, 467, 1353

\bibitem[{{Reiners} {et~al.}(2009){Reiners}, {Basri}, \&
  {Christensen}}]{2009ApJ...697..373R}
{Reiners}, A., {Basri}, G., \& {Christensen}, U.~R. 2009, \apj, 697, 373

\bibitem[{{Rodr{\'{\i}}guez-Ledesma} {et~al.}(2009){Rodr{\'{\i}}guez-Ledesma},
  {Mundt}, \& {Eisl{\"o}ffel}}]{2009A&A...502..883R}
{Rodr{\'{\i}}guez-Ledesma}, M.~V., {Mundt}, R., \& {Eisl{\"o}ffel}, J. 2009,
  \aap, 502, 883

\bibitem[{{Sacco} {et~al.}(2008){Sacco}, {Franciosini}, {Randich}, \&
  {Pallavicini}}]{2008A&A...488..167S}
{Sacco}, G.~G., {Franciosini}, E., {Randich}, S., \& {Pallavicini}, R. 2008,
  \aap, 488, 167

\bibitem[{{Scargle}(1982)}]{1982ApJ...263..835S}
{Scargle}, J.~D. 1982, \apj, 263, 835

\bibitem[{{Scholz}(2009)}]{2009AIPC.1094...61S}
{Scholz}, A. 2009, in American Institute of Physics Conference Series, Vol.
  1094, Proceedings of the 15th Cambridge Workshop on Cool Stars, Stellar
  Systems and the Sun, ed. E.~{Stempels}, 61

\bibitem[{{Scholz} \& {Eisl{\"o}ffel}(2004)}]{2004A&A...419..249S}
{Scholz}, A., \& {Eisl{\"o}ffel}, J. 2004, \aap, 419, 249

\bibitem[{{Scholz} \& {Eisl{\"o}ffel}(2005)}]{2005A&A...429.1007S}
---. 2005, \aap, 429, 1007

\bibitem[{{Scholz} \& {Jayawardhana}(2008)}]{2008ApJ...672L..49S}
{Scholz}, A., \& {Jayawardhana}, R. 2008, \apjl, 672, L49

\bibitem[{{Scholz} {et~al.}(2009){Scholz}, {Xu}, {Jayawardhana}, {Wood},
  {Eisl{\"o}ffel}, \& {Quinn}}]{2009MNRAS.398..873S}
{Scholz}, A., {Xu}, X., {Jayawardhana}, R., {Wood}, K., {Eisl{\"o}ffel}, J., \&
  {Quinn}, C. 2009, \mnras, 398, 873

\bibitem[{{Sherry} {et~al.}(2004){Sherry}, {Walter}, \&
  {Wolk}}]{2004AJ....128.2316S}
{Sherry}, W.~H., {Walter}, F.~M., \& {Wolk}, S.~J. 2004, \aj, 128, 2316

\bibitem[{{Sherry} {et~al.}(2008){Sherry}, {Walter}, {Wolk}, \&
  {Adams}}]{2008AJ....135.1616S}
{Sherry}, W.~H., {Walter}, F.~M., {Wolk}, S.~J., \& {Adams}, N.~R. 2008, \aj,
  135, 1616

\bibitem[{{Sokoloski} {et~al.}(2001){Sokoloski}, {Bildsten}, \&
  {Ho}}]{2001MNRAS.326..553S}
{Sokoloski}, J.~L., {Bildsten}, L., \& {Ho}, W.~C.~G. 2001, \mnras, 326, 553

\bibitem[{{Sperl}(1998)}]{1998CoAst.111....1S}
{Sperl}, M. 1998, Communications in Asteroseismology, 111, 1

\bibitem[{{Stassun} {et~al.}(1999){Stassun}, {Mathieu}, {Mazeh}, \&
  {Vrba}}]{1999AJ....117.2941S}
{Stassun}, K.~G., {Mathieu}, R.~D., {Mazeh}, T., \& {Vrba}, F.~J. 1999, \aj,
  117, 2941

\bibitem[{{Stetson}(1996)}]{1996PASP..108..851S}
{Stetson}, P.~B. 1996, \pasp, 108, 851

\bibitem[{{Stetson}(2000)}]{2000PASP..112..925S}
---. 2000, \pasp, 112, 925

\bibitem[{{Vivas} {et~al.}(2004){Vivas}, {Zinn}, {Abad}, {Andrews}, {Bailyn},
  {Baltay}, {Bongiovanni}, {Brice{\~n}o}, {Bruzual}, {Coppi}, {Della Prugna},
  {Ellman}, {Ferr{\'{\i}}n}, {Gebhard}, {Girard}, {Hernandez}, {Herrera},
  {Honeycutt}, {Magris}, {Mufson}, {Musser}, {Naranjo}, {Rabinowitz},
  {Rengstorf}, {Rosenzweig}, {S{\'a}nchez}, {S{\'a}nchez}, {Schaefer},
  {Schenner}, {Snyder}, {Sofia}, {Stock}, {van Altena}, {Vicente}, \&
  {Vieira}}]{2004AJ....127.1158V}
{Vivas}, A.~K., {Zinn}, R., {Abad}, C., {Andrews}, P., {Bailyn}, C., {Baltay},
  C., {Bongiovanni}, A., {Brice{\~n}o}, C., {Bruzual}, G., {Coppi}, P., {Della
  Prugna}, F., {Ellman}, N., {Ferr{\'{\i}}n}, I., {Gebhard}, M., {Girard}, T.,
  {Hernandez}, J., {Herrera}, D., {Honeycutt}, R., {Magris}, G., {Mufson}, S.,
  {Musser}, J., {Naranjo}, O., {Rabinowitz}, D., {Rengstorf}, A., {Rosenzweig},
  P., {S{\'a}nchez}, G., {S{\'a}nchez}, G., {Schaefer}, B., {Schenner}, H.,
  {Snyder}, J.~A., {Sofia}, S., {Stock}, J., {van Altena}, W., {Vicente}, B.,
  \& {Vieira}, K. 2004, \aj, 127, 1158

\bibitem[{{Walter} {et~al.}(1997){Walter}, {Wolk}, {Freyberg}, \&
  {Schmitt}}]{1997MmSAI..68.1081W}
{Walter}, F.~M., {Wolk}, S.~J., {Freyberg}, M., \& {Schmitt}, J.~H.~M.~M. 1997,
  Memorie della Societa Astronomica Italiana, 68, 1081

\bibitem[{{Wolk}(1996)}]{1996PhDT........63W}
{Wolk}, S.~J. 1996, PhD thesis, State Univ. New York, Stony Brook

\bibitem[{{Young}(1967)}]{1967AJ.....72..747Y}
{Young}, A.~T. 1967, \aj, 72, 747

\bibitem[{{Young} {et~al.}(1991){Young}, {Genet}, {Boyd}, {Borucki},
  {Lockwood}, {Henry}, {Hall}, {Smith}, {Baliumas}, {Donahue}, \&
  {Epand}}]{1991PASP..103..221Y}
{Young}, A.~T., {Genet}, R.~M., {Boyd}, L.~J., {Borucki}, W.~J., {Lockwood},
  G.~W., {Henry}, G.~W., {Hall}, D.~S., {Smith}, D.~P., {Baliumas}, S.~L.,
  {Donahue}, R., \& {Epand}, D.~H. 1991, \pasp, 103, 221

\bibitem[{{Zapatero Osorio} {et~al.}(2000){Zapatero Osorio}, {B{\'e}jar},
  {Mart{\'{\i}}n}, {Rebolo}, {Barrado y Navascu{\'e}s}, {Bailer-Jones}, \&
  {Mundt}}]{2000Sci...290..103Z}
{Zapatero Osorio}, M.~R., {B{\'e}jar}, V.~J.~S., {Mart{\'{\i}}n}, E.~L.,
  {Rebolo}, R., {Barrado y Navascu{\'e}s}, D., {Bailer-Jones}, C.~A.~L., \&
  {Mundt}, R. 2000, Science, 290, 103

\bibitem[{{Zapatero Osorio} {et~al.}(2002){Zapatero Osorio}, {B{\'e}jar},
  {Pavlenko}, {Rebolo}, {Allende Prieto}, {Mart{\'{\i}}n}, \& {Garc{\'{\i}}a
  L{\'o}pez}}]{2002A&A...384..937Z}
{Zapatero Osorio}, M.~R., {B{\'e}jar}, V.~J.~S., {Pavlenko}, Y., {Rebolo}, R.,
  {Allende Prieto}, C., {Mart{\'{\i}}n}, E.~L., \& {Garc{\'{\i}}a L{\'o}pez},
  R.~J. 2002, \aap, 384, 937

\bibitem[{{Zapatero Osorio} {et~al.}(2003){Zapatero Osorio}, {Caballero},
  {B{\'e}jar}, \& {Rebolo}}]{2003A&A...408..663Z}
{Zapatero Osorio}, M.~R., {Caballero}, J.~A., {B{\'e}jar}, V.~J.~S., \&
  {Rebolo}, R. 2003, \aap, 408, 663

\end{thebibliography}

\clearpage

\LongTables
\begin{landscape}
\begin{deluxetable}{clcccccc}
\tabletypesize{\scriptsize}
\tablecolumns{10}
\tablewidth{0pt}
\tablecaption{\bf $\sigma$ Orionis: confirmed and candidate members in our photometric sample}
\tablehead{
\colhead{Object} & \colhead{Other IDs} & \colhead{SpT}&
\colhead{Variable?} & \colhead{} & \colhead{} &
\colhead{Membership evidence} & \colhead{Refs}
}
\startdata 
2MASS J05372806-0236065 & SO59           &    &   &   &   &  & 13\\
2MASS J05373648-0241567 & S Ori 40, KJN75, SO116 & M7 &  &  &  & $v_r$, H$\alpha$, Li, Na & 1,4,5\\
2MASS J05373784-0245442 & SWW184, SO123 &   & Y$^{13}$  &   &    &  (PM) & 12\\
2MASS J05373790-0236085 &                &    &   &   &   &  & \\
2MASS J05374413-0235198 &                &    &   &   &   &  & \\
2MASS J05375161-0235257 & SWW125, F1, Mayrit 797272, SO214 & M1-3 &   &   &    & H$\alpha$, Li & 7\\
2MASS J05375206-0236046 & KJN62, M182, Mayrit 790270  &   &   &   &    & ($v_r$ NM?$^9$), Li, Na, (PM)  & 4,9,12\\
2MASS J05375398-0249545 & SWW221, Mayrit 1129222  &   & Y$^{12}$ &   &    & D, (PM) & 12,14 \\
2MASS J05375404-0244407 & SWW68, SO240 &   &   &   &    & (PM) & 12\\
2MASS J05375486-0241092 & SWW174, B237, SO247, Mayrit 809248  &   &   &   &    & $v_r$, (Na NM?$^{10}$), D, (PM) & 10,12,13\\
2MASS J05375745-0238444 & S Ori 12, KJN39, M162, SO271, Mayrit 728257 & M6  &   &   &   & $v_r$, Li, Na, D, (PM)  & 1,4,5,9,12,13\\
2MASS J05375840-0241262 & SWW53, KJN18, M118, SO283, Mayrit 767245 &   &   &   &    & $v_r$, Li, Na, (PM) & 4,9,12\\
2MASS J05375970-0251033 & SO293  &   &   &   &    &    &\\
2MASS J05380055-0245097 & SWW140, M178, F4,SO297, Mayrit 861230  &   &   &   &    & $v_r$, Na, (PM) & 9,12\\
2MASS J05380107-0245379 & SWW180, M85, SO300, Mayrit 873229  &   & Y$^{13}$ &   &    & Na, D, (PM) & 9,12,13\\
2MASS J05380552-0235571 & S Ori J053805.5, M186, SO327, Mayrit 588270   &   &   &   &    & $v_r$, Na, D, (PM) & 9,12,13\\
2MASS J05380826-0235562 & SWW41, F9, SO362, Mayrit 547270  &   & Y$^{13}$ &   &   & H$\alpha$, Li, D, (PM) & 2,12,13\\
2MASS J05380994-0251377 & SWW52, M133, SO374, Mayrit 1073209  &   & Y$^{13}$ &   &    & $v_r$, Na, D, (PM) & 9,12,13\\
2MASS J05381175-0245012 & SO385  &   &   &   &    &    &\\
2MASS J05381265-0236378 &                &    &   &   &   &  & \\
2MASS J05381315-0245509 & SWW98, SO396, Mayrit 757219  &   & Y$^{13}$  &   &    & D, (PM) & 12,13\\
2MASS J05381330-0251329 & KJN48, M137, SO401, Mayrit 1045207 &   &   &   &    &  $v_r$, Li, Na, (PM) & 4,9,12\\
2MASS J05381589-0234412 & SO424  &   &   &   &    &    &\\
2MASS J05381610-0238049 & S Ori J053816.0, SWW12, KJN11, M167, Mayrit 447254 &   &   &   &    & $v_r$, Li, Na, (PM) & 4,9,12\\
2MASS J05381741-0240242 & S Ori 27, KJN60, M146, Mayrit 488237  & M7 (M6.5$^3$)  & Y$^{15}$ &   &    &  $v_r$, Li, Na, (PM) & 1,3,4,9,12\\
2MASS J05381778-0240500 & S Ori J053817.8-024050, SWW5, F17, SO435, Mayrit 498234   &   & Y$^{13}$ &  &    & D, (PM) & 12,13\\
2MASS J05381824-0248143 & SWW40, M174, SO444, Mayrit 835208   &   &   &   &    & $v_r$, Na, D(EV), (PM) & 9,12,13\\
2MASS J05381834-0235385 & S Ori J053818.2-023539, KJN76, M203, F19, SO446, Mayrit 396273   &   &   &   &    & $v_r$, (Na NM?$^4$), (PM) & 4,9,12\\
2MASS J05381886-0251388 & SWW39, M136, SO451, Mayrit 1016202  &   &   &   &    & $v_r$, Na, D, (PM) & 9,12,13\\
2MASS J05381914-0235279 & SO454  &   &   &   &    & (PM) & 12\\
2MASS J05382021-0238016 & S Ori J053820.1-023802, SWW131, M168, F20, SO460, Mayrit 387252 & M4 &   &  &    &  $v_r$, H$\alpha$, Li, Na, (PM)  & 1,2,3,9,12\\
2MASS J05382050-0234089 & r053820-0234, SWW124, M106, SO462, Mayrit 380287 & M4 & Y$^{12,13}$ &   &    & $v_r$, H$\alpha$, Li, Na, D, (PM)  & 1,3,9,12,13\\
2MASS J05382088-0246132 & S Ori 31, SO465, Mayrit 710210   & M7  & Y$^{15,16}$ &   &    & (PM) & 1,12\\
2MASS J05382089-0251280 & M138, SO466, Mayrit 994201  &   &   &   &    & $v_r$, Na, (PM) & 9,12\\
2MASS J05382307-0236493 & SWW103, B51, SO482, Mayrit 329261 &   & Y$^{13}$ &   &    & $v_r$, (Na NM?$^{10}$), D, (PM) & 10,12,13\\
2MASS J05382332-0244142 & S Ori J053823.3-024414, SWW139, KJN15, M52, F25, Mayrit 589213   &   &   &   &    & $v_r$, Li, (PM) & 4,9,12\\
2MASS J05382354-0241317 & S Ori J053823.6-024132, SWW3, B229, M121, F26, SO489, Mayrit 459224  &   &   &   &    & $v_r$, Na, (PM) & 9,10,12\\
S Ori J053825.1-024802  & S Ori 53 &   &   &   &    &    &\\
2MASS J05382543-0242412 & S Ori J053825.4-024241, SO500, Mayrit 495216  & M6  & Y$^{6,12}$  &   &  & $v_r$, H$\alpha$, FL, D, (PM) & 6,8,12,13\\
2MASS J05382557-0248370 & S Ori 45 & M8.5  & Y$^{15,16,17}$ &   &    &  $v_r$, H$\alpha$, Li, FL  & 1,3,5\\
2MASS J05382623-0240413 & S Ori J053826.1-024041, KJN58, M141, SO509, Mayrit 395225   & M8 (M5,M6$^6$)  & Y$^{15}$ &   &    & $v_r$, Li, Na, (PM) & 1,4,6,9,12\\
2MASS J05382684-0238460 & S Ori J053826.8-022846, B368, M163, SO514, Mayrit 316238   &   &   &   &    & $v_r$, H$\alpha$, Li, Na, D, (PM) & 2,9,10,12,13\\
2MASS J05382725-0245096 & 4771-41, F32, KJN7, SO518, Mayrit 609206  &  &   &   &   &  $v_r$, H$\alpha$, Li, FL, D & 1,3,13\\
2MASS J05382750-0235041 & S Ori J053827.5-023504, SWW67, M96, F33, SO520, Mayrit 265282   & M3.5  &   &   &    & $v_r$, H$\alpha$, Li, Na, D, (PM) & 2,9,12,13\\
2MASS J05382774-0243009 & SWW87, F34, SO525 &   &   &    &    & $v_r$, H$\alpha$, Li, (PM)   & 2,12\\
2MASS J05382848-0246170 & SWW188   &   &   &   &    &    &\\
2MASS J05382896-0248473 & S Ori J053829.0-024847, M170, SO537, Mayrit 803197  &  M6 &  &   &   & $v_r$, Na, D & 1,8,9,13\\
2MASS J05383141-0236338 & SWW50, SO562, Mayrit 203260  &   &   &   &  &  $v_r$, H$\alpha$, Li, D, (PM)  & 2,12,13\\
2MASS J05383157-0235148 & r053831-0235, SWW49, F44, SO536, Mayrit 203283 & M0  &   &   &  & $v_r$, H$\alpha$, Li, D, (PM)  & 1,2,3,12,13\\
2MASS J05383160-0251268 & SWW178, SO564, Mayrit 947192   &   &   &   &    & (PM) & 12\\
2MASS J05383284-0235392 & r053832-0235b, SO572, F54   &   &   &   &    & $v_r$, H$\alpha$, Li   & 2\\
2MASS J05383302-0239279 & F50, SO576   &   &   &   &    & (PM NM?$^{12}$) &\\
2MASS J05383335-0236176 & SWW130, F52, SO582 &   &   &   &    & (PM) & 12\\
2MASS J05383388-0245078 & S Ori J053833.9, KJN36, M202, Mayrit 571197 &   &   &   &  & ($v_r$ NM?$^9$), Li, Na, D, (PM) & 4,8,9,12\\
2MASS J05383405-0236375 & r053833-0236, SWW66, F54, SO587, Mayrit 165257  & M3.5  &   &  &    & $v_r$, H$\alpha$, Li, FL, D & 1,2,3,13\\
2MASS J05383460-0241087 & S Ori J053834.5-024109, SWW80, SO598, Mayrit 344206 &   & Y$^{13}$ &   &    & D, (PM) & 12,13\\
2MASS J05383669-0244136 & S Ori J053836.7-024414, SWW16, M63, SO621, Mayrit 508194 &   &   &   &    & $v_r$, H$\alpha$, Li, (PM) & 2,9,12\\
2MASS J05383745-0250236 & SWW11, M155, SO628, Mayrit 870187 &   &   &   &    & $v_r$, Na, (PM) & 9,12\\
2MASS J05383858-0241558 & S Ori J053838.6, KJN44, B215, M114, SO641, Mayrit 368195 & M5.5 &   &   &    & $v_r$, Li, Na, (PM) & 4,6,9,10,12\\
2MASS J05383902-0245321 & SWW31, M156, SO646, Mayrit 578189   &   & Y$^{13}$ &  &    &  $v_r$, H$\alpha$, Li, D, (PM) & 2,9,12,13\\
2MASS J05383922-0253084 & SO648  &   &   &   &    & (PM NM?$^{12}$) &\\
2MASS J05385317-0243528 & SWW47, F106, SO785, Mayrit 489165  &   & Y$^{13}$  &   &    & $v_r$, H$\alpha$, Li, (PM) & 2,12\\
2MASS J05385382-0244588 & S Ori J053853.8-024459   &   &   &   &    & (PM) & 12\\
2MASS J05385492-0228583 & SWW10, SE77, KJN21, SO797, Mayrit 449020  &   & Y$^{11}$  &   &    & $v_r$, Li, Na, (PM) & 4,12\\
2MASS J05385492-0240337 & S Ori J053854.9-024034  &   &   &   &  & D & 8\\
2MASS J05385542-0241208 & S Ori J053855.4-024121, Mayrit 358154 & M5  & Y$^{12}$ &   &  & H$\alpha$, FL, D, (PM) & 7,8,12\\
2MASS J05385623-0231153 & K1.02-91  &   &   &   &    &    &\\
2MASS J05385922-0233514 & SO827, SWW227, F118, Mayrit 252059   &   &  Y$^{13}$ &   &   &  $v_r$, H$\alpha$, Li, D, (PM) & 2,12,13\\
2MASS J05385946-0242198 &                &    &   &   &   &  & \\
2MASS J05390052-0239390 & 4771-1056, F122  &   &   &   &    &    &\\
2MASS J05390115-0236388 & KJN9, M213, F124, SO841, Mayrit 249099   &   &   &   &    &  $v_r$, Li, Na,(PM) & 4,9,12\\
2MASS J05390193-0235029 & SO848, S Ori J053902.1-023501, Mayrit 264077 & M3 & Y$^{13}$ &  &    & H$\alpha$, FL, D, (PM) & 7,8,12,13\\
2MASS J05390276-0229558 & SWW28, F126, SO855, Mayrit 453037  &   &   &   &    &  $v_r$, H$\alpha$, Li, (PM) & 2,12\\
S Ori J053903.2-023020  & S Ori 51 &   &   &   &     &   &    \\
2MASS J05390357-0246269 & SWW122, SO865, Mayrit 687156  &   & Y$^{13}$ &   &    & D, (PM) & 12,13\\
2MASS J05390449-0238353 & S Ori 17, SO870, Mayrit 334118  & M6  &   &   &    &  Li & 1,5\\
2MASS J05390458-0241493 & SO871, Mayrit 458140 &   & Y$^{12}$  &   &  & D, (PM) & 12,13\\
2MASS J05390524-0233005 & SWW175, KJN4, F131, SO877, Mayrit 355060   &   &   &   &    & $v_r$, Li, Na, (PM) & 4,12\\
2MASS J05390540-0232303 & 4771-1075, KJN7, F132, SO879, Mayrit 374056  &   & Y$^{13}$ &   &    &  $v_r$, H$\alpha$, Li  & 1,2,3\\
CTIO J05390664-0238050  &                                              &   &          &   &    &                        &      \\
2MASS J05390759-0228234 & r053907-0228, SWW121, SE82, F137, SO896, Mayrit 571037  & M3  &   &   &    &  $v_r$, H$\alpha$, Li, (PM) & 1,2,3,12\\
2MASS J05390760-0232391 & 4771-1092, F138, SO897, Mayrit 397060  &   &   &  &    &  $v_r$, H$\alpha$, Li, D(TD) & 2,13\\
2MASS J05390808-0228447 & S Ori 8, SE83, SO901, Mayrit 558039  &   &   &   &    & D(EV), (PM) & 12,13\\
2MASS J05390821-0232284 & S Ori 7, SWW108, SO902, Mayrit 410059  &   &   &   &    & (PM) & 12\\
2MASS J05390878-0231115 & SWW129, SO908, Mayrit 461051  &   &   &   &    & D, (PM) & 12,13\\
2MASS J05390894-0239579 & S Ori 25, F140, SO911, Mayrit 433123  & M7.5 (M6.5$^5$) & Y$^{15}$ &  &  & $v_r$, H$\alpha$, Li, (PM) & 1,5,12\\
2MASS J05391001-0228116 & S Ori J053909.9-022814, KJN33, SO917 & M5  &   &   &    & NM?$^{12,4}$, D(EV) & 1,4,12,13\\
2MASS J05391003-0242425 & SO918, Mayrit 552137  &   &   &   &    & (PM) & 12\\
S Ori J053910.8-023715  & S Ori 50 &    &    &    &     &   &    \\
2MASS J05391139-0233327 & SOri J053911.4-023333, KJN42, SO925, Mayrit 425070  &  M5 &   &   &    & $v_r$, Li, Na, (PM) & 1,4,12\\
2MASS J05391151-0231065 & SWW195, F144, SO927, Mayrit 497054  &   &   &   &    &  $v_r$, H$\alpha$, Li, D, (PM) & 2,12,13\\
2MASS J05391163-0236028 & 4771-1038, KJN8, SWW153, F145, SO929, Mayrit 403090   &   &   &   &    &  $v_r$, H$\alpha$, Li  & 1,2,3\\
2MASS J05391232-0230064 & SWW203, F147, SO933, Mayrit 544049  &   &   &   &    & (PM) & 12\\
2MASS J05391308-0237509 & SOri 30, SO936, Mayrit 438105   & M6  &   &   &    & D, (PM) & 1,8,12,13\\
2MASS J05391346-0237391 & F148, SO940, Mayrit 441103   &   &   &   &    & (PM) & 12\\
2MASS J05391447-0228333 & SOri J053914.5-022834, SWW95, SE88, F149, SO946, Mayrit 631045   & M3.5 &   &   &    &  $v_r$, Li, (PM) & 1,3,12\\
2MASS J05391510-0240475 & SOri 16, SO957, Mayrit 538122   &   &   &   &    & (PM) & 12\\
2MASS J05391576-0238262 & SOri 26  & M4.5  &   &   &    & (PM) & 1,12\\
2MASS J05391582-0236507 & SO967, K1.02-4, F151, Mayrit 468096 &   &   &   &    & D, (PM) & 12,13\\
2MASS J05391699-0241171 & F153, SO976, M578123 &   & Y$^{13}$ &   &    & (PM) & 12\\
2MASS J05391883-0230531 & 4771-0910, SO984, F157, Mayrit 596059   &   &   &  &  &  $v_r$, H$\alpha$, Li, D & 2,13\\
2MASS J05392023-0238258 & S Ori 5, SWW60, SO999, Mayrit 551105  &   &   &   &    & (PM) & 12\\
2MASS J05392097-0230334 & S Ori 3, KJN20, F160, SO1005, Mayrit 633059   &   &   &   &    & $v_r$, Li, Na, (PM) & 4,12\\
2MASS J05392174-0244038 & SO1009, Mayrit 735131  &   &   &   &   & D(EV) & 13\\
2MASS J05392224-0245524 & S Ori J053922.2-024552, SO1013  &  &   &   &    &    &\\
2MASS J05392286-0233330 & r053923-0233, SWW185, F161, SO1017, Mayrit 590076  & M2  & Y$^{13}$ &  &    & $v_r$, H$\alpha$, Li, (PM) & 1,2,3,12\\
2MASS J05392307-0228112 &                &    &   &   &   &  & \\
2MASS J05392319-0246557 & S Ori 28, KJN64, Mayrit 872139  &  & Y$^{15}$  &   &    & ($v_r$ NM?$^9$), Li, Na, (PM) & 4,9,12\\
2MASS J05392341-0240575 & S Ori 42 & M7.5 & Y$^{15}$ &   &  & H$\alpha$, D & 1,8\\
2MASS J05392435-0234013 & SWW127, M191, F164, SO1027 &   &   &   &    & $v_r$, H$\alpha$, Na, (PM NM?$^{12}$) & 2,9\\
2MASS J05392519-0238220 & SWW135, F165, SO1036, Mayrit 622103 &  & Y$^{13}$  &   &    & $v_r$, H$\alpha$, Li, D & 2,13\\
2MASS J05392524-0227479 & B157, SO1037 &   &   &   &    & $v_r$, Na, (PM NM?$^{12}$) & 10\\
2MASS J05392560-0238436 & HH446, Mayrit 633105   &   &   &   &  & (PM) & 12\\
2MASS J05392561-0234042 & SWW7, SO1043, Mayrit 623079  &   &   &   &    &    &\\
2MASS J05392633-0228376 & SOri 2, SWW164, SE93, SO1050, Mayrit 764055 &    & Y$^{11}$ &   &    & D, (PM) & 12,13\\
2MASS J05392677-0242583 & SWW45, SO1057, Mayrit 756124 &  & Y$^{13}$  &   &    & D(EV), (PM) & 12,13\\
2MASS J05392685-0236561 & S Ori 36, KJN74, M177, SO1059 &   &   &   &  & $v_r$, Li, Na (bin?), D & 4,8,9,13\\
2MASS J05393056-0238270 & SO1081, SWW222, B260, F169  &   &   &   &    & H$\alpha$ (NM?$^{2,10}$)  & 2,10\\
2MASS J05393234-0227571 & SO1092, Mayrit 861056  &   &   &   &    &    &\\
2MASS J05393432-0238468 & S Ori 21, KJN61, M126, SO1108, Mayrit 761103   &   &   &   &    & $v_r$, Li, Na, (PM) & 4,9,12\\
2MASS J05393673-0231588 & B237   &   &   &   &    & $v_r$, (Na NM?$^{10}$) & 10\\
2MASS J05393759-0244304 & S Ori 14, KJN49, M169, SO1135, Mayrit 942123 &   &   &   &    & $v_r$, Li, Na, (PM) & 4,9,12\\
2MASS J05393931-0232252 & S Ori 4, SWW107, M117, SO1151, Mayrit 839077 &    &   &   &    & (PM) & 12\\
2MASS J05393982-0231217 & SO1153, Mayrit 871071 &  & Y$^{13}$  &   &    & D & 13\\
2MASS J05393982-0233159 & F174, SO1154, Mayrit 841079  &   & Y$^{13}$ &   &  & D, (PM) & 12,13\\
2MASS J05393998-0243097 & F175, SO1155 &   &   &   &    & D & 13\\
2MASS J05394057-0239123 & SO1162, B233   &   &   &   &    & $v_r$, Na, (PM NM?$^{12}$) & 10\\
2MASS J05394318-0232433 & S Ori J053943.2-023243, SWW75, SO1182, Mayrit 897077 &   & Y$^{13}$ &   &    & D, (PM) & 12,13\\
2MASS J05394411-0231092 & SO1189, Mayrit 936072   &   &   &   &    &    &\\
2MASS J05394433-0233027 & S Ori 11, M110, SO1191, Mayrit 910079   & M6  &   &   &    & $v_r$, Na, (PM) & 1,9,12\\
2MASS J05394725-0241359 & SWW192  &   &   &   &    &    &\\
2MASS J05394770-0236230 & B179, , SO1216 &   &   &   &    & $v_r$, Na, (PM) & 9,10,12\\
2MASS J05394784-0232248 & SO1217, Mayrit 969077   &   &   &   &    &    &\\
2MASS J05394799-0240320 & SWW32, SO1219, Mayrit 986106   &   &   &   &    & (PM) & 12\\
2MASS J05394806-0245571 & S Ori J053948.1-024557, SWW92, SO1220 &  &   &   &    & (PM) & 12\\
2MASS J05394826-0229144 & S Ori J053948.1-022914, SE108  & M7 & Y$^{11}$  &   &    & NM?$^{8,12}$ & 1\\
2MASS J05394891-0229110 & SWW126, B319   &   &   &   &    & ($v_r$ NM?$^{10}$), Na & 10\\
2MASS J05395038-0243307 & SO1235, Mayrit 1082115   &   &  Y$^{13}$ &   &    &    &\\
2MASS J05395056-0234137 & S Ori J053950.6-023414, KJN19, M115, SO1238, Mayrit 992084 &   &   &   &    & $v_r$, Li, Na, (PM) & 4,9,12\\
2MASS J05395236-0236147 & S Ori J053952.3-023615, M104  &   &   &   &    & Na, ($v_r$ NM?$^9$), (PM) & 9,12\\
2MASS J05395248-0232023 & SO1250  &   &   &   &    &    &\\
2MASS J05395313-0243083 & SO1256, Mayrit 1110113  &   &   &   &    &    &\\
2MASS J05395313-0230294 & M209   &   &   &   &    & $v_r$, Na & 9\\
2MASS J05395362-0233426 & SO1260, Mayrit 1041082  &   & Y$^{13}$  &   &  & D, (PM) & 12,13\\
2MASS J05395433-0237189 & S Ori J053954.3-023720, M98, SO1268, Mayrit 1045094 & M6 &  &   &   & $v_r$, Na, D(TD), (PM) & 6,8,9,12,13\\
2MASS J05395645-0238034 & SOriJ053956.4-023804, B143, M93, SO1285, Mayrit 1081097 &   &   &   &  & $v_r$, Na, D, (PM) & 10,12,13 \\
2MASS J05395753-0232120 & S Ori J053957.5-023212, M131, SO1295, Mayrit 1114078 &   &   &   &    & $v_r$, Na, (PM) & 9,12\\
2MASS J05400338-0229014 & SO1337, Mayrit 1250070   &   &   &   &    &    &\\
2MASS J05400453-0236421 & S Ori J054004.5-023642, KJN73, M102, SO1338, Mayrit 1196092 &  & Y$^{15}$  &   &  & $v_r$, Na, D, (PM) & 4,8,9,12,13\\
2MASS J05400525-0230522 & S Ori J054005.1-023052, M143, SO1344, Mayrit 1245076   & M5 &   &   &    & $v_r$, H$\alpha$, Li, Na, D & 1,3,9,13\\
2MASS J05400708-0232446 & S Ori J054007.1-023245, M125, SO1353, Mayrit 1249081  &   &   &   &    & $v_r$, Na, (PM) & 9,12\\
2MASS J05400867-0232432 & SO1359, Mayrit 1273081 &   &   &   &    &    &\\
2MASS J05400889-0233336 & SO1361, Mayrit 1269083 &   & Y$^{13}$ &   &   & D & 13\\
\enddata
\tablecomments{We list confirmed and candidate $\sigma$~Ori members with available photometry available from 
our monitoring campaign. Column two provides alternate identifications based on previous membership surveys of 
the cluster (we omit studies that are primarily follow-up). ``S Ori,'' objects are from  
\citet{1999ApJ...521..671B,2001ApJ...556..830B}, ``r'' and ``4771-'' ids are from 
\citet{1996PhDT........63W}'s x-ray-selected source list, ``SO'' objects are from 
\citet{2007ApJ...662.1067H}'s list of candidate cluster members, and Mayrit numbers are from the Mayrit 
catalog compiled by \citet{2008A&A...478..667C}. All other ids correspond to the author(s)'s initial followed 
by their own numbering system: SWW numbers refer to the survey of \citet{2004AJ....128.2316S}; KJN is the 
survey of \citet{2005MNRAS.356...89K}, SE is \citet{2004A&A...419..249S}, M refers to 
\citet{2008MNRAS.385.2210M}, B is for \citet{2005MNRAS.356.1583B}, and F is \citet{2006A&A...446..501F}. 
Source HH446 is from from \citet{2004ApJ...606..353A}. The six objects without ids were found in this work 
(see $\S$7.1.1). We also note that several of the objects identified in \citet{2004AJ....128.2316S} are 
duplicated in their list and thus only included once here (SWW103 is SWW207; SWW126 is SWW162). Based on the 
finder chart provided by \citet{1999ApJ...521..671B}, we also conclude that S Ori 26 is incorrectly 
identified by \citet{2009A&A...505.1115L}; the actual object is their UGCS~J05:39:15.76-02:38:26.3, a 
proper-motion selected $\sigma$~Ori member. The membership evidence column refers to photometric and 
spectroscopic measurements that confirm the object's youth and/or cluster membership, e.g., H$\alpha$ or Na 
emission lines indicative of low gravity, forbidden emission lines(OI, NII, SII; ``FL''), presence of Li 
absorption, radial velocity (``$v_r$'') consistent with the $\sigma$ Ori mean \citep[$27<v_r<37$ km~s$^{-1}$; 
][]{2006MNRAS.371L...6J}, infrared excess from {\em Spitzer} indicative of a disk (``D''), and proper motion 
(``(PM)'') consistent with $sigma$~Ori membership (we have applied parentheses since this latter criterion is 
not enough to definitively select members but is useful for eliminating some non-members).  Disks noted as 
``EV'' or ``TD'' refer to evolved and transitional disks, respectively, as classified by 
\citet{2007ApJ...662.1067H}. We note that while \citet{2008ApJ...688..362L} did not explicitly list which 
stars have infrared excesses indicative of disks, we have used their photometry (derived from Spitzer images 
acquired by \citet{2007ApJ...662.1067H} and \citet{2008ApJ...672L..49S}) to identify disk-bearing candidates 
($\S$7.4). Unsurprisingly, we recover all but one of the disks already identified by 
\citet{2007A&A...470..903C} and \citet{2007ApJ...662.1067H} from the same images. We therefore do not include
\citet{2008ApJ...688..362L} in our disk references, except in the case of the one newly-identified 
disk-bearing object, 2MASS J05375398-0249545. We do not list objects that are saturated in our photometry or 
were presented in the above references but later determined to be non-members. Objects with evidence both for 
and against membership are listed with an ``NM'' along with the the specific criterion suggesting 
non-membership. References for this information are as follows:
$^1$\citet{2003A&A...404..171B}, $^2$\citet{2008A&A...488..167S}, $^3$\citet{2002A&A...384..937Z}, $^4$\citet{2005MNRAS.356...89K},
$^5$\citet{2003ApJ...592..266M}, $^6$\citet{2006A&A...445..143C}, $^7$\citet{2008A&A...491..515C}, $^8$\citet{2007A&A...470..903C},
$^9$\citet{2008MNRAS.385.2210M}, $^{10}$\citet{2005MNRAS.356.1583B}, $^{11}$\citet{2004A&A...419..249S},
$^{12}$\citet{2009A&A...505.1115L}, $^{13}$\citet{2007ApJ...662.1067H}, $^{14}$\citet{2008ApJ...688..362L}, $^{15}$\citet{2004A&A...424..857C},
$^{16}$\citet{2001A&A...367..218B}, $^{17}$\citet{2003A&A...408..663Z}. }
\end{deluxetable}
\clearpage
\end{landscape}

\tabletypesize{\normalsize}
\LongTables
\begin{deluxetable}{cccccc}
\tabletypesize{\scriptsize}
\tablecolumns{10}
\tablewidth{0pt}
\tablecaption{\bf Photometry of confirmed and candidate cluster members in the sample}
\tablehead{
\colhead{Object}  & \colhead{$R$} & \colhead{$I$} & \colhead{$J$} & \colhead{$H$} & \colhead{$K$}  \\
}
\startdata
2MASS J05372806-0236065 & 16.37$\pm$0.03 & 15.10$\pm$0.03 & 13.74$\pm$0.03 & 13.08$\pm$0.03 & 12.80$\pm$0.03 \\
2MASS J05373648-0241567 & 19.88$\pm$0.07 & 17.90$\pm$0.05 & 15.47$\pm$0.05 & 14.94$\pm$0.05 & 14.56$\pm$0.10 \\
2MASS J05373784-0245442 & 15.22$\pm$0.03 & 14.00$\pm$0.03 & 12.69$\pm$0.03 & 11.95$\pm$0.02 & 11.72$\pm$0.03 \\
2MASS J05375161-0235257 & 14.49$\pm$0.03 & 13.27$\pm$0.03 & 11.89$\pm$0.03 & 11.17$\pm$0.02 & 10.98$\pm$0.02 \\
2MASS J05375206-0236046 & 19.23$\pm$0.05 & 17.26$\pm$0.04 & 15.14$\pm$0.04 & 14.55$\pm$0.04 & 14.20$\pm$0.06 \\
2MASS J05375398-0249545 & 18.17$\pm$0.04 & 16.77$\pm$0.03 & 14.52$\pm$0.03 & 13.25$\pm$0.02 & 12.46$\pm$0.03 \\
2MASS J05375404-0244407 & 15.85$\pm$0.03 & 14.49$\pm$0.03 & 13.02$\pm$0.03 & 12.34$\pm$0.03 & 12.10$\pm$0.02 \\
2MASS J05375486-0241092 & 17.08$\pm$0.04 & 15.36$\pm$0.04 & 13.50$\pm$0.03 & 12.90$\pm$0.03 & 12.64$\pm$0.03 \\
2MASS J05375745-0238444 & 18.12$\pm$0.05 & 16.25$\pm$0.04 & 14.23$\pm$0.03 & 13.63$\pm$0.03 & 13.29$\pm$0.03 \\
2MASS J05375840-0241262 & 17.19$\pm$0.04 & 15.32$\pm$0.04 & 13.29$\pm$0.03 & 12.70$\pm$0.02 & 12.42$\pm$0.03 \\
2MASS J05375970-0251033 & 12.80$\pm$0.10 & 12.05$\pm$0.03 & 10.69$\pm$0.03 &  9.87$\pm$0.02 &  9.71$\pm$0.02 \\
2MASS J05380055-0245097 & 16.23$\pm$0.04 & 14.52$\pm$0.04 & 12.73$\pm$0.03 & 12.08$\pm$0.02 & 11.82$\pm$0.02 \\
2MASS J05380107-0245379 & 16.16$\pm$0.04 & 14.47$\pm$0.04 & 12.41$\pm$0.03 & 11.62$\pm$0.02 & 11.12$\pm$0.02 \\
2MASS J05380552-0235571 & 19.61$\pm$0.06 & 17.69$\pm$0.04 & 15.28$\pm$0.04 & 14.77$\pm$0.06 & 14.24$\pm$0.07 \\
2MASS J05380826-0235562 & 15.18$\pm$0.03 & 13.86$\pm$0.03 & 12.14$\pm$0.03 & 11.38$\pm$0.02 & 11.05$\pm$0.02 \\
2MASS J05380994-0251377 & 15.24$\pm$0.03 & 13.88$\pm$0.03 & 12.34$\pm$0.02 & 11.57$\pm$0.02 & 11.24$\pm$0.02 \\
2MASS J05381175-0245012 & 13.16$\pm$0.12 & 12.22$\pm$0.03 & 10.47$\pm$0.03 &  9.72$\pm$0.02 &  9.43$\pm$0.02 \\
2MASS J05381315-0245509 & 14.66$\pm$0.03 & 13.51$\pm$0.03 & 12.07$\pm$0.03 & 11.26$\pm$0.02 & 10.77$\pm$0.02 \\
2MASS J05381330-0251329 & 18.54$\pm$0.05 & 16.62$\pm$0.04 & 14.57$\pm$0.03 & 14.00$\pm$0.03 & 13.63$\pm$0.04 \\
2MASS J05381589-0234412 & 14.06$\pm$0.02 & 13.37$\pm$0.02 & 12.37$\pm$0.03 & 11.75$\pm$0.02 & 11.59$\pm$0.02 \\
2MASS J05381610-0238049 & 16.85$\pm$0.04 & 15.22$\pm$0.04 & 13.58$\pm$0.03 & 12.88$\pm$0.02 & 12.61$\pm$0.03 \\
2MASS J05381741-0240242 & 19.24$\pm$0.05 & 17.22$\pm$0.05 & 14.83$\pm$0.03 & 14.31$\pm$0.04 & 14.09$\pm$0.05 \\
2MASS J05381778-0240500 & 16.77$\pm$0.04 & 15.00$\pm$0.04 & 13.20$\pm$0.03 & 12.58$\pm$0.02 & 12.24$\pm$0.02 \\
2MASS J05381824-0248143 & 15.23$\pm$0.03 & 14.18$\pm$0.03 & 12.76$\pm$0.03 & 12.02$\pm$0.02 & 11.80$\pm$0.02 \\
2MASS J05381834-0235385 & 20.39$\pm$0.08 & 18.24$\pm$0.05 & 15.45$\pm$0.04 & 14.83$\pm$0.05 & 14.49$\pm$0.08 \\
2MASS J05381886-0251388 & 15.71$\pm$0.03 & 14.25$\pm$0.03 & 12.81$\pm$0.02 & 12.04$\pm$0.02 & 11.73$\pm$0.02 \\
2MASS J05381914-0235279 & 14.26$\pm$0.02 & 13.46$\pm$0.02 & 12.31$\pm$0.03 & 11.57$\pm$0.02 & 11.39$\pm$0.02 \\
2MASS J05382021-0238016 & 16.06$\pm$0.04 & 14.33$\pm$0.04 & 12.58$\pm$0.03 & 11.86$\pm$0.02 & 11.61$\pm$0.02 \\
2MASS J05382050-0234089 & 17.00$\pm$0.06 & 14.55$\pm$0.05 & 12.65$\pm$0.03 & 11.92$\pm$0.02 & 11.65$\pm$0.02 \\
2MASS J05382088-0246132 & 19.43$\pm$0.06 & 17.46$\pm$0.04 & 15.19$\pm$0.04 & 14.57$\pm$0.05 & 14.16$\pm$0.08 \\
2MASS J05382089-0251280 & 19.13$\pm$0.05 & 17.09$\pm$0.05 & 14.78$\pm$0.03 & 14.21$\pm$0.03 & 13.87$\pm$0.05 \\
2MASS J05382307-0236493 & 17.14$\pm$0.04 & 15.65$\pm$0.03 & 13.80$\pm$0.03 & 13.17$\pm$0.03 & 12.78$\pm$0.02 \\
2MASS J05382332-0244142 & 16.86$\pm$0.04 & 15.17$\pm$0.04 & 13.46$\pm$0.03 & 12.85$\pm$0.02 & 12.56$\pm$0.02 \\
2MASS J05382354-0241317 & 16.89$\pm$0.04 & 15.13$\pm$0.04 & 13.29$\pm$0.03 & 12.74$\pm$0.03 & 12.40$\pm$0.02 \\
S Ori J053825.1-024802 & 21.64$\pm$0.29 & 20.31$\pm$0.09 &      -           &        -         &        -         \\
2MASS J05382543-0242412 & 18.77$\pm$0.05 & 16.96$\pm$0.04 & 14.88$\pm$0.03 & 14.16$\pm$0.04 & 13.57$\pm$0.03 \\
2MASS J05382557-0248370 & 22.38$\pm$0.38 & 20.03$\pm$0.09 & 16.67$\pm$0.11 & 16.02$\pm$0.13 & 15.59$\pm$0.21 \\
2MASS J05382623-0240413 & 19.03$\pm$0.05 & 17.05$\pm$0.04 & 14.91$\pm$0.04 & 14.28$\pm$0.04 & 13.92$\pm$0.06 \\
2MASS J05382684-0238460 & 18.12$\pm$0.05 & 16.17$\pm$0.04 & 14.11$\pm$0.04 & 13.48$\pm$0.03 & 13.21$\pm$0.04 \\
2MASS J05382725-0245096 & 13.85$\pm$0.03 & 12.95$\pm$0.02 & 11.96$\pm$0.03 & 10.79$\pm$0.03 &  9.94$\pm$0.03 \\
2MASS J05382750-0235041 & 15.99$\pm$0.04 & 14.45$\pm$0.04 & 12.83$\pm$0.03 & 12.11$\pm$0.02 & 11.86$\pm$0.03 \\
2MASS J05382774-0243009 & 15.04$\pm$0.03 & 13.67$\pm$0.03 & 12.19$\pm$0.03 & 11.45$\pm$0.02 & 11.29$\pm$0.02 \\
2MASS J05382848-0246170 & 16.33$\pm$0.03 & 15.06$\pm$0.03 & 13.82$\pm$0.03 & 13.20$\pm$0.03 & 12.94$\pm$0.03 \\
2MASS J05382896-0248473 & 19.05$\pm$0.05 & 17.06$\pm$0.05 & 14.82$\pm$0.04 & 14.28$\pm$0.04 & 13.88$\pm$0.06 \\
2MASS J05383141-0236338 & 15.31$\pm$0.04 & 13.89$\pm$0.03 & 12.17$\pm$0.03 & 11.47$\pm$0.02 & 10.99$\pm$0.03 \\
2MASS J05383157-0235148 & 14.98$\pm$0.03 & 13.83$\pm$0.03 & 11.52$\pm$0.03 & 10.71$\pm$0.02 & 10.35$\pm$0.02 \\
2MASS J05383160-0251268 & 14.54$\pm$0.03 & 13.53$\pm$0.02 & 12.11$\pm$0.03 & 11.18$\pm$0.02 & 10.98$\pm$0.02 \\
2MASS J05383284-0235392 & 13.60$\pm$0.04 & 12.71$\pm$0.02 & 11.54$\pm$0.03 & 10.90$\pm$0.02 & 10.73$\pm$0.03 \\
2MASS J05383302-0239279 & 17.84$\pm$0.04 & 16.23$\pm$0.04 & 14.59$\pm$0.03 & 14.02$\pm$0.03 & 13.70$\pm$0.04 \\
2MASS J05383335-0236176 & 14.77$\pm$0.03 & 13.45$\pm$0.03 & 12.05$\pm$0.03 & 11.29$\pm$0.02 & 11.11$\pm$0.03 \\
2MASS J05383388-0245078 & 18.01$\pm$0.04 & 16.15$\pm$0.04 & 14.25$\pm$0.03 & 13.68$\pm$0.03 & 13.35$\pm$0.04 \\
2MASS J05383405-0236375 & 15.37$\pm$0.04 & 13.77$\pm$0.04 & 11.98$\pm$0.03 & 11.33$\pm$0.02 & 11.08$\pm$0.03 \\
2MASS J05383460-0241087 & 16.38$\pm$0.04 & 14.86$\pm$0.04 & 13.10$\pm$0.03 & 12.45$\pm$0.02 & 12.12$\pm$0.03 \\
2MASS J05383669-0244136 & 16.13$\pm$0.04 & 14.35$\pm$0.04 & 12.54$\pm$0.03 & 11.89$\pm$0.03 & 11.62$\pm$0.03 \\
2MASS J05383745-0250236 & 16.43$\pm$0.04 & 14.63$\pm$0.04 & 12.81$\pm$0.03 & 12.18$\pm$0.02 & 11.92$\pm$0.02 \\
2MASS J05383858-0241558 & 18.33$\pm$0.05 & 16.48$\pm$0.04 & 14.56$\pm$0.03 & 13.96$\pm$0.03 & 13.65$\pm$0.04 \\
2MASS J05383902-0245321 & 15.77$\pm$0.04 & 14.39$\pm$0.03 & 12.91$\pm$0.03 & 12.20$\pm$0.02 & 11.89$\pm$0.03 \\
2MASS J05383922-0253084 & 14.72$\pm$0.03 & 13.83$\pm$0.02 & 12.70$\pm$0.03 & 12.04$\pm$0.03 & 11.87$\pm$0.02 \\
2MASS J05385317-0243528 & 14.93$\pm$0.03 & 13.78$\pm$0.02 & 12.23$\pm$0.03 & 11.51$\pm$0.03 & 11.30$\pm$0.03 \\
2MASS J05385382-0244588 & 20.09$\pm$0.06 & 17.93$\pm$0.04 & 15.45$\pm$0.04 & 14.94$\pm$0.05 & 14.59$\pm$0.09 \\
2MASS J05385492-0228583 & 17.18$\pm$0.04 & 15.51$\pm$0.03 & 13.80$\pm$0.03 & 13.20$\pm$0.03 & 12.87$\pm$0.03 \\
2MASS J05385492-0240337 & 20.90$\pm$0.09 & 18.75$\pm$0.04 & 15.92$\pm$0.07 & 15.17$\pm$0.06 & 14.71$\pm$0.11 \\
2MASS J05385542-0241208 & 19.94$\pm$0.06 & 18.09$\pm$0.04 & 15.62$\pm$0.10 & 14.84$\pm$0.05 & 13.97$\pm$0.06 \\
2MASS J05385623-0231153 & 15.36$\pm$0.02 & 14.58$\pm$0.02 & 13.42$\pm$0.03 & 12.77$\pm$0.02 & 12.52$\pm$0.03 \\
2MASS J05385922-0233514 & 16.31$\pm$0.03 & 14.95$\pm$0.03 & 12.89$\pm$0.03 & 11.98$\pm$0.02 & 11.40$\pm$0.03 \\
2MASS J05390052-0239390 & 12.79$\pm$0.02 & 12.46$\pm$0.01 & 11.66$\pm$0.03 & 11.22$\pm$0.02 & 11.11$\pm$0.02 \\
2MASS J05390115-0236388 & 16.73$\pm$0.03 & 15.17$\pm$0.03 & 13.52$\pm$0.03 & 12.89$\pm$0.03 & 12.61$\pm$0.03 \\
2MASS J05390193-0235029 & 17.51$\pm$0.03 & 16.13$\pm$0.03 & 14.45$\pm$0.04 & 13.38$\pm$0.03 & 12.61$\pm$0.03 \\
2MASS J05390276-0229558 & 15.80$\pm$0.03 & 14.27$\pm$0.03 & 12.61$\pm$0.03 & 12.00$\pm$0.02 & 11.69$\pm$0.02 \\
S Ori J053903.2-023020 & 22.49$\pm$0.35 & 20.68$\pm$0.06 &      -           &        -         &        -         \\
2MASS J05390357-0246269 & 15.86$\pm$0.03 & 14.34$\pm$0.03 & 12.84$\pm$0.03 & 12.12$\pm$0.02 & 11.86$\pm$0.03 \\
2MASS J05390449-0238353 & 18.95$\pm$0.04 & 16.99$\pm$0.04 & 14.77$\pm$0.04 & 14.19$\pm$0.03 & 13.80$\pm$0.04 \\
2MASS J05390458-0241493 & 15.93$\pm$0.02 & 14.87$\pm$0.02 & 13.96$\pm$0.04 & 12.91$\pm$0.04 & 12.22$\pm$0.04 \\
2MASS J05390524-0233005 & 16.56$\pm$0.03 & 15.01$\pm$0.03 & 13.39$\pm$0.03 & 12.72$\pm$0.02 & 12.46$\pm$0.03 \\
2MASS J05390540-0232303 & 13.15$\pm$0.02 & 12.55$\pm$0.01 & 11.55$\pm$0.03 & 10.86$\pm$0.02 & 10.67$\pm$0.02 \\
2MASS J05390759-0228234 & 15.83$\pm$0.03 & 14.42$\pm$0.03 & 12.88$\pm$0.03 & 12.14$\pm$0.02 & 11.96$\pm$0.03 \\
2MASS J05390760-0232391 & 13.54$\pm$0.09 & 12.82$\pm$0.03 & 11.30$\pm$0.03 & 10.57$\pm$0.02 & 10.26$\pm$0.02 \\
2MASS J05390808-0228447 & 17.59$\pm$0.04 & 15.89$\pm$0.03 & 14.14$\pm$0.03 & 13.52$\pm$0.03 & 13.25$\pm$0.04 \\
2MASS J05390821-0232284 & 17.59$\pm$0.04 & 15.80$\pm$0.04 & 13.80$\pm$0.03 & 13.25$\pm$0.03 & 12.92$\pm$0.03 \\
2MASS J05390878-0231115 & 16.62$\pm$0.03 & 15.04$\pm$0.03 & 13.04$\pm$0.03 & 12.16$\pm$0.02 & 11.70$\pm$0.02 \\
2MASS J05390894-0239579 & 19.53$\pm$0.05 & 17.39$\pm$0.04 & 14.65$\pm$0.03 & 14.13$\pm$0.04 & 13.74$\pm$0.05 \\
2MASS J05391001-0228116 & 17.68$\pm$0.03 & 16.13$\pm$0.03 & 14.60$\pm$0.03 & 14.00$\pm$0.04 & 13.78$\pm$0.05 \\
2MASS J05391003-0242425 & 15.18$\pm$0.02 & 14.30$\pm$0.02 & 12.97$\pm$0.03 & 12.21$\pm$0.03 & 11.97$\pm$0.02 \\
S Ori J053910.8-023715 & 22.60$\pm$0.37 & 20.82$\pm$0.06 &      -           &        -         &        -         \\
2MASS J05391139-0233327 & 18.31$\pm$0.04 & 16.48$\pm$0.04 & 14.45$\pm$0.03 & 13.93$\pm$0.03 & 13.57$\pm$0.04 \\
2MASS J05391151-0231065 & 14.04$\pm$0.02 & 13.11$\pm$0.02 & 11.99$\pm$0.03 & 11.19$\pm$0.02 & 10.73$\pm$0.02 \\
2MASS J05391163-0236028 & 13.71$\pm$0.09 & 12.93$\pm$0.03 & 11.62$\pm$0.03 & 10.97$\pm$0.03 & 10.75$\pm$0.02 \\
2MASS J05391232-0230064 & 16.50$\pm$0.04 & 14.66$\pm$0.04 & 12.61$\pm$0.03 & 12.05$\pm$0.03 & 11.73$\pm$0.02 \\
2MASS J05391308-0237509 & 19.44$\pm$0.05 & 17.52$\pm$0.04 & 15.24$\pm$0.04 & 14.75$\pm$0.04 & 14.31$\pm$0.07 \\
2MASS J05391346-0237391 & 16.89$\pm$0.04 & 15.22$\pm$0.03 & 13.41$\pm$0.03 & 12.77$\pm$0.02 & 12.50$\pm$0.03 \\
2MASS J05391447-0228333 & 16.37$\pm$0.03 & 14.89$\pm$0.03 & 13.34$\pm$0.03 & 12.65$\pm$0.03 & 12.34$\pm$0.03 \\
2MASS J05391510-0240475 & 18.85$\pm$0.04 & 16.88$\pm$0.04 & 14.67$\pm$0.03 & 14.04$\pm$0.03 & 13.66$\pm$0.04 \\
2MASS J05391576-0238262 & 19.09$\pm$0.08 & 17.21$\pm$0.01 & 14.95$\pm$0.06 & 14.38$\pm$0.06 & 14.09$\pm$0.06 \\
2MASS J05391582-0236507 & 16.45$\pm$0.03 & 14.93$\pm$0.03 & 13.25$\pm$0.03 & 12.54$\pm$0.03 & 12.22$\pm$0.03 \\
2MASS J05391699-0241171 & 17.56$\pm$0.03 & 15.99$\pm$0.03 & 14.29$\pm$0.03 & 13.63$\pm$0.02 & 13.37$\pm$0.04 \\
2MASS J05391883-0230531 & 13.23$\pm$0.02 & 12.55$\pm$0.02 & 11.40$\pm$0.03 & 10.64$\pm$0.03 & 10.34$\pm$0.02 \\
2MASS J05392023-0238258 & 17.44$\pm$0.04 & 15.61$\pm$0.04 & 13.61$\pm$0.03 & 13.04$\pm$0.03 & 12.78$\pm$0.02 \\
2MASS J05392097-0230334 & 17.52$\pm$0.04 & 15.59$\pm$0.04 & 13.29$\pm$0.03 & 12.75$\pm$0.03 & 12.44$\pm$0.03 \\
2MASS J05392174-0244038 & 13.25$\pm$0.09 & 12.58$\pm$0.03 & 11.10$\pm$0.03 & 10.40$\pm$0.02 & 10.22$\pm$0.02 \\
2MASS J05392224-0245524 & 19.03$\pm$0.04 & 17.22$\pm$0.04 & 15.32$\pm$0.04 & 14.84$\pm$0.05 & 14.41$\pm$0.08 \\
2MASS J05392286-0233330 & 15.36$\pm$0.03 & 14.16$\pm$0.03 & 12.83$\pm$0.03 & 12.13$\pm$0.02 & 11.87$\pm$0.03 \\
2MASS J05392319-0246557 & 19.31$\pm$0.05 & 17.35$\pm$0.04 & 15.33$\pm$0.04 & 14.78$\pm$0.04 & 14.34$\pm$0.07 \\
2MASS J05392341-0240575 & 21.92$\pm$0.20 & 19.47$\pm$0.05 & 16.73$\pm$0.13 & 15.92$\pm$0.12 & 15.55$\pm$0.21 \\
2MASS J05392435-0234013 & 15.52$\pm$0.03 & 14.27$\pm$0.03 & 12.98$\pm$0.03 & 12.27$\pm$0.03 & 12.06$\pm$0.02 \\
2MASS J05392519-0238220 & 13.84$\pm$0.07 & 13.08$\pm$0.02 & 11.31$\pm$0.03 & 10.45$\pm$0.02 & 10.00$\pm$0.02 \\
2MASS J05392524-0227479 & 18.42$\pm$0.03 & 16.94$\pm$0.03 & 15.55$\pm$0.04 & 14.79$\pm$0.05 & 14.56$\pm$0.08 \\
2MASS J05392560-0238436 & 18.23$\pm$0.03 & 17.29$\pm$0.02 & 15.25$\pm$0.04 & 14.28$\pm$0.03 & 13.65$\pm$0.04 \\
2MASS J05392561-0234042 & 16.71$\pm$0.04 & 15.00$\pm$0.04 & 13.20$\pm$0.03 & 12.54$\pm$0.02 & 12.25$\pm$0.05 \\
2MASS J05392633-0228376 & 16.94$\pm$0.04 & 15.28$\pm$0.03 & 13.50$\pm$0.03 & 12.84$\pm$0.02 & 12.56$\pm$0.02 \\
2MASS J05392677-0242583 & 17.03$\pm$0.03 & 15.46$\pm$0.03 & 13.18$\pm$0.03 & 12.40$\pm$0.03 & 12.12$\pm$0.02 \\
2MASS J05392685-0236561 & 20.00$\pm$0.06 & 17.97$\pm$0.04 & 15.46$\pm$0.04 & 14.84$\pm$0.05 & 14.49$\pm$0.07 \\
2MASS J05393056-0238270 & 16.66$\pm$0.03 & 15.29$\pm$0.03 & 13.81$\pm$0.03 & 13.18$\pm$0.03 & 12.95$\pm$0.03 \\
2MASS J05393234-0227571 & 13.25$\pm$0.04 & 12.50$\pm$0.02 & 11.18$\pm$0.02 & 10.50$\pm$0.02 & 10.33$\pm$0.02 \\
2MASS J05393432-0238468 & 19.20$\pm$0.05 & 17.19$\pm$0.04 & 14.76$\pm$0.03 & 14.19$\pm$0.04 & 13.79$\pm$0.05 \\
2MASS J05393673-0231588 & 18.73$\pm$0.03 & 17.26$\pm$0.03 & 15.71$\pm$0.05 & 15.04$\pm$0.06 & 14.76$\pm$0.09 \\
2MASS J05393759-0244304 & 18.63$\pm$0.05 & 16.63$\pm$0.04 & 14.38$\pm$0.03 & 13.82$\pm$0.03 & 13.38$\pm$0.03 \\
2MASS J05393931-0232252 & 17.37$\pm$0.04 & 15.52$\pm$0.04 & 13.44$\pm$0.03 & 12.90$\pm$0.02 & 12.53$\pm$0.03 \\
2MASS J05393982-0231217 & 13.79$\pm$0.07 & 13.06$\pm$0.02 & 11.84$\pm$0.03 & 10.90$\pm$0.02 & 10.22$\pm$0.02 \\
2MASS J05393982-0233159 & 15.90$\pm$0.03 & 14.84$\pm$0.02 & 12.22$\pm$0.03 & 10.96$\pm$0.02 & 10.07$\pm$0.02 \\
2MASS J05393998-0243097 & 12.52$\pm$0.02 & 12.29$\pm$0.01 & 10.65$\pm$0.03 &  9.92$\pm$0.02 &  9.53$\pm$0.02 \\
2MASS J05394057-0239123 & 18.87$\pm$0.04 & 17.27$\pm$0.03 & 15.40$\pm$0.05 & 14.67$\pm$0.05 & 14.41$\pm$0.08 \\
2MASS J05394318-0232433 & 16.31$\pm$0.03 & 14.74$\pm$0.03 & 13.03$\pm$0.03 & 12.30$\pm$0.02 & 11.91$\pm$0.02 \\
2MASS J05394411-0231092 & 13.18$\pm$0.11 & 12.61$\pm$0.03 & 11.21$\pm$0.03 & 10.51$\pm$0.02 & 10.33$\pm$0.02 \\
2MASS J05394433-0233027 & 18.33$\pm$0.04 & 16.47$\pm$0.04 & 14.29$\pm$0.03 & 13.72$\pm$0.03 & 13.37$\pm$0.04 \\
2MASS J05394725-0241359 & 17.27$\pm$0.02 & 16.37$\pm$0.02 & 15.09$\pm$0.04 & 14.24$\pm$0.03 & 14.00$\pm$0.06 \\
2MASS J05394770-0236230 & 16.67$\pm$0.03 & 15.12$\pm$0.03 & 13.47$\pm$0.03 & 12.77$\pm$0.02 & 12.53$\pm$0.03 \\
2MASS J05394784-0232248 & 13.16$\pm$0.12 & 12.62$\pm$0.03 & 10.97$\pm$0.03 & 10.29$\pm$0.02 & 10.08$\pm$0.02 \\
2MASS J05394799-0240320 & 15.21$\pm$0.03 & 13.85$\pm$0.03 & 12.43$\pm$0.03 & 11.65$\pm$0.02 & 11.43$\pm$0.02 \\
2MASS J05394806-0245571 & 15.45$\pm$0.03 & 14.15$\pm$0.03 & 12.92$\pm$0.03 & 12.28$\pm$0.02 & 12.03$\pm$0.02 \\
2MASS J05394826-0229144 & 20.81$\pm$0.10 & 18.79$\pm$0.04 & 16.42$\pm$0.09 & 15.59$\pm$0.10 & 15.19$\pm$0.14 \\
2MASS J05394891-0229110 & 16.04$\pm$0.04 & 14.61$\pm$0.03 & 13.28$\pm$0.03 & 12.59$\pm$0.03 & 12.30$\pm$0.03 \\
2MASS J05395038-0243307 & 14.03$\pm$0.06 & 12.99$\pm$0.03 & 11.77$\pm$0.03 & 10.98$\pm$0.02 & 10.77$\pm$0.02 \\
2MASS J05395056-0234137 & 17.18$\pm$0.04 & 15.48$\pm$0.03 & 13.68$\pm$0.03 & 13.00$\pm$0.03 & 12.73$\pm$0.03 \\
2MASS J05395236-0236147 & 15.70$\pm$0.03 & 14.34$\pm$0.03 & 12.89$\pm$0.03 & 12.19$\pm$0.02 & 11.94$\pm$0.03 \\
2MASS J05395248-0232023 & 13.32$\pm$0.02 & 12.65$\pm$0.01 & 11.51$\pm$0.03 & 10.88$\pm$0.02 & 10.66$\pm$0.03 \\
2MASS J05395313-0243083 & 13.14$\pm$0.05 & 12.24$\pm$0.02 & 11.13$\pm$0.03 & 10.47$\pm$0.02 & 10.27$\pm$0.03 \\
2MASS J05395313-0230294 & 20.33$\pm$0.07 & 18.41$\pm$0.04 & 16.20$\pm$0.08 & 15.82$\pm$0.12 & 15.56$\pm$0.23 \\
2MASS J05395362-0233426 & 15.59$\pm$0.03 & 14.39$\pm$0.03 & 12.82$\pm$0.03 & 12.06$\pm$0.03 & 11.59$\pm$0.03 \\
2MASS J05395433-0237189 & 19.13$\pm$0.05 & 17.14$\pm$0.04 & 14.75$\pm$0.03 & 14.21$\pm$0.04 & 13.80$\pm$0.05 \\
2MASS J05395645-0238034 & 17.01$\pm$0.04 & 15.28$\pm$0.04 & 13.35$\pm$0.03 & 12.79$\pm$0.02 & 12.43$\pm$0.03 \\
2MASS J05395753-0232120 & 16.82$\pm$0.04 & 15.10$\pm$0.04 & 13.31$\pm$0.03 & 12.69$\pm$0.02 & 12.36$\pm$0.02 \\
2MASS J05400338-0229014 & 13.94$\pm$0.05 & 12.94$\pm$0.02 & 11.72$\pm$0.03 & 11.03$\pm$0.02 & 10.81$\pm$0.02 \\
2MASS J05400453-0236421 & 19.95$\pm$0.05 & 17.92$\pm$0.04 & 15.30$\pm$0.05 & 14.81$\pm$0.05 & 14.27$\pm$0.07 \\
2MASS J05400525-0230522 & 17.70$\pm$0.04 & 15.92$\pm$0.04 & 13.95$\pm$0.03 & 13.37$\pm$0.03 & 13.07$\pm$0.03 \\
2MASS J05400708-0232446 & 16.84$\pm$0.04 & 15.17$\pm$0.03 & 13.42$\pm$0.03 & 12.81$\pm$0.02 & 12.54$\pm$0.03 \\
2MASS J05400867-0232432 & 15.66$\pm$0.04 & 13.78$\pm$0.04 & 11.77$\pm$0.03 & 11.15$\pm$0.02 & 10.85$\pm$0.02 \\
2MASS J05400889-0233336 & 14.49$\pm$0.28 & 13.39$\pm$0.11 & 11.50$\pm$0.03 & 10.55$\pm$0.02 &  9.91$\pm$0.02 \\
\enddata
\tablecomments{We list $R$ and $I$-band photometry derived from our data and calibrated to the Cousins band, 
along with $J$, $H$, and $K$ magnitudes taken from the 2MASS survey. Several brown dwarfs were too faint to be 
detected in 2MASS and hence we do not list values for these longer wavelength bands.}
\end{deluxetable}

\clearpage
\begin{deluxetable}{ccccccc}
\tabletypesize{\scriptsize}
\tablecolumns{10}
\tablewidth{0pt}
\tablecaption{\bf Objects with detected periodic variability}
\tablehead{
\colhead{Object} & \colhead{Period [d]} & \colhead{error} & \colhead{Amplitude [mag]} & \colhead{error} & 
\colhead{Variable Type} & \colhead{Member?} 
}
\startdata
2MASS J05372806-0236065 & 10.47 & 1.12 & 0.007 & 0.001 & S & M \\
2MASS J05373648-0241567 & 0.79  & 0.01 & 0.035 & 0.004 & S & Y \\
2MASS J05373784-0245442 & 11.52 & 0.20 & 0.021 & 0.001 & S & M \\
2MASS J05373790-0236085 & 10.00 & 0.53 & 0.004 & 0.001 & S & M$^1$\\
 CTIO J05373835-0243516 & 0.13  & 0.01 & 0.275 & 0.007 & EB? & N \\
 CTIO J05373954-0238446 & 0.61  & 0.01 & 0.036 & 0.006 & S  & N \\
2MASS J05374413-0235198 & 0.63  & 0.01 & 0.028 & 0.005 & U & M$^2$\\
 CTIO J05374598-0238011 & 0.12  & 0.01 & 0.101 & 0.005 & O & N \\
2MASS J05375206-0236046 & 2.03  & 0.05 & 0.022 & 0.002 & U & M \\
2MASS J05375285-0251096 & 10.78 & 0.64 & 0.007 & 0.001 & S & N \\
2MASS J05375404-0244407 & 1.90  & 0.02 & 0.010 & 0.001 & S & M\\
2MASS J05375486-0241092 & 2.98  & 0.01 & 0.028 & 0.001 & S & M\\
2MASS J05375745-0238444 & 0.61  & 0.01 & 0.036 & 0.014 & U & Y\\
2MASS J05380055-0245097 & 1.28  & 0.01 & 0.025 & 0.001 & S & Y\\
2MASS J05380655-0250280 & 0.05  & 0.01 & 0.006 & 0.003 & S & N \\
2MASS J05380678-0245400 & 8.17  & 0.33 & 0.008 & 0.001 & S & N \\
2MASS J05381265-0236378 & 2.31  & 0.06 & 0.023 & 0.005 & S & M$^3$ \\
2MASS J05381330-0251329 & 2.58  & 0.03 & 0.017 & 0.001 & S & Y\\
 CTIO J05381348-0236118 & 2.10  & 0.01 & 0.310 & 0.001 & EB & N \\
2MASS J05381367-0235385 & 3.64  & 0.01 & 0.450 & 0.001 & EB & N \\
2MASS J05381522-0236491 & 9.70  & 0.63 & 0.007 & 0.001 & S & N \\
2MASS J05381610-0238049 & 0.76  & 0.01 & 0.003 & 0.001 & U & Y\\
2MASS J05381680-0246567 & 2.38  & 0.03 & 0.014 & 0.002 & S & N \\
2MASS J05381778-0240500 & 2.41  & 0.03 & 0.008 & 0.001 & U & Y\\
2MASS J05381824-0248143 & 4.47  & 0.05 & 0.013 & 0.001 & S & Y\\
 CTIO J05381870-0246582 & 0.25  & 0.01 & 0.760 & 0.001 & EB & N \\
2MASS J05381886-0251388 & 6.62  & 0.09 & 0.038 & 0.002 & S/U & Y\\
2MASS J05381949-0241224 & 0.11  & 0.01 & 0.275 & 0.026 & S & N \\
2MASS J05382021-0238016 & 0.96  & 0.01 & 0.014 & 0.004 & U & Y\\
 CTIO J05382129-0240318 & 4.64  & 0.36 & 0.350 & 0.036 & EB & N \\
2MASS J05382188-0241039 & 1.00  & 0.01 & 0.650 & 0.001 & O & N \\
2MASS J05382332-0244142 & 0.83  & 0.01 & 0.010 & 0.001 & S & Y\\
2MASS J05382354-0241317 & 1.71  & 0.01 & 0.017 & 0.001 & S & Y\\
2MASS J05382557-0248370 & 0.30  & 0.01 & 0.034 & 0.014 & S & Y\\
2MASS J05382750-0235041 & 2.70  & 0.02 & 0.021 & 0.001 & S & Y\\
2MASS J05382773-0250050 & 10.94 & 1.03 & 0.005 & 0.001 & S & N \\
2MASS J05383284-0235392 & 6.34  & 0.36 & 0.005 & 0.001 & U & Y\\
2MASS J05383302-0239279 & 1.11  & 0.01 & 0.014 & 0.001 & S & M\\
2MASS J05383335-0236176 & 4.41  & 0.07 & 0.011 & 0.001 & U & M\\
2MASS J05383405-0236375 & 1.13  & 0.01 & 0.014 & 0.001 & U & Y\\
2MASS J05383745-0250236 & 1.72  & 0.01 & 0.021 & 0.001 & S & Y\\
2MASS J05383858-0241558 & 1.75  & 0.01 & 0.028 & 0.002 & S & Y\\
 CTIO J05390031-0237059 & 1.34  & 0.01 & 0.253 & 0.039 & S & N \\
2MASS J05390052-0239390 & 3.11  & 0.01 & 0.078 & 0.002 & S & M\\
2MASS J05390524-0233005 & 1.92  & 0.03 & 0.017 & 0.002 & U & Y\\
 CTIO J05390664-0238050 & 0.88  & 0.01 & 0.020 & 0.003 & S & M$^4$ \\
2MASS J05390759-0228234 & 4.92  & 0.05 & 0.025 & 0.001 & S & Y\\
2MASS J05390808-0228447 & 1.68  & 0.02 & 0.016 & 0.002 & S & Y\\
2MASS J05390821-0232284 & 1.79  & 0.01 & 0.019 & 0.001 & S & M\\
2MASS J05390894-0239579 & 2.64  & 0.05 & 0.024 & 0.003 & U & Y\\
2MASS J05390988-0238164 & 9.62  & 0.59 & 0.123 & 0.010 & S & N\\
2MASS J05391139-0233327 & 1.79  & 0.01 & 0.025 & 0.002 & S & Y\\
2MASS J05391163-0236028 & 11.29 & 0.26 & 0.066 & 0.002 & S & Y\\
2MASS J05391232-0230064 & 2.08  & 0.02 & 0.012 & 0.001 & S & M\\
2MASS J05391308-0237509 & 1.96  & 0.04 & 0.024 & 0.004 & U & Y\\
2MASS J05391346-0237391 & 1.42  & 0.01 & 0.009 & 0.001 & S & M\\
2MASS J05391447-0228333 & 3.01  & 0.02 & 0.032 & 0.001 & S & Y\\
2MASS J05391576-0238262 & 0.64  & 0.01 & 0.042 & 0.001 & S & M\\
2MASS J05391582-0236507 & 2.55  & 0.02 & 0.034 & 0.002 & S & Y\\
2MASS J05391699-0241171 & 2.97  & 0.06 & 0.021 & 0.002 & U & M\\
2MASS J05391883-0230531 & 1.82  & 0.01 & 0.051 & 0.001 & S/U & Y\\
2MASS J05392023-0238258 & 0.95  & 0.01 & 0.007 & 0.002 & U & M\\ 
2MASS J05392097-0230334 & 2.92  & 0.04 & 0.036 & 0.003 & S & Y\\  
2MASS J05392286-0233330 & 7.21  & 0.05 & 0.059 & 0.001 & S & Y\\
2MASS J05392435-0234013 & 4.73  & 0.15 & 0.005 & 0.001 & U & M\\
2MASS J05392560-0238436 & 8.18  & 0.42 & 0.124 & 0.014 & U & M\\
2MASS J05392561-0234042 & 3.56  & 0.10 & 0.011 & 0.002 & U & M\\
2MASS J05392633-0228376 & 2.27  & 0.01 & 0.019 & 0.002 & U & Y\\ 
2MASS J05393056-0238270 & 6.28  & 0.19 & 0.008 & 0.001 & S & M\\
2MASS J05393670-0228162 & 0.10  & 0.01 & 2.055 & 0.001 & EB & N\\
2MASS J05393759-0244304 & 2.24  & 0.01 & 0.035 & 0.002 & S & Y\\
2MASS J05393833-0235196 & 1.72  & 0.04 & 0.037 & 0.009 & U & N\\
2MASS J05393931-0232252 & 2.18  & 0.02 & 0.015 & 0.001 & S & M\\
2MASS J05394433-0233027 & 0.90  & 0.01 & 0.050 & 0.002 & S & Y\\
2MASS J05394770-0236230 & 0.93  & 0.01 & 0.029 & 0.001 & S & Y\\
2MASS J05394799-0240320 & 2.76  & 0.01 & 0.065 & 0.001 & S/U & M\\
2MASS J05395038-0243307 & 7.79  & 0.15 & 0.023 & 0.001 & S & M\\
2MASS J05395056-0234137 & 3.17  & 0.02 & 0.023 & 0.001 & S & Y\\
2MASS J05395236-0236147 & 0.93  & 0.01 & 0.015 & 0.001 & S & M\\
2MASS J05395645-0238034 & 1.67  & 0.01 & 0.010 & 0.001 & S & Y\\ 
2MASS J05395753-0232120 & 0.93  & 0.01 & 0.010 & 0.002 & U & Y\\
2MASS J05400338-0229014 & 8.15  & 0.16 & 0.009 & 0.001 & S & M\\
2MASS J05400453-0236421 & 0.76  & 0.01 & 0.027 & 0.010 & S & Y\\
2MASS J05400708-0232446 & 1.55  & 0.01 & 0.014 & 0.001 & S & Y\\
\enddata

\tablecomments{Periodic variables and their 3--$\sigma$ uncertainties. We categorize variability type into 
several types based on light curve appearance (refer to Fig.\ \ref{lightcurves}): likely eclipsing binaries 
(EB); fairly sinusoidal (S), periodic but specific shape unknown due to noise or other features (U), or other 
distinct shape, such as that of a pulsator (O). A few stars marked ``S/U'' are mostly sinusoidal but have 
interesting blip-like features over short time scale. We consider objects to be confirmed cluster members 
(``yes''- Y) if they have either broad H$\alpha$ in emission, Li in absorption, weak alkali absorption lines 
(e.g., Na), forbidden emission lines (e.g., OI, NII, SII), or infrared excess indicative of a disk, as listed 
in Table~1. Objects with only proper motions, only variability, no spectroscopic data, or conflicting 
membership indicators are listed as possible members (``maybe''- M). Non-member classification (N) is reserved 
for targets whose colors are too blue to be sufficiently young for $\sigma$~Ori and whose variability type is 
indicative of a field eclipsing binary or pulsator. The following table entries represent new candidate 
cluster members based on our photometry, with our astrometrically determined coordinates listed in the object 
name: $^1$With $I=13.43\pm0.01$ and $R=13.96\pm0.02$, and a simple periodic light curve, this object is a 
candidate $\sigma$~Ori member; but since its colors fall at the blue edge of the cluster sequence, we 
emphasize that this is a {\em tentative} identification. $^2$This object is a new candidate brown dwarf, with 
$I=18.37\pm0.04$ and $R=20.19\pm0.08$. $^3$This object is also a new candidate brown dwarf, with 
$I=18.27\pm0.05$ and $R=20.25\pm0.08$. $^4$We identify this object as a new candidate $\sigma$ Ori member, 
with $I=17.04\pm0.03$ and $R=18.72\pm0.04$.}
\end{deluxetable}

\clearpage
\begin{deluxetable}{ccccc}
\tabletypesize{\scriptsize}
\tablecolumns{10}
\tablewidth{0pt}
\tablecaption{\bf Objects with detected aperiodic variability}
\tablehead{
\colhead{Object} & \colhead{Peak-to-peak amplitude [mag]}  & \colhead{RMS [mag]}
& \colhead{Member?} & pEW H$\alpha$ [$\AA$]
}
\startdata
2MASS J05375161-0235257 & 0.10 & 0.02 & Y & -4.5$\pm$0.5$^1$\\
2MASS J05375398-0249545 & 1.95 & 0.48 & Y & - \\
2MASS J05380107-0245379 & 0.41 & 0.10 & Y & - \\
2MASS J05380826-0235562 & 0.29 & 0.08 & Y & -27.43$\pm$2.36$^2$ \\
2MASS J05380994-0251377 & 0.16 & 0.04 & Y & - \\
2MASS J05381315-0245509 & 0.13 & 0.03 & Y & - \\
2MASS J05382050-0234089 & 0.61 & 0.12 & Y & -28.0$\pm$4.0$^3$\\
2MASS J05382307-0236493 & 0.07 & 0.01 & M & - \\
2MASS J05382543-0242412 & 0.55 & 0.16 & Y & -260$\pm$30$^4$\\
2MASS J05382725-0245096 & 0.83 & 0.23 & Y & -53.5$\pm$9.0$^3$\\
2MASS J05382774-0243009 & 0.13 & 0.04 & Y & -5.02$\pm$0.30$^2$\\
2MASS J05383141-0236338 & 0.19 & 0.04 & Y & -197.57$\pm$11.64$^2$\\
2MASS J05383157-0235148 & 0.13 & 0.04 & Y & -10.18$\pm$0.92$^2$\\
2MASS J05383388-0245078 & 0.29 & 0.06 & M & - \\
2MASS J05383460-0241087 & 0.18 & 0.04 & Y & - \\
2MASS J05383902-0245321 & 0.64 & 0.15 & Y & -10.63$\pm$0.65$^2$\\
2MASS J05383922-0253084 & 0.06 & 0.01 & M & - \\
2MASS J05385542-0241208 & 0.87 & 0.19 & Y & -190$\pm$20$^1$\\
2MASS J05385922-0233514 & 0.82 & 0.17 & Y & - \\
2MASS J05385946-0242198 & 0.05 & 0.01 & M$^1$ & - \\
2MASS J05390193-0235029 & 0.93 & 0.28 & Y & -72$\pm4$$^1$\\
2MASS J05390276-0229558 & 0.10 & 0.02 & Y & -4.45$\pm$0.27$^2$\\
2MASS J05390357-0246269 & 0.10 & 0.03 & Y & - \\
2MASS J05390458-0241493 & 1.00 & 0.20 & Y & - \\
2MASS J05390540-0232303 & 0.10 & 0.02 & Y & -0.94$\pm$0.05$^2$\\
2MASS J05390760-0232391 & 0.61 & 0.17 & Y & -13.19$\pm$1.38$^2$\\
2MASS J05390878-0231115 & 0.73 & 0.18 & Y & - \\
2MASS J05391151-0231065 & 0.55 & 0.13 & Y & -25.76$\pm$0.79$^2$\\
2MASS J05392307-0228112 & 0.12 & 0.02 & M$^2$ & - \\
2MASS J05392519-0238220 & 0.55 & 0.14 & Y & -40.03$\pm$2.80$^2$ \\
2MASS J05392677-0242583 & 0.93 & 0.28 & Y & - \\
2MASS J05393982-0231217 & 0.53 & 0.15 & Y & - \\
2MASS J05393982-0233159 & 1.72 & 0.41 & Y & - \\
2MASS J05393998-0243097 & 0.34 & 0.09 & Y & - \\
2MASS J05394318-0232433 & 0.38 & 0.09 & Y & - \\
2MASS J05394784-0232248 & 0.17 & 0.04 & M & - \\
2MASS J05394891-0229110 & 0.08 & 0.01 & M & - \\
2MASS J05395248-0232023 & 0.05 & 0.01 & M & - \\
2MASS J05395362-0233426 & 0.17 & 0.04 & Y & - \\
2MASS J05400525-0230522 & 0.16 & 0.03 & Y & -20.5$\pm$6.0$^3$\\
2MASS J05400867-0232432 & 0.05 & 0.02 & M & - \\
2MASS J05400889-0233336 & 0.97 & 0.28 & M & - \\
\enddata
\tablecomments{We list the key features of our aperiodic variables detected in the $I$ band. Membership and 
H$\alpha$ values were determined by other groups; thus H$\alpha$ measurements are {\em not} simultaneous with 
our photometric data. Membership criteria are the same as in Table~3, with ``Y'' for definitive $\sigma$~Ori 
members and ``M'' for possible members (no non-members exhibited high-RMS light curve fluctuations). The two 
objects with numbered notes represent new candidate cluster members based on their position in the 
color-magnitude diagram and light curve RMS indicative of variability. Their magnitudes are $^1I\sim$12.6 
(2MASS J05385946-0242198) and $^2I\sim$12.9 (2MASS J05392307-0228112). References are as follows: 
$^1$\citet{2008A&A...491..515C}, $^2$\citet{2008A&A...488..167S}, $^3$\citet{2002A&A...384..937Z}, 
$^4$\citet{2007A&A...470..903C}} 
\end{deluxetable}

\end{document}